\documentclass[acmsmall]{acmart}

\usepackage{todonotes}
\usepackage{enumitem}
\usepackage{multirow}
\usepackage{subcaption}

\title{How Consistent Are Humans When Grading Programming Assignments?}
\author{Marcus Messer}
\orcid{0000-0001-5915-9153}
\affiliation{\department{Department of Informatics} \institution{King's College London} \city{London} \country{UK}}
\email{marcus.messer@kcl.ac.uk}

\author{Neil C. C. Brown}
\orcid{0000-0001-6086-2479}
\affiliation{\department{Department of Informatics} \institution{King's College London} \city{London} \country{UK}}
\email{neil.c.c.brown@kcl.ac.uk}

\author{Michael K\"olling}
\orcid{0000-0003-0544-2003}
\affiliation{\department{Department of Informatics} \institution{King's College London} \city{London} \country{UK}}
\email{michael.kolling@kcl.ac.uk}

\author{Miaojing Shi}
\orcid{0000-0002-4933-0073}
\affiliation{\department{College of Electronic and Information Engineering} \institution{Tongji University} \city{Shanghai} \country{China}}
\email{mshi@tongji.edu.cn}

\ccsdesc[500]{Social and professional topics~Student assessment}
\ccsdesc[300]{General and reference~Empirical studies}

\begin{document}

\begin{abstract}
Providing consistent summative assessment to students is important, as the grades they are awarded affect their progression through university and future career prospects. 
While small cohorts are typically assessed by a single assessor, such as the module/class leader, larger cohorts are often assessed by multiple assessors, typically teaching assistants, which increases the risk of inconsistent grading.

To investigate the consistency of human grading of programming assignments, we asked 28 participants to each grade 40 CS1 introductory Java assignments, providing grades and feedback for correctness, code elegance, readability and documentation; the 40 assignments were split into two batches of 20. 
The 28 participants were divided into seven groups of four (where each group graded the same 40 assignments) to allow us to investigate the consistency of a group of assessors. 
In the second batch of 20, we duplicated one assignment from the first to analyse the internal consistency of individual assessors.

We measured the inter-rater reliability of the groups using Krippendorff's $\alpha$ -- an $\alpha > 0.667$ is recommended to make tentative conclusions based on the rating.
Our groups were inconsistent, with an average $\alpha = 0.2$ when grading correctness and an average $\alpha < 0.1$ for code elegance, readability and documentation. 

To measure the individual consistency of graders, we measured the distance between the grades they awarded for the duplicated assignment in batch one and batch two. 
Only one participant of the 22 who didn't notice that the assignment was a duplicate was awarded the same grade for correctness, code elegance, readability and documentation. 
The average grade difference was 1.79 for correctness and less than 1.6 for code elegance, readability and documentation.

Our results show that human graders in our study can not agree on the grade to give a piece of student work and are often individually inconsistent, suggesting that the idea of a ``gold standard'' of human grading might be flawed. This highlights that a shared rubric alone is not enough to ensure consistency, and other aspects such as assessor training and alternative grading practices should be explored to improve the consistency of human grading further when grading programming assignments.
\end{abstract}

\maketitle

\keywords{Computer Science Education; Assessment; Programming Assessment; Consistency}

\section{Introduction}
As many computer science cohorts are becoming larger~\cite{Liao2017,Lehman2021,Brodley2022}, delivering high-quality grades and feedback for student assignments within a timeframe that maximises learning is becoming increasingly difficult~\cite{Eicher2021,Kay2022}. 
Promptly providing consistently applied assessment criteria is critical to maintaining student satisfaction \citep{Kane2008}.

Summative assessment, where the student is evaluated on what they have learnt during the semester or chapter \cite{Glazer2014}, majorly impacts the students' outcomes throughout and after their degree \cite{Knight2002}. 
During their degree, their grades in the previous year's courses can determine the students' progression to the next year. 
Students' final degree classification or grades are often factors in their employment prospects outside of education~\cite{Stepanova2021, Knight2002}. 
Having processes to ensure consistent or fair marking is crucial to maintaining the perceived value of summative grades and students' perceptions of justice\cite{Nesbit2006}.

While small cohorts are often assessed by a single assessor, such as the module/class leader, larger cohorts require multiple assessors, typically graduate teaching assistants, to provide grades and meaningful feedback in a short timeframe. 
These assessors often use a rubric comprising the evaluation criteria, quality definitions, and a scoring strategy \cite{Reddy2010} to grade and provide meaningful feedback on the students' work. 
The rubrics can be written at varying granularities, from generic criteria encompassing the final grades, to individual rubric items allocating individual marks. 
However, using multiple assessors can lead to issues with how consistently the rubric is applied \cite{Mustapha2016}, especially for subjective elements of an assignment like code quality. 
Though rubrics provide a scoring strategy for teaching assistants, many experience uncertainty when assessing, even those who are experienced teaching assistants \cite{Riese2018}.

Instead of using multiple graders to assess large cohorts, many large programming courses often use automated assessment tools to grade their assignments. 
While the automated approaches can provide a consistent grade based on provided criteria, such as unit tests, they often introduce other issues. 
For example, using automated assessment tools that implement tests requires a stringent structure and a well-tested and developed test suite to produce grades accurately \cite{Messer2024a}. 
These test suite approaches often take time to perfect and cannot search for additional marks for students on the grade boundary or grade those that submit uncompilable code.  Nor can they grade issues such as code structure or readability.

Neither human grading nor automated assessment tools are infallible. 
As the grades awarded are of such high importance, many higher education institutions have policies or practices to minimise the impact of mistakes within the assessment. 
Examples of these policies or practices in our country and our institution specifically include: 
\begin{itemize}
    \item Assessment boards and sub-boards often oversee all aspects of assessment, from the design of the exam or coursework to the marking scheme to the awarded marks.
    \item Second marking is often used to validate a single marker's work and can be done blind, without the grades and feedback from the first marker, or they can be provided with the grades and feedback from the first marker.
    \item Scaling marks, especially if a new exam or coursework has performed significantly lower than previous course iterations.
    \item Stepped marking is where assessors can only select certain marks steps, such as for a first, 72, 75, or 78 can be selected, limiting issues with students getting final marks at the classification bounds.
    \item Student appeals is where the students can request a regrade request on coursework if they disagree with a grade or feedback they have been given.
\end{itemize}

Our study aims to investigate the consistency of human graders when assessing a CS1 programming assignment, including how consistently the rubric is applied between multiple and individual graders, by answering the following research questions:
\begin{enumerate}[label={\textbf{RQ\arabic*}}, align=left]
    \item How consistently does a group of typical graders apply a grading rubric? Specifically, what is the inter-rater agreement of multiple assessors grading the same submissions?
    \item Are individual graders consistent at applying a grading rubric? Specifically, do they provide the same marks for the same work when marked a second time?
\end{enumerate}

These questions will have implications for CS1 courses: how consistent is the grading on large courses with multiple markers; how consistent are single graders (relevant to small courses); and how consistent is human grading in general, given that many course leaders will contemplate the use of automated or AI-driven grading?

We pre-registered our study on the Open Science Foundation before conducting the study~\cite{Messer2024c}. 
Section \ref{sect:related_work} discusses existing work on consistency of grading practices and grading with groups of assessors. 
Section \ref{sect:method} introduces our data collection and analysis methodology.
Section \ref{sect:results} presents the results from our pilot study and our full study, and Section \ref{sect:discussion} discusses our results and answers our research questions. 
We describe threats to validity in Section \ref{sect:threats} and conclude our results and discussion in Section \ref{sect:conclusion}.  Our contributions in this paper include: 

\begin{itemize}
    \item A new ``\anon{Menagerie}'' dataset: a publicly available graded dataset of students' assignments, available to the research community, and suitable as a data source for further studies concerning grading accuracy and consistency.
    \item An in-depth analysis of the consistency of awarded grades when utilising multiple graders.
    \item A comprehensive investigation into the awarded grades for individual graders when marked the same assessment a second time.
\end{itemize}

\section{Related Work}\label{sect:related_work}
\subsection{Consistency in Grading}
Applying assessment criteria consistently with detailed feedback is a key factor in student satisfaction \citep{Kane2008}. 
However, evaluating the consistency of grading and feedback is not often researched.
\citet{Migut2020} presented a poster showing their preliminary study on whether grading an online exam matches traditional paper grading and found a small but significant difference between online and paper grading.  
\citet{Borela2023} also presented a poster showing the outcome of a workshop to train TAs with a rubric, which showed a reduction in grade variance after the workshop. 

Both \citet{Ahoniemi2008} and \citet{Auvinen2011} developed rubric-based grading tools to improve grading consistency when grading large classes.
\citeauthor{Auvinen2011}'s tool, Rubyric, provides an interface for instructors to create rubrics and add feedback phrases. 
They conducted user satisfaction surveys and found that the six undergraduate teaching assistants who graded the assignments thought that Rubyric made them more consistent. 
Students tended to agree that the feedback was useful and detailed \cite{Auvinen2011}. 
Similarly, \citeauthor{Ahoniemi2008} developed a tool called ALOHA, which allows instructors to construct a rubric and add feedback phrases for common mistakes. 
They conducted both an instructor survey and a statistical evaluation of the tool. 
The survey found that the instructors found the semi-automated feedback phrasing particularly useful. 
The statistical analysis, which compared grades when graded with a paper rubric and those graded with the tool, found that those using the tool had no significant difference between the graders, and when grading without the tool, had a significant difference between the grades given by the most lenient grader and two stricter graders.

Evaluating the consistency of assessment has been conducted in numerous other disciplines. 
\citet{Bloxham2016} took a multidisciplinary approach and investigated 24 assessors from various disciplines: psychology, nursing, chemistry and history. 
Each assessor was given five assignments to grade, each on a typical task for their discipline. 
They used Kelly's repository grid exercise \citep{Fransella2004} to elicit constructs and grades by first comparing combinations of three assignments, stating which two are the same and how one is different, and applying a score between 1-5 for these constructs. 
This study aimed to determine the consistency of the individual constructs, and they found no consistency in the rankings of constructs within any of the subjects.

Within medical education, \citet{Dunbar2018} evaluated the consistency of nursing assessment. 
They ran a study asking seven educators to assess a reenacted physical examination in person and a recording of the same examination a month later. 
They found that the grading in person and the recording had an inter-rater agreement of approximately 84\% between the participants, with the inter-rater measurement calculated by percentage agreement.
\citet{McManus2006} explored the variance of examiner leniency and stringency within clinical examination. 
They used three years of exam grades, approximately 10,000 candidates, each graded by two examiners, and applied the Rasch model, a latent trait model used to determine the probability of a person succeeding on an item \cite{Wright1977}. 
Their results suggest that greater examiner stringency was associated with greater examiner experience. 

\citet{Henderson2004} examined the grading practices of 30 assessors when grading exams. 
The participants, who were from multiple institutions, were provided with five answers to an introductory calculus exam, with two answers being of interest for investigating the conflict between valuing reasoning and correctness. 
Overall, the scoring of these two solutions differed significantly among individual instructors, with 12 instructors awarding higher grades for one solution than the other, 13 awarding lower grades, and 5 awarding equal grades for both submissions.

\citet{Willey2010a}, investigated methods to ensure marking consistency when using multiple tutors to grade a second-year engineering course in design fundamentals. 
Each tutor was asked to grade their session tutorial of 32 students, and to achieve a consistent standard of marking between different tutors, they used double-blind marking and remarking. 
As part of the grading process, the tutors were arranged into small groups to discuss and regrade the assignments collaboratively. 
The course coordinator also provided two benchmark assignments from previous years that the tutors subsequently graded using specified criteria, effectively reducing the variability in grading between different tutors. 
While the study found only small variations between markings from different tutors, students still complained of a perceived lack of consistency, specifically being exacerbated by inconsistencies in how different tutors delivered feedback.

\citet{Hicks2017} examined the consistency of teaching assistants when grading a first-year engineering course's engineering MATLAB assignments. 
The course was split into 15 sections, with five undergraduate and one graduate teaching assistant, where the undergraduate teaching assistants graded the homework assignments while supervised by the graduate teaching assistant. 
The undergraduate teaching assistants were asked to grade against a rubric, with most criteria having four achievement levels: no evidence, under-achieved, partially achieved and fully achieved. 
As a baseline grade, the course leader graded 172 submissions. 
The authors utilised percentage agreement, Cohen's kappa and Krippendorff's alpha to measure the consensus and consistency between the course leader and the undergraduate teaching assistants. 
They found that the course leaders and undergraduate teaching assistants had an agreement of 49.4\% and that all metrics indicated that the undergraduate teaching assistants were applying the same interpretations of the rubrics that the course leaders intended.

\citet{Passonneau2023} as part of their investigation into whether analytic grading rubrics, which have more specifics than the rubrics provided to the students to define expectations, produce reliable assessments, they analyse the inter-rater reliability of four undergraduate teaching assistants when grading physics lab reports. 
They found that grades administered by the teaching assistants did not correlate to posthoc assessments from trained raters, identified missed learning opportunities for students and that instructors were misled about student progress.

While investigating the consistency of the grading process is conducted within other domains, from the best of our knowledge, this research is rarely conducted within computer science education. 
Multiple and individual graders consistently applying a grading rubric is critical in providing a fair and meaningful assessment of students' work. 
In turn, it improves a student's learning and overall satisfaction with the course.

\subsection{Rubric Design}\label{sect:related_rubric}
Rubrics are often used to make assessment expectations transparent \cite{Brookhart2018}, and they describe desirable qualities and common pitfalls in student work against a gradation of quality \cite{Andrade2005}. 
A rubric has three essential features: the evaluation criteria, the definitions of quality and a scoring strategy \cite{Reddy2010}. 

A review by \citet{Reddy2010} discussed rubric use in higher education. They present 20 articles on rubrics and include topics on student and instructor perceptions, use and academic performance, and rubrics' validity and reliability. They found that instructors can have contrasting perceptions of using rubrics, with some believing that rubrics provide an objective basis for evaluation, while others resist integrating rubrics into their assessment. 
The review also investigated how a well-designed scoring rubric should alleviate inconsistencies in the grading process, both when rubrics are used by a single grader or a group of grader. A well-designed rubric should minimise errors due to grader training, grader feedback, and clarity in the criteria description. They further discuss how there is ample evidence of disagreement between assessors when using rubrics and stress the importance of grader training.

\citet{Jonsson2007} reviewed the reliability, validity and educational consequences of using rubrics in assessment. 
They analysed 75 studies and found that rubrics can improve the reliability of assessment if they are analytic, topic-specific, and complemented with exemplars or rater training, especially with open-ended assessment, as it is not always possible to limit the format of the assessment to achieve high levels of reliability without sacrificing the validity.

\citet{Rinne2024}'s recent study aimed to identify sources of inconsistency in assessing undergraduate dissertations from a primary school teacher education programme. 
They identify that rubrics are often short and open to interpretation, resulting in inconsistent assessment, and discuss sources of inconsistency when grading, which include examiners' interpretations of quality, and found that the same grade could be awarded for different reasons, and the same judgement could result in different grades.

To increase the agreement in teachers' grading, \citet{Jonsson2021} conducted a study into whether analytic or holistic rubrics enabled consistent assessment. 
Analytic assessment involves assessing different aspects of student performance, such as mechanics, grammar, style and organisation. On the other hand, holistic assessment means making an overall assessment that considers all criteria simultaneously \cite{Jonsson2021}.
\citeauthor{Jonsson2021}'s study investigated analytic and holistic grading in both an English as a foreign language (EFL) and a mathematics high-school course. They found that analytic grading is slightly preferable to holistic assessment regarding teacher agreement for both EFL and mathematics. They further conclude that the results support the argument that teachers' grading is `complex, intuitive and tacit' \cite{Bloxham2016} and that any attempts to achieve grading consistency are likely futile. However, their results also indicate a small step towards increasing agreement.

In addition to the active area of research investigating how to design rubrics to maximise consistency while maintaining validity, within computer science, defining rubrics for aspects of programming, especially code quality, is still a developing field.
\citet{Stegeman2016} defined a code quality rubric for introductory programming, including names, comments, formatting and expressions. They tested this rubric against several programming assignments and found that no substantial parts of code quality appeared to be missing. 
Recently, \citet{Kirk2024}, motivated by the difficulty for programming educators to find appropriate guidance on teaching style concepts, have defined a literature-informed model for code style, which includes many of the same topics introduced by \citeauthor{Stegeman2016} but provides further depth and rationale behind the elements.  The model provides an abstract level of the main style aspects that map to understanding and changing code and include principles that relate to language, layout, constructs, design, redundancy, implementation, repetition and structure.

\subsection{Teaching Assistants and Assessment}\label{sec:TAs}
Teaching assistants are often crucial in supporting students' learning; they typically consist of undergraduate, master's and PhD students \cite{Riese2021, Wald2020, Riese2022}. 

The three most common duties for teaching assistants include assisting in programming labs, leading sections to reinforce concepts, and grading; many institutions utilise undergraduate and graduate teaching assistants \cite{Mirza2019, Willey2010a, Wald2020}.

\citet{Mirza2019} conducted a systematic literature review into how undergraduate teaching assistants are used within computer science courses and found that UTAs often grade assignments, whether they provide the final grades for the assignment, assign coarse primary grades or are one of the multiple undergraduate teaching assistants to grade each student assignment.

\citet{Riese2021} explored the challenges faced by teaching assistants in computer science education across Europe. 
At all the institutions involved in their analysis, teaching assistants were involved in assessment in some form, whether that be assigning points or pass/fail grades. 
They found that teaching assistants often faced challenges when assessing student work, including understanding the assessment criteria, providing meaningful feedback and failing students.

\citet{Wald2020} discuss the trade-offs of employing student teaching assistants in their position paper, including how they can reduce teaching loads for academics and provide graduate students with teaching experience and financial support. 
However, teaching experience can differ between teaching assistants, especially undergraduate teaching assistants, who typically have limited academic and pedagogical knowledge, which are both essential for student learning. 
They suggest limiting the role of graduate and late-stage undergraduate teaching assistants as graders, tutors, and demonstrators, as they have yet to find any sound pedagogical arguments supporting undergraduate teaching assistants as graders. Instead, they suggest that they facilitate discussion and review course material. 

\citet{Kristiansen2024} also highlight the importance of properly trained teaching assistants in their study comparing manual and automated feedback. 
They found in a controlled study of undergraduate students in a programming assignment that there was a need for both automated and manual feedback, providing superior results in task effectiveness and student preferences than just automated or manual assessment.

\citet{Riese2022} conducted a qualitative analysis into how course coordinators use assessment throughout their course and what they expect from TAs.
They found that the seven experienced course coordinators in CS1 courses typically use TAs to provide continuous assessments for large cohorts. 
They discuss how some TAs set the bar too high while others set it too low, which can lead to inconsistency and that some TAs can view their students as friends, leading to ethical dilemmas. 
Similarly, to \citeauthor{Kristiansen2024}, \citeauthor{Riese2022} suggest that not all tasks are suitable for TAs and that assessing students' competency and knowledge can be difficult for experienced teachers.

\subsection{Educational Datasets}
For our study, we required a dataset of student submissions for the grader participants to grade. 
While several datasets of students' programming assignments exist, many are not publicly accessible \cite{Messer2024a}, consist of small-scale programs \cite{Freitas2023, Azcona2020} or do not capture contextual details \cite{Brown2014}.

Some project-scale datasets exist, such as the Scratch website scenarios\footnote{Scratch Scenarios: \url{https://scratch.mit.edu/ideas}} or the Blackbox dataset~\cite{Brown2014} that are public or available on request. 
However, they have the disadvantage that the context of the users is unknown (they may be students at any stage of education or not even students at all), and the aim of their project is unknown, preventing judgment against any criteria for progress or success.

In smaller-scale datasets, including FalconCode \cite{Freitas2023}, Hour of Code\footnote{Hour Of Code: \url{https://code.org/research}}, Dublin City University's Programming Submission Dataset \cite{Azcona2020}, and online judge tools, such as HackerRank\footnote{HackerRank: \url{https://www.hackerrank.com/}}, the task is known. 
However, all of these datasets are closed-ended and focus on small-scale tasks, including highly structured template code, specific names of classes and functions defined in the requirements, or simple one-function tasks, such as \texttt{FizzBuzz} or the Rainfall Problem \cite{Soloway1986}. 
Therefore, we would not be able to evaluate the graders consistently when evaluating aspects of code quality, as many of the students' decisions regarding code design are already made for them, and many of these datasets use documentation to explain the task.

To the best of our knowledge, no large project-scale publicly available datasets exist. 
Therefore, we developed our dataset, \anon{Menagerie}. 
It consists of real CS1 Java programming submissions for a simulation-based assignment and the associated grades and feedback that arose during this study; more details can be found in Section \ref{sect:dataset}.

\section{Research Context}
\subsection{The \anon{Menagerie} Dataset}\label{sect:dataset}
As discussed in Section \ref{sect:related_work}, most publicly available programming datasets do not contain detailed grading or feedback data. 
If they contain grades, they are often limited to whether the program passes or fails a set of automated tests, as most automated assessment tools utilise unit testing to assess the submissions \cite{Messer2024a}. 
As such, we developed our own publicly available programming dataset containing human graders' assessments.

We introduce the \anon{Menagerie} dataset, which comprises 667 real student submissions for \anon{King's College London}'s second semester CS1 Introduction to Programming Java course and is available publicly, see Section \ref{sect:data_ava}. 
The same assignment was given to students between 2017 and 2021 inclusive, with minor changes each year. 
The changes each year included updating the assignment instructions to clarify points of misunderstanding from the previous years' assignments, with the tasks and the provided template code remaining the same year on year. 

The dataset consists of the following:
\begin{itemize}
    \item The assignment requirements and grading rubric.
    \item The project's starter code.
    \item The students' Java files.
    \item Grades and associated feedback for 272 submissions (produced by the current study).
\end{itemize}

The dataset does not include the students' original awarded grades, as we did not receive ethical approval to release them publicly. 
However, as part of the current study, we asked seven groups of four to grade and give feedback on 40 submissions per group. We produced grades and feedback for 272 submissions; seven assignments were duplicated to evaluate grader self-consistency.
Details about the grades and feedback can be found in Section \ref{sect:grades}. 

We have received ethical approval from our institution\footnote{Ethics review reference: LRS/DP-22/23-35578} to conduct this study and publicly release the summary of the grader demographics and the individual grades and feedback from each grader.

\subsubsection{Assignment Outline}\label{sect:assignment}
\begin{figure}
    \centering
    \includegraphics[width=0.5\linewidth]{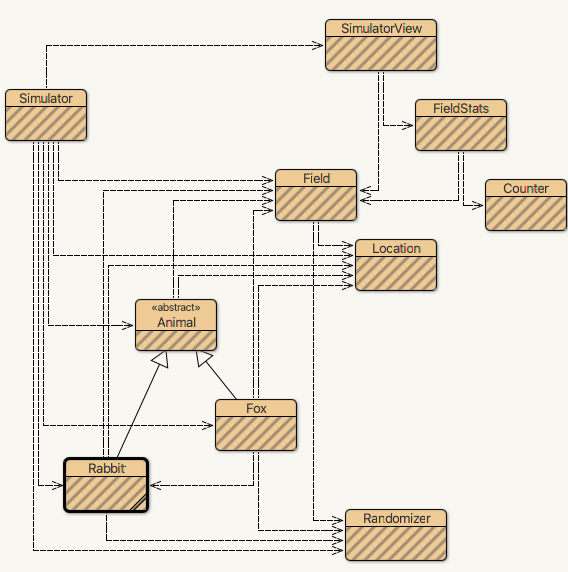}
    \caption{The UML diagram of the provided template code.}
    \label{fig:template_url}
\end{figure}

The assignment was a small-group, open-ended paired programming assignment to utilise object-oriented programming concepts to develop a predator/prey simulator with groups of two or three. 
The students were asked to conduct pair programming with explicit instructions on how to pair program. 
One student was the ``driver'', who wrote the code, and the other was the ``navigator'', who reviewed the code as it was written, and they were instructed to switch roles throughout the assignment.
The students were provided with a template project based on the ``foxes-and-rabbit'' project from the sixth edition of \citet{Barnes2006}'s book on ``Objects First with Java''.
The template includes a graphical user interface (GUI), a Field class, which contains a two-dimensional array for the simulation environment, and two animals, a Fox and a Rabbit;  \autoref{fig:template_url} shows the class diagram of the template code. 
Students were asked to extend the template code with the following base tasks:
\begin{itemize}
    \item The simulation should have at least five species, with at least two being predators and at least two not being predators.
    \item At least two predators should compete for the same food source.
    \item Some or all species should distinguish between male and female animals and only propagate when male and female species are in a neighbouring cell of the two-dimensional array.
    \item The simulation should keep track of the time of the day, and some species should exhibit different behaviours at some time of the day.
\end{itemize}

After completing the base tasks, the students were asked to implement one or more challenge tasks. 
The students could choose to invent their own tasks or to use one or more of the following suggestions:
\begin{itemize}
    \item Simulate the lifecycle of plants, including growth and being a food source for at least one animal.
    \item Simulate changing weather states and how they affect other simulation aspects.
    \item Simulate disease within the species, including the spread of the disease.
\end{itemize}

During the assignment, students could choose which species they implemented, how those interacted with other species, and how their implementation completed the provided tasks. 
These opportunities to decide what they implemented allowed students to apply their creativity, choose how they designed their code, implement object-oriented principles, and name classes and functions. 
The ability of the students to design their solution to a problem and the open-ended nature of the assignment allows instructors to assess a wider variety of skills than they typically could in a close-ended small-scale assignment like the Rainfall problem \cite{Soloway1986,Fisler2014}.
In this assignment, the students were assessed on the correctness of the implementation, how elegantly they implemented the solution, the overall readability of the source code, and how well they documented their code. 
A detailed example submission and the details on the data processing pipeline for cleaning the dataset can be found in Appendix \ref{appx:Menagerie}.
\subsubsection{Dataset Limitations}
The students were asked to submit a report discussing their implementation as part of the assignments. 
However, we have excluded these from the dataset, and they were not used to grade the assignments, as they can not be suitably anonymised for public release. 
While we automated the de-duplication of the group submissions and removed further duplicate submissions during the anonymisation, there is a remote chance that some repeated assignments within the dataset could still exist.

While the students were given instructions on how to pair program, the assessment did not capture if and how the students conducted pair programming. 
Anecdotally, some groups started pair programming, especially in the associated lab sessions. 
However, as the submission deadline approached, groups tended to separate the work, with both students writing code in parallel.

As these are historical assignments that we received \textit{post hoc} permission to release, the students' demographics were not captured. 
However, all students were in their first year at \anon{King's College London} and were a mixture of domestic and international students.

\subsection{Assessment}\label{sect:grades}
The participants graded the submissions from the \anon{Menagerie} dataset, as discussed in Section \ref{sect:dataset}, and used the rubric from the course as guidelines on how to assign grades, discussed in Section \ref{sect:rubric} and provided in Appendix \ref{appx:rubric}.
The grades and feedback were captured using Gradescope~\citep{Singh2017}, a commercial grading platform designed for STEM assignments, including programming assessments, and allows for rubrics to be uploaded and for graders to provide line-level or overall feedback. 
While we captured the feedback, which is available in the dataset, this study focuses on the grades provided.

\subsection{Rubric}\label{sect:rubric}
The rubric we provided participants can be found in Appendix \ref{appx:rubric}. This is the original rubric from the assignment from which the submissions were captured. 
The participants were asked to supply letter grades between A and F, with `+' and `-' grades for all letters; for exceptional submissions, participants could award A++. The rubric in Appendix \ref{appx:rubric} includes number bounds for each letter; this is due to Gradescope only supporting numerical grades. Within Gradescope, we supplied options that corresponded to the letter grades, with `+' and `-'. The numbers were not used for this study, and participants were instructed to select letter grades in Gradescope.

Participants were asked to grade the following programming skills:
\begin{itemize}
    \item Correctness -- Assess if a student has comprehended and completed the assignments in a way that complies with the assignment requirements \cite{Gerdes2010}.
    \item Code Elegance -- Investigate the overall code design. Have the students use Java design principles correctly, including inheritance and polymorphism, methods to reduce duplicated code, and their overall approach to implementing the required features \cite{Borstler2018, Kirk2024}.
    \item Readability -- Evaluate whether a student's submission is easy to understand. This can include code style, whether formatting, indentation and naming conventions \cite{Borstler2018, Kirk2024}.
    \item Documentation -- Assess if and how well students have documented their code, including inline comments and the natural language code summary (docstring) \cite{Borstler2018, Rai2022, Kirk2024}. 
\end{itemize}

While code elegance, readability, and documentation are all aspects of code quality, we asked participants to provide grades for each aspect, allowing for richer data capture of different aspects of code quality. 
Similar aspects have been utilised in multiple different studies \cite{Messer2024a, Neutens2022}. 
Typically, these three aspects of code quality and correctness would be aggregated into a single final grade; we chose not to aggregate the grades, as different assignments will provide different weightings on these aspects. Thus, we analyse the consistency of grading code elegance, readability, documentation, and correctness, individually. We leave the methods of combining the grades for educators, but it does not affect this investigation into the consistency of grading the constituent components.

This rubric consists of the three essential features of a rubric, as discussed in Section \ref{sect:related_rubric}, including the evaluation criteria, the definitions of quality, and a scoring strategy. While some improvements could be made to the clarity of the definitions of quality and the scoring strategy, the rubric we provided our participants was the original rubric from the course, which provided the following:
\begin{itemize}
    \item A typical representation of a rubric, which has seen many years of use at \anon{King's College London}.
    \item The same rubric that historical students provided alongside the assignment requirements guided the design and development of their submissions. 
\end{itemize}

\section{Methodology}\label{sect:method}
\subsection{Group Consistency (RQ1)}\label{sect:inter-rater-method}
\begin{figure}[b]
    \centering
    \includegraphics[width=0.6\textwidth]{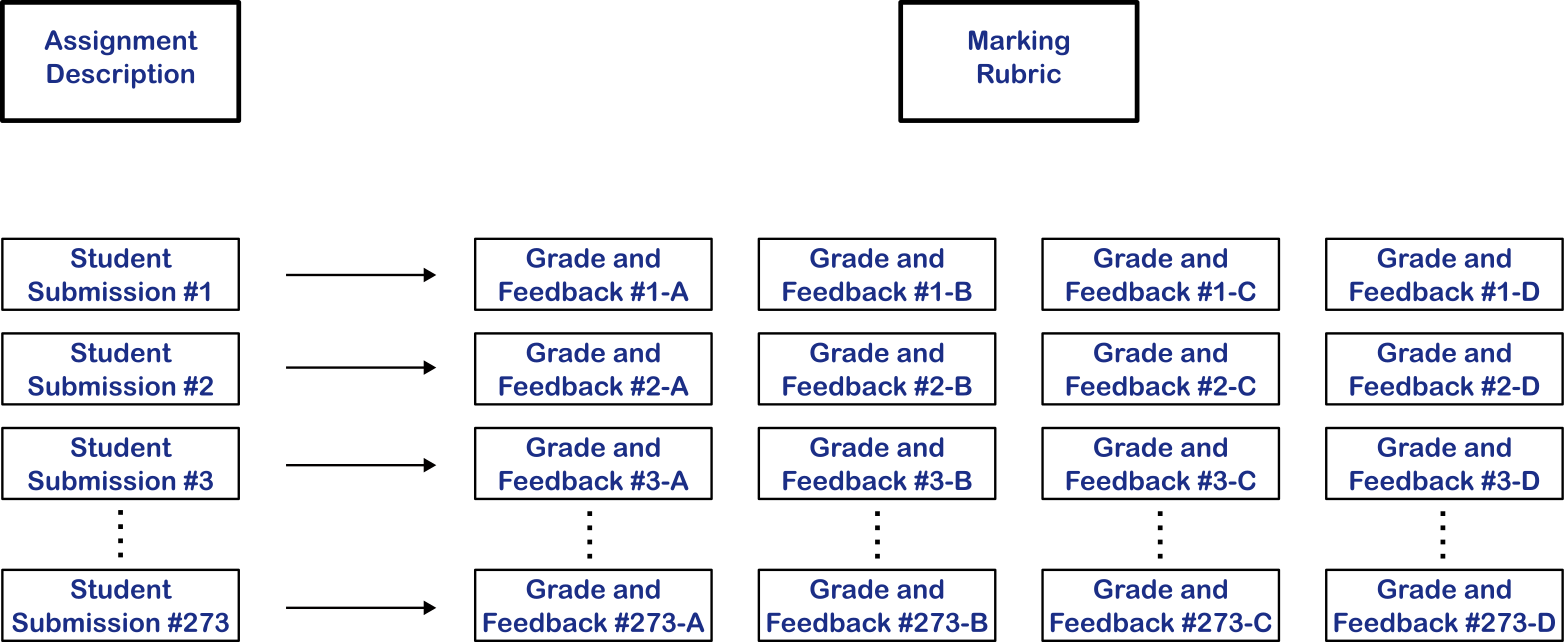}
    \caption{The data includes an assignment description, a marking rubric, a set of student submissions attempting the assignment, plus a set of grades and feedback from multiple human markers (A--D here, although more markers are involved) for each student submission.}
    \label{fig:grading}
\end{figure}

To explore the consistency between graders, each assignment was graded four times, each time by a different grader. 
Figure \ref{fig:grading} shows an example of what the data includes after the grading is complete. 
Grading each assignment multiple times by different graders allowed us to investigate how consistently a group of graders apply the grading rubric to individual assignments.

To investigate how consistently a group of graders apply a grading rubric, we used Krippendorff's $\alpha$ (implemented using the Fast Krippendorff library~\cite{castro-2017-fast-Krippendorff}) to assess the reliability of agreement between the graders on the awarded grades. 
Krippendorff's $\alpha$ is a percent chance-based reliability coefficient, similar to Cohen's $\kappa$ and Fleiss' $\kappa$. 
It evaluates the deviation from perfect reliability by the proportion of observed to expected disagreement and allows any number of raters and categories while allowing for missing data \cite{Krippendorff2013}. 
It also has several weightings that can be applied to different types of data, including ordinal, which is data that is both categorical and has an order, such as letter grades. 

As many percent chance agreement-based coefficients, including Krippendorff's $\alpha$  and Fleiss' $\kappa$, underestimate the agreement when one of the labels is dominant \cite{Gwet2008,Doughty2024, Wongpakaran2013}, we also compute Gwet's $AC_2$, which is a weighted variant to Gwet's $AC_1$, which was designed to be robust in the high prevalence of a single class \cite{Gwet2008}. 
Gwet's $AC_2$ is more robust in instances of high prevalence, as it removes from consideration all agreements that occurred by chance \cite{Gwet2008}. 

As we planned, each assignment was graded by four of the graders rather than all graders, making Krippendorff's $\alpha$ and Gwet's $AC_2$ the ideal measures. 
Using Krippendorff's $\alpha$ and Gwet's $AC_2$ varies from our pre-registration, as our original choice of metric (Fliess' $\kappa$) required a complete set of grades and did not consider the ordinal nature of grading, so we realised that this was not the best choice. 

\subsection{Individual Consistency (RQ2)}\label{sect:intra-rater-method}
To aid in the research into self-consistency and the study pacing, we supplied the graders with batches of submissions, allowing us to repeat a few chosen submissions for the graders to regrade, resulting in some submissions being graded eight times. 
We chose to repeat assignments representative of the dataset, specifically assignments of the median size, a common theme, and submissions that fall near the mean of the grades awarded within the previous batch. 
The original assignment for \anon{Menagerie} was a pair-programming assignment, which was de-duplicated; however, duplicates could still exist and we told our graders this as a cover story for why they might see multiple similar assignments. 
The possibility of duplicates, specifically choosing assignments and the time between batches, helped mitigate the likelihood of the graders noticing that they were grading the same assignment (which might have led to the graders going back to the first grading attempt and copying and pasting their grades instead of grading it fresh a second time). 

To evaluate the consistency of individual participants, we calculate the distance between the grade awarded in the first and second batches for each participant and for each skill. 
We could not use typical inter-rater analysis to evaluate self-consistency, as the two grades are too few for inter-rater reliability metrics, including Krippendorff's $\alpha$, Gwet's $AC_2$ and Fliess' $\kappa$. 
We chose to limit the number of repeated assignments to ensure that the participants were unaware that they were grading a duplicate and to maximise the number of annotations for the dataset.

To provide further information on the participants' experience grading these assignments, we conducted a semi-structured interview. 
These interviews were conducted in English by one of the members of the research team and lasted approximately 30 minutes; Microsoft Teams was used to provide automatic transcription. 
During the interview, the participants were asked the following questions:

\begin{enumerate}
    \item Can you tell me about your experience grading these assignments?
    \item Can you tell me about your grading process?
    \item How well do you think the rubric aided you when grading the assignment? 
    \item Did you review your grades to validate your consistency?
    \item How do you think the environment you were in while grading affected your grading?
    \item How do you think the mood you were in while grading affected your grading?
    \item How do you think the time of day you graded per session affected your grading?
    \item Were there any specific submissions that stuck in your mind?
    \item Did you notice any duplicates or near-duplicates in the data?
    \item Do you have any further comments on your experience?
\end{enumerate}

Unfortunately, many participants did not provide detailed answers for most of the questions asked, even when prompted for more information. 
This is likely due to the interviewer's lack of experience conducting semi-structured interviews. 
As such, we have decided not to provide an in-depth analysis of the semi-structured interviews and only use the answers from questions four and nine in our results. These questions were coded as ``yes'' or ``no'' by one member of the research team and used to provide further context to our analysis of participant's individual consistency.

\subsection{Pilot Study}
To validate our data collection process, we conducted a pilot study with two assessors: one who was a final-year computer science PhD student with numerous years of grading this specific assignment and a first-year computer science PhD student with one year of grading experience and had previously completed this assignment as an undergraduate. 
They were provided the pre-study survey, asked to grade two batches of twenty assignments each, and given two weeks to complete each batch, with a two-week break. 
After grading, both participants were asked to attend the post-study semi-structured interview, which we expanded for the pilot study to ask for any improvements that could be made for the full study.

The participants completed the grading within the supplied time frame and found Gradescope helpful when grading. 
However, they both struggled to remember to complete the grading diary and requested more information about the assignment to be supplied.

For the full study, we provided detailed instructions on using Gradescope, a brief overview of what the students had learned before submitting the assignment, the coursework specification, rubric, and template code.  We added a link to this information and the diary to each submission in Gradescope, so that participants were automatically reminded at the end of grading each submission.

\subsection{Participant Selection}
We conducted a pre-study survey to identify suitable grader participants and collect demographic data. 
Our criteria for our participants follow the typical demographics of teaching assistants performing grading in many institutions \cite{Mirza2019, Riese2021}; and are as follows:

\begin{itemize}
    \item Students at our institutions in their third year of undergraduate, a master's student or a PhD student.
    \item They identify as proficient in Java programming and have at least three years of programming experience.
\end{itemize}

All selected participants were remunerated £280 GBP (\$340 USD) for participating; this rate was concurrent with our institution's hourly rate for teaching assistants. We decided it was important that our participants receive pay similar to what they would receive when grading for a live assignment to ensure they take the study seriously and complete all their grading.

\subsection{Position Statement}
The researchers in this study are three academic staff members and a graduate student at a large UK university. 
The dataset is captured from one of the academic staff's courses, CS1 Java courses that ran between 2017 and 2021 inclusive, with permission from the college's ethics committee (LRS/DP-22/23-34272), the students' permission was not required for this data, and all data was validated by research assistants to ensure the data was anonymised, as prescribed by the college's data handling policy. 
The participants who took part in the study are current students at our institution but are not current students of any members of the research team. 
The consistency study was further approved by our college's ethics committee to use the existing dataset and to provide remuneration to our participants (LRS/DP-22/23-35578).

\section{Results}\label{sect:results}
\subsection{Participant Demographics}
Through our pre-study survey, we selected 28 participants who matched our criteria to complete the study. 25 of our participants had over five years of programming experience, with three having between three and five years, and all identified as proficient in Java. 
Seven participants were PhD students, 13 were Masters students, and eight were third-year undergraduate students. 
Two participants graded for between two and three years, one participant graded between one and two years, six participants graded for less than 1 year, and 21 participants had not been assessed before. 
These are a relatively typical profile of those recruited to act as graders at our institution, and as described in \autoref{sec:TAs}, typical of graders at many other institutions as well.

Most of our participants identified as male, with three identifying as female, which follows a similar distribution of students enrolled in the computer science courses at our institution. 

\begin{figure}
    \centering
    \includegraphics[width=0.7\textwidth]{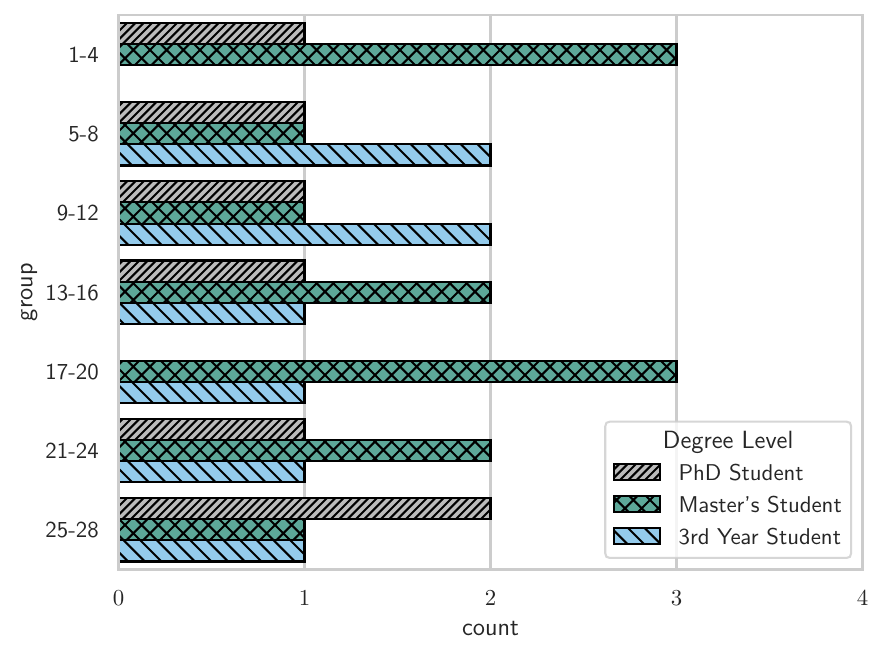}
    \caption{The degree level composition for each group of participants.}
    \label{fig:degree_level_by_group}
\end{figure}

Figure \ref{fig:degree_level_by_group} shows the degree level composition for each group of four. 
Nearly every group contained one or more PhD students, over half the groups contained two or more master's students, and most groups had one or two third-year students.

\subsection{Grade Overview}
\begin{figure}[h!]
    \centering
    \includegraphics[width=0.65\linewidth]{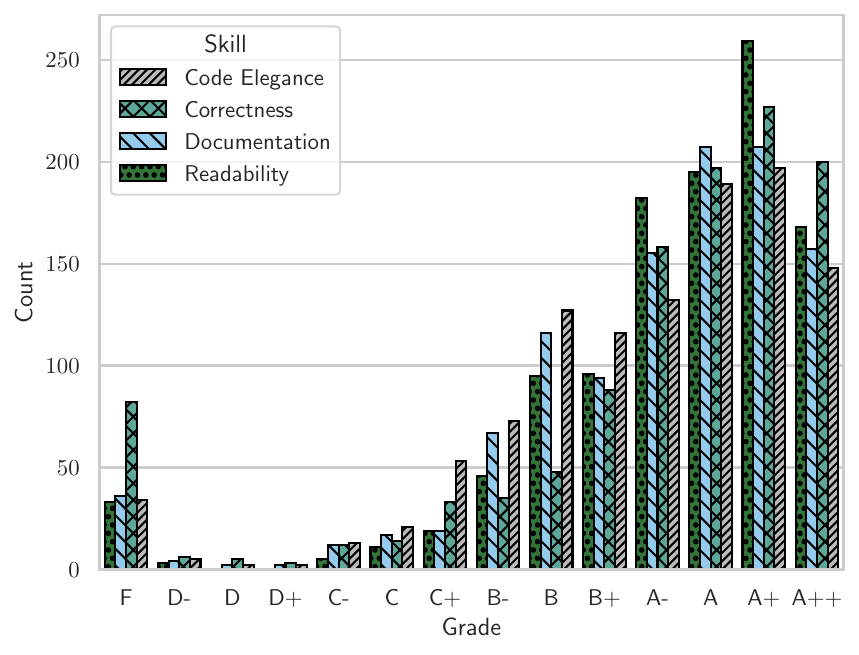}
    \caption{The count of awarded grades.}
    \label{fig:grade_count}
\end{figure}

Figure \ref{fig:grade_count} shows the grade distribution for each skill.
Most assignments achieved high grades, with approximately 10\% achieving below a B-.
Those graded F for correctness include assignments that failed to compile. 
Some participants failed to select a grade for certain skills within certain submissions; these have been omitted from our results. 
While there is not a lot of variance in the data, anecdotally this is a similar distribution of grades that we saw while conducting the original summative grading.

\subsection{Grading Consistency of Multiple Graders (RQ1)}
\subsubsection{Inter-Rater Metrics} 
Each batch of 20 assignments was graded by a group of four graders. 
We discuss our inter-rater reliability results, using Krippendorff's Alpha ($\alpha$) for each group of four graders and each skill, the spread of the awarded grades, and the reliability of individual grades. 

We opted to use the ordinal metric within Krippendorff's $\alpha$, as letter grades are inherently ordinal; they are categorical and have an order. 
The ordinal metric applies weights to the $\alpha$ calculation by utilising the ordinal difference between ranks \cite{Krippendorff2013}.

When using Krippendorff's alpha, if the raters completely agree, $\alpha = 1$; if there is no agreement, then $\alpha \leq 0$.
Krippendorff also suggests that $\alpha > 0.800$ is acceptable and $0.800 > \alpha \geq 0.667$ are only for drawing tentative conclusions \cite{Krippendorff2013}.

Due to an accidental duplicate when setting up the second batch, participants 21-24 graded assignment 105 twice, these grades have been omited from our results.

\begin{figure}[t]
\captionsetup{labelsep=space,justification=justified,singlelinecheck=off,margin=1ex}
\begin{minipage}[t]{0.35\textwidth}
     \centering
    \includegraphics[width=\linewidth]{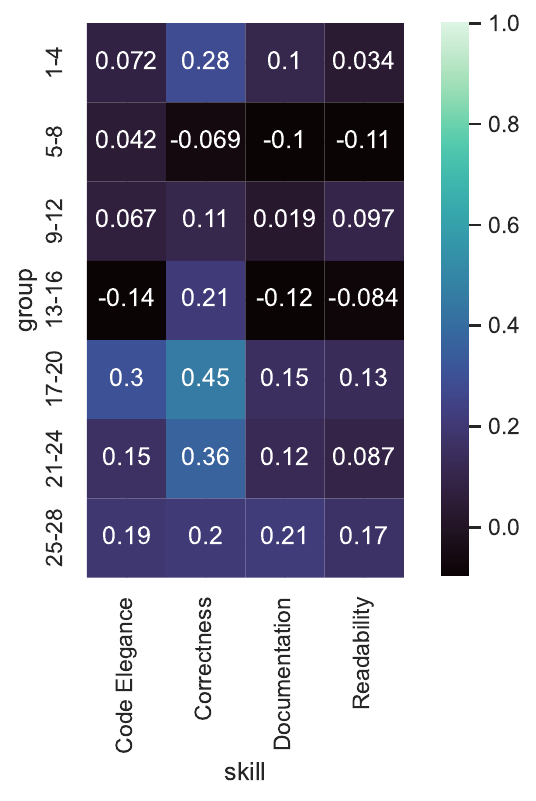} 
    \caption{This heatmap shows the consistency, measured by Krippendorff's Alpha ($\alpha$) with the ordinal metric, for each group of participants and each skill. Grades Awarded (A++, A+, A, A-, B+, B, B-, C+, C, C-, D+, D, D-, F, Not Graded)}
    \label{fig:ir_heatmap_fine}
\end{minipage}%
\begin{minipage}[t]{0.35\textwidth}
     \centering
    \includegraphics[width=\linewidth]{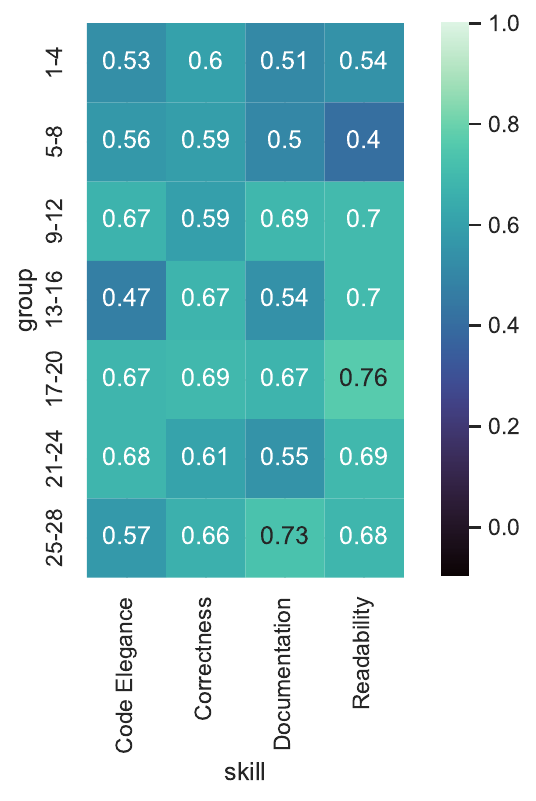} 
    \caption{This heatmap shows Gwet's $AC_2$ with the ordinal metric applied for each group of participants and each skill.}
    \label{fig:ac2_heatmap}
\end{minipage}
   
\end{figure}

Figure \ref{fig:ir_heatmap_fine} shows the $\alpha$ for each of the seven groups and each skill. 
These results show that in our study, consistency was low, but were slightly more likely to agree on the grades for correctness and code elegance but did not agree on the grades for readability or documentation, with the average $\alpha$ being 0.220, 0.099, 0.046, and 0.055, respectively.

Figure \ref{fig:ac2_heatmap} shows the results of Gwet's $AC_2$.
Following the same bounds defined by Krippendorff, no groups achieve acceptable agreement ($AC_2 > 0.800$) for any of the skills evaluated. 
However, some groups show partial agreement with $AC_2 \geq 0.667$ for specific skills, though on average, no skill had an $AC_2 \geq 0.667$, with the average $AC_2$ for correctness being 0.629, for code elegance being 0.592, for readability being 0.638 and for documentation being 0.596.

Out of the four skills that the participants graded, correctness was most likely to be graded more consistently, with five of the seven groups achieving a higher $\alpha$ than the other four skills, with groups 5-8 and 25-28 achieving the second most consistently graded skill for their respective groups. 
In comparison, all participant groups were similarly inconsistent when grading code elegance, readability and documentation, all skills achieving an average $\alpha < 0.1$.

\subsubsection{Grading Experience} 
\begin{figure}
     \centering 
    \includegraphics[width=0.55\textwidth]{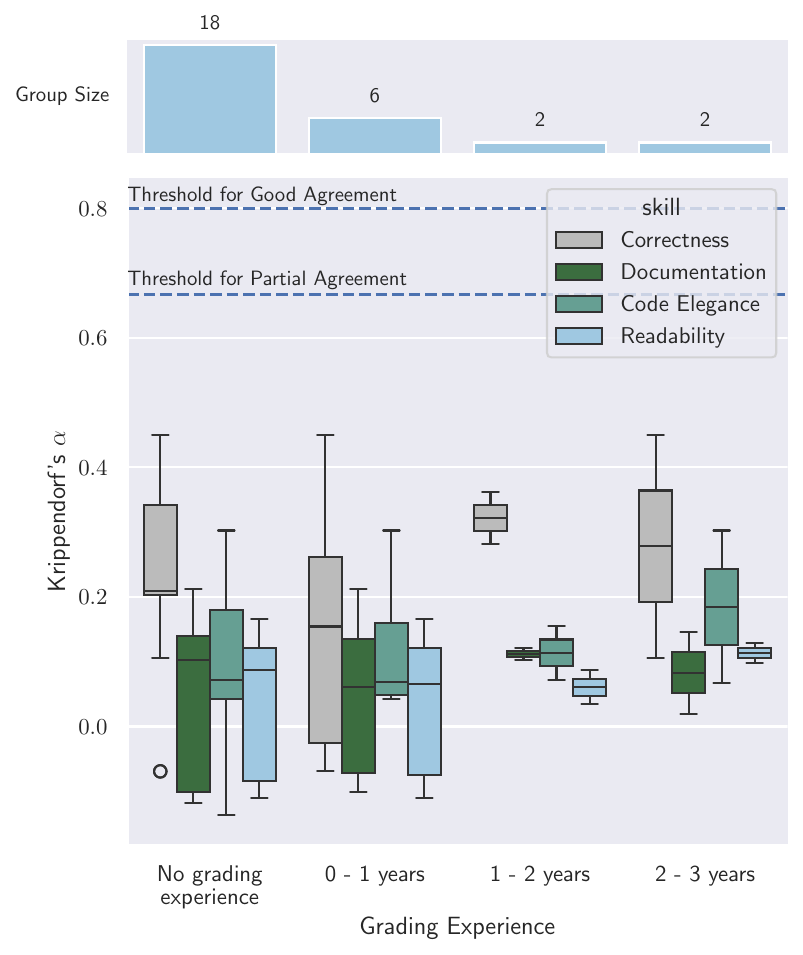}    
    \caption{This figure shows the distribution of Krippendorff's $\alpha$ for the participant's grading experience and all four skills. The box plots' box shows the inter-quartile range, while the whiskers extend to points within 1.5 of the inter-quartile range. The thresholds for good and partial agreement are the thresholds provided by Krippendorff \cite{Krippendorff2013}. The bar plot shows the proportion of participants with varying levels of grading experience.}
    \label{fig:inter_exp_dist}
\end{figure}

Our participants had varying grading experience levels, which could have affected the consistency of their grading. 
Figure \ref{fig:inter_exp_dist} shows the distribution of Krippendorff's $\alpha$ for each of the four skills. 
Independent of the years of experience, no groups achieved an agreement close to the threshold for partial agreement. 
Our results for 1-2 years and 2-3 years of experience show varying levels of agreement, as the sample size of participants who had more than one year of grading experience was small, which is typical at many institutions where undergraduate, master's and PhD students all assess student work \cite{Mirza2019}, as many students typically graduate after one or two years of working as teaching assistants. 
Those with one year of experience in grading show no significant improvement in inter-rate agreement compared to those with no grading experience.

Group 17-20, which had the highest average $\alpha = 0.257$ consisted of three master's students and one third-year student. 
Another group with a similar composition was group 1-4, consisting of three master's students and one PhD student; however, they had an average $\alpha = 0.123$, the fourth lowest average $\alpha$. 

Group 25-28, which had the highest proportion of PhD students, at two, with one master's and one third-year, had the second highest average $\alpha = 0.192$. 
Those that primarily consisted of third-year students, at two, with one master's and one-third year, performed the worst in terms of inter-rater reliability, with Group 13-16 having an average $\alpha = -0.032$ and Group 5-8 having an average $\alpha = -0.0598$.
In most cases, being of a higher degree level when grading CS1 introductory programming does not increase the consistency when using multiple graders. 

\newpage

\subsubsection{Grade Range}
\begin{figure}
    \begin{subfigure}[t]{0.4\textwidth}
         \centering
        \includegraphics[width=\linewidth]{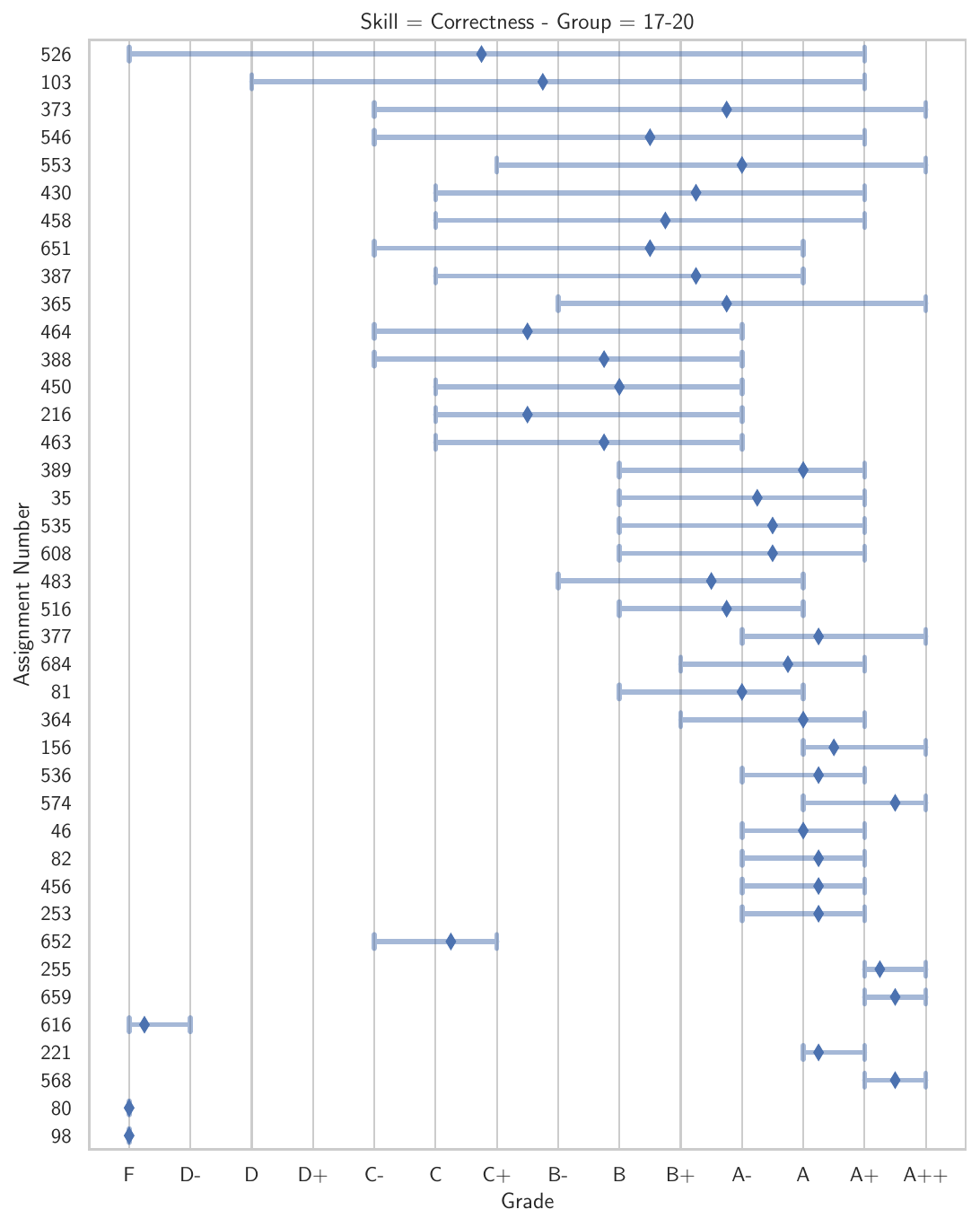}
        \caption{Highest Correctness $\alpha$ - Group 17-20}
        \label{fig:highest_correctness}
    \end{subfigure}%
    \begin{subfigure}[t]{0.4\textwidth}
       \centering
        \includegraphics[width=\linewidth]{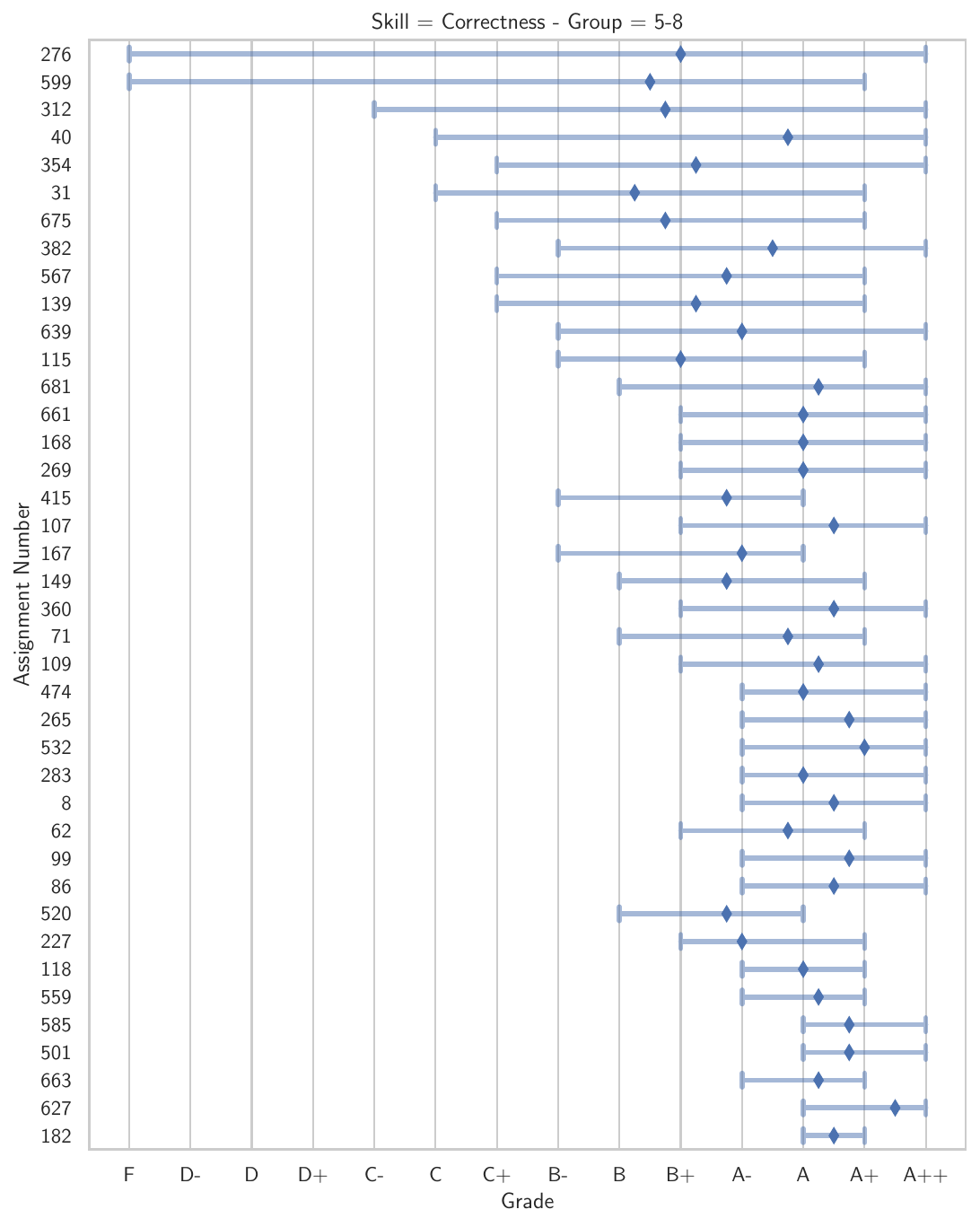}
        \caption{Lowest Correctness $\alpha$ - Group 5-8}
    \end{subfigure}
     \caption{These figures are for the highest and lowest $\alpha$ and show the minimum, maximum and mean correctness grade awarded by the participants for each assignment. Plots for all groups and all grades can be found in Appendix \ref{appx:grades_per_assignment}.}
    \label{fig:best_worst_irr}
\end{figure}

To further explore the inter-rater reliability, we evaluated the spread of grades awarded by each group. 
Figure \ref{fig:best_worst_irr} shows the minimum, maximum, and mean correctness grades for group 17-20, which was the most consistent when grading correctness, with an $\alpha = 0.45$, and for group 5-8, which was the least consistent when grading correctness, with an $\alpha = -0.069$. 
Appendix \ref{appx:grades_per_assignment} shows the distribution of grades for all groups and all grades.

For our most consistent group (group 17-20), only assignments awarded an F (assignments 80 and 98) – since they either did not compile or did not include any student code – were assessed with complete agreement. 
The original authors of assignment 80 submitted the template code they provided, and assignment 98 did not compile or, if the assessors fixed the compilation error, had an unhandled \texttt{NumberFormatException}, causing the simulator not to execute. 

To further investigate the reliability when awarding each grade, we stipulate that assignments grade ranges less than or equal to two (e.g. A to B+ is two grades away, and A to A- is one grade away) have good reliability; those with a range of three (e.g. A+ to B+) have mediocre reliability, and those with a range greater than three (e.g. A- to C+) have poor reliability; where the grade range is the maximum awarded grade minus the minimum awarded grade for each assignment. 

\begin{figure}
    \def\figsize{0.37}
    \begin{subfigure}[t]{\figsize\textwidth}
         \centering
        \includegraphics[width=\linewidth]{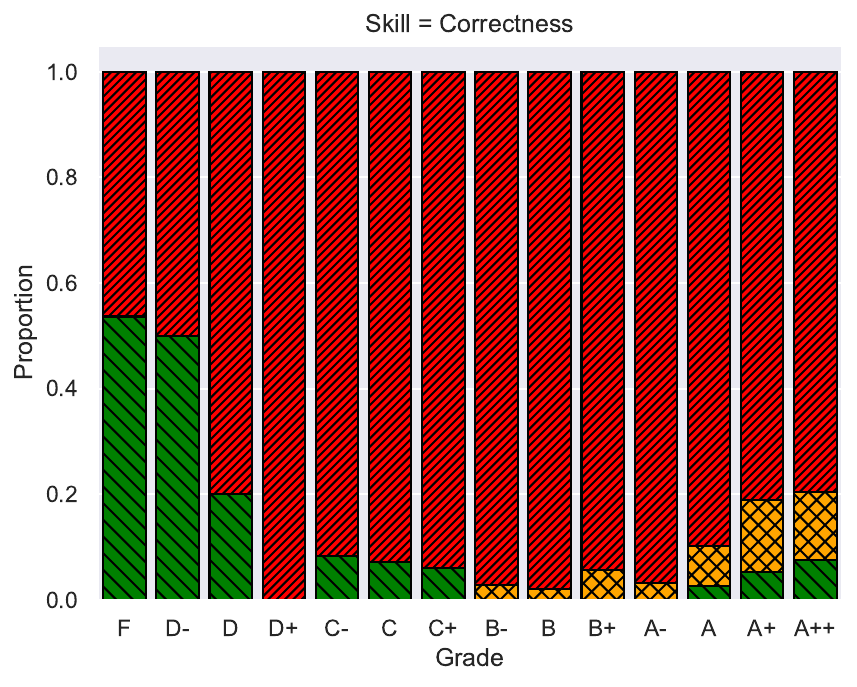}
    \end{subfigure}%
    \begin{subfigure}[t]{\figsize\textwidth}
       \centering
        \includegraphics[width=\linewidth]{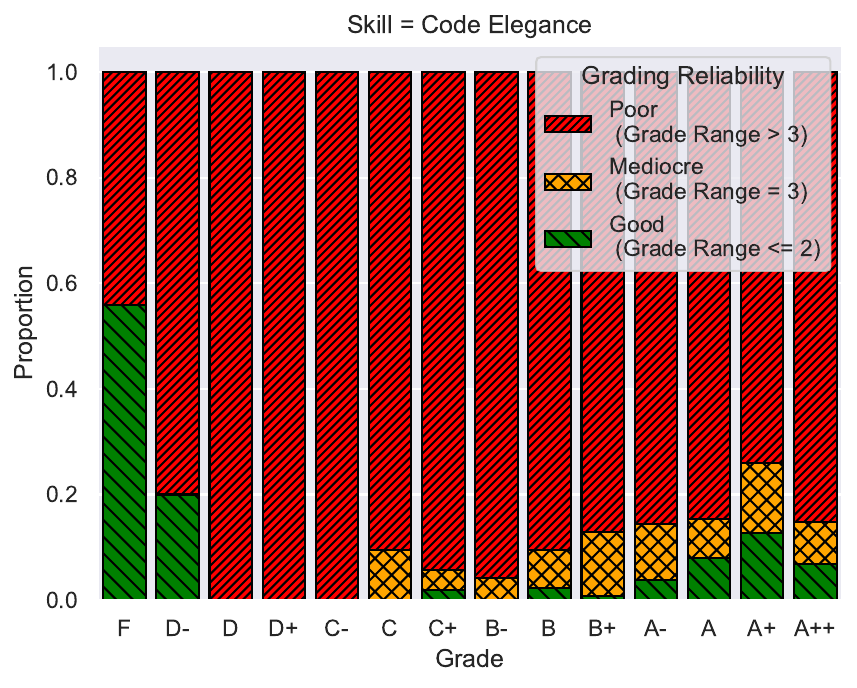}
    \end{subfigure}
    \begin{subfigure}[t]{\figsize\textwidth}
         \centering
        \includegraphics[width=\linewidth]{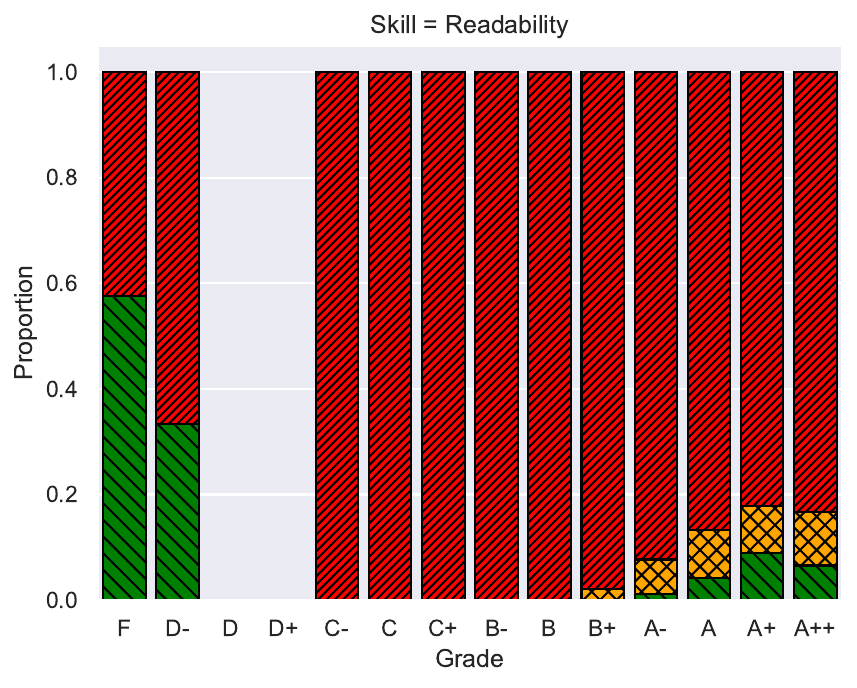}
    \end{subfigure}%
    \begin{subfigure}[t]{\figsize\textwidth}
         \centering
        \includegraphics[width=\linewidth]{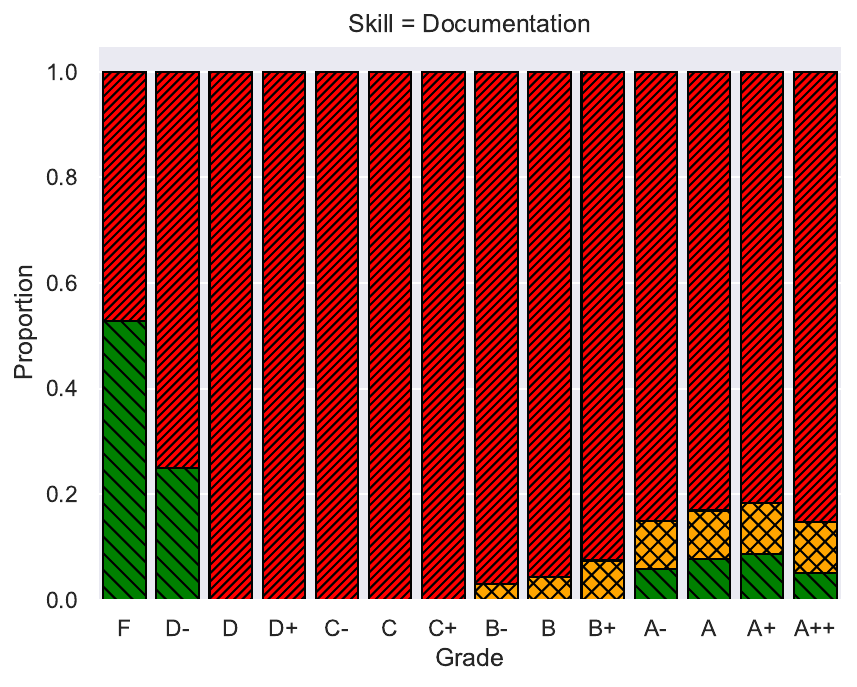}
    \end{subfigure}%
     \caption{This figure shows the proportion of how reliable each grade is for each skill across the grading scale. Reliability is defined by the grade range for each assignment, with `Good' as a grade range  $\leq 2$, `Mediocre'  $= 3$, and `Poor' $\geq 3$. }
    \label{fig:grade_rel}
\end{figure}

Figure \ref{fig:grade_rel} shows the proportion of the reliability for each grade and skill and shows that our participants were more reliable at awarding F and D- grades across all four skills than any other, with the middle grades (C- to B+) being the most unreliably graded across all four skills. The higher grades (A- to A++) were reliable or partiality reliable in less than 20\% of assignments graded across all four skills.

\subsubsection{An In-Depth Analysis of the Grades Awarded by Group 17-20}
Several assignments were awarded grades that were one grade bound away, with assignments 568, 211, 659, and 255 all being awarded an A++ or an A+ and assignment 616 being awarded a D- or an F. 
With these assignments being at the tails of the potential grades, it can be difficult to differentiate between the two furthest bounds, especially with the subjectiveness of open-ended assignments. 
For assignment 616, three of the four assessors graded the assignment an F for correctness, as the submission only contained the base code. 
However, participant 18 opted to assign the grade D- and gave the feedback, ``The submission does not cover any of the requirements in the coursework''.
We do not know why Participant 18 opted to grade assignment 616 with a D- and assignment 80, another assignment that did not meet the coursework requirements, with an F. 
However, this does highlight the potential issues with the internal consistency of human graders, which we discuss in more detail in Section \ref{sect:rq2_disc}.

Assignment 526 was the most inconsistently graded within the group 17-20, with grades ranging from A+ to F, with two awarding the assignment F for failing to compile, one awarded the assignment an A, and the other awarding the assignment an A+. 
As part of the anonymisation process, some base classes were copied into the submission. 
This was brought to our attention early within batch one; all participants were instructed to delete the problem classes and regrade any assignments that could not be compiled. 
However, some may not have completed this step.

The second most inconsistently graded assignment was 103, with the four participants awarding a D, a C-, a B+ and an A+. 
Participants 19 and 20, who awarded the assignment a D and a C-, respectively, highlighted that the assignment spawned three windows, with one having no visible functionality, but completed all the base tasks and some of the challenge tasks. Participants 17 and 18, who awarded the assignment a B+ and an A+, respectively, chose to focus on completing the challenge tasks and decided not to penalise the submission for spawning three windows as heavily. 
Participant 17 stated in their feedback, ``Whilst the base tasks were correctly implemented, the challenge tasks were incorrectly configured...''.
This is an excellent example of the hawk and dove effect, where some examiners are more stringent and require higher performance, the hawks, than others, the doves \cite{McManus2006}.
In this case, participants 19 and 20 are the hawks and require the program to have minimal errors, while participants 17 and 18 are the doves, and chose to focus on the highlights of what the student has submitted and provide a more lenient grade by overlooking the bugs within the application.

While correctness is mostly objective, code elegance, readability, and documentation are inherently subjective. 
Even if aspects of these are defined in the course, such as the specific code style or code design practice that the students must use, the examiners have their personal preferences on what makes code readable, well-designed or well-documented.
In this assignment, students were asked to use the code style demonstrated in \citet{Barnes2006} ``Objects First with Java'' and to use object-oriented programming, including inheritance and polymorphism, which they covered in class in the few weeks before the coursework was set. 

Out of our seven groups, group 17-20 had the highest $\alpha$ while grading code elegance with an $\alpha = 0.3$, less than the tentative acceptable value of $\alpha > 0.667$.
When grading code elegance, Group 17-20 did not agree for any of the 40 assignments they graded. 
Two assignments, 568 and 616, out of the 40 that group 17-20 graded were one grade bound away. 
Assignment 616 was awarded the same grades as correctness, with three awarding an F and the fourth awarding a D-. 
Assignment 568 was awarded grades at the opposite end of the spectrum, with three assessors awarding an A+ and one awarding an A. 
The assessor that awarded an A gave the feedback that ``... some functions are very long...'', whereas two of the assessors who awarded the A+ praised the student for their implementation, while the third did not provide any feedback with their grade.

The largest variance in grades awarded by group 17-20 for code elegance was for assignment 458; the grades awarded were C, B-, A- and A++. 
Participant 17, who awarded the A-, gave the feedback ``Solid implementation on both core and challenge class.''.
Participant 20, who was awarded the C, commented on the repetitiveness of the code and how functions could be refactored to the superclass. 
These differences in awarded grades and the associated feedback can suggest how grading subjective elements of assignments can introduce a higher variance in the awarded grades. 

This trend of a few assignments being somewhat consistently graded and many being inconsistently graded holds for readability and documentation, with Appendix \ref{appx:grades_per_assignment} showing the grade range for each skill. 
Figure \ref{fig:grade_rel} shows this trend for each grade, with the middle grades (D - B+) more likely to have a larger variance in awarded grades across all four skills. 
While the middle grades showed no cases of any grading achieving acceptable levels of consistency, the high (A++ - A) and low ranges (F - D-) displayed at least some instances of grader agreement. 
Grades F and D- have the largest portions of grade ranges, being less than or equal to two. 
At the other end of the grade spectrum, grades A, A+ and A++  have a smaller proportion of grade ranges that are less than or equal to two. 

\subsection{Grading Consistency of Individual Graders (RQ2)}
\subsubsection{Grade Range}
\begin{figure}
    \begin{minipage}[t]{0.5\textwidth}
         \centering
        \includegraphics[width=\linewidth]{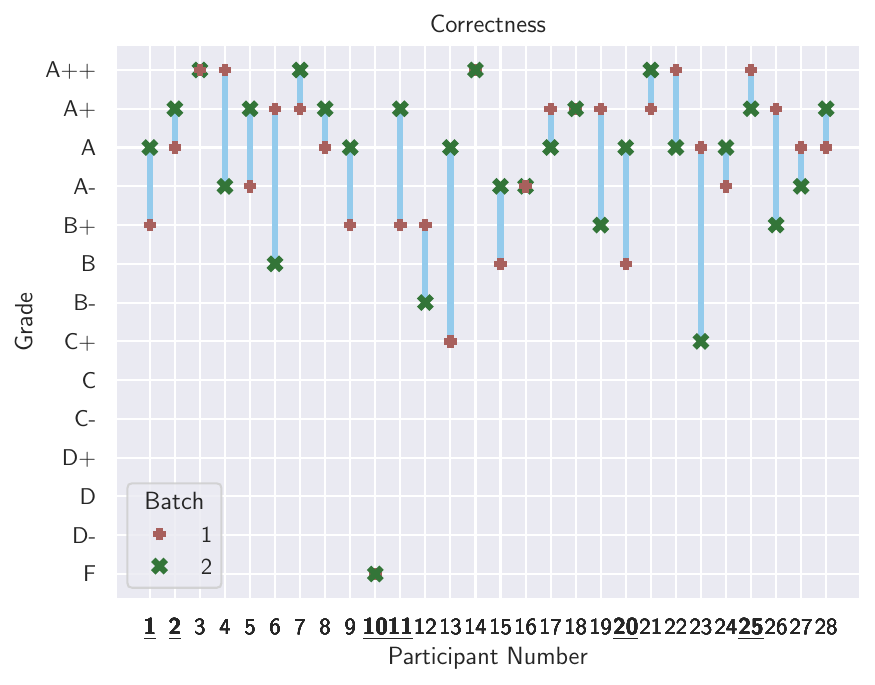}
    \end{minipage}%
    \begin{minipage}[t]{0.5\textwidth}
       \centering
        \includegraphics[width=\linewidth]{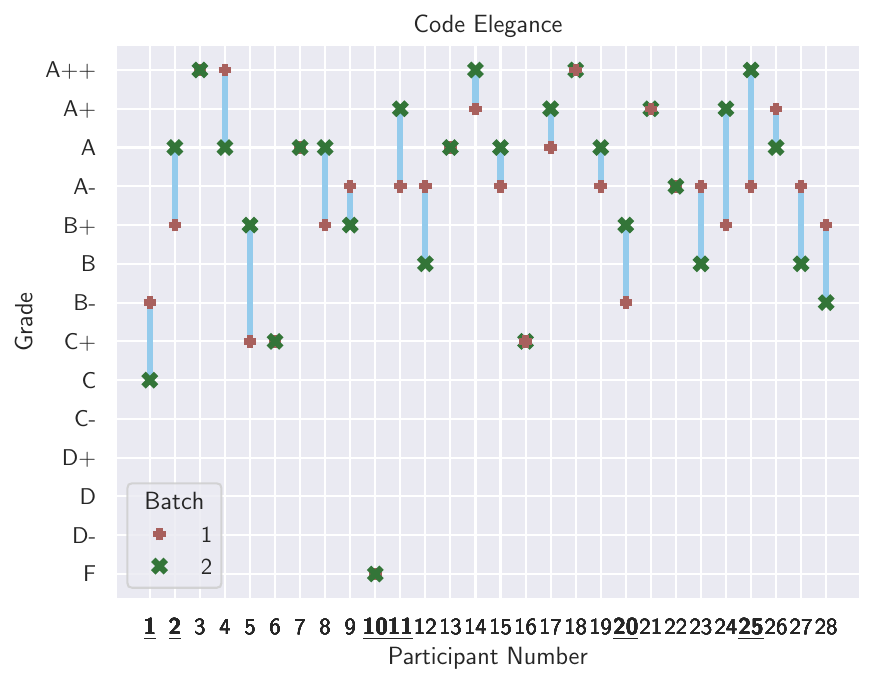}
    \end{minipage}
    \begin{minipage}[t]{0.5\textwidth}
        \centering
        \includegraphics[width=\linewidth]{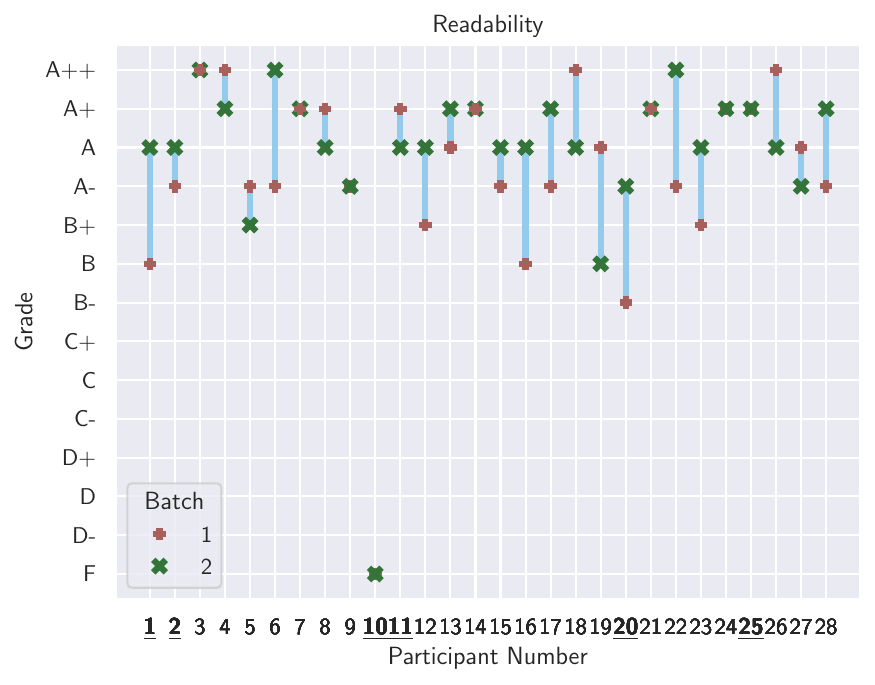}
    \end{minipage}%
    \begin{minipage}[t]{0.5\textwidth}
        \centering
        \includegraphics[width=\linewidth]{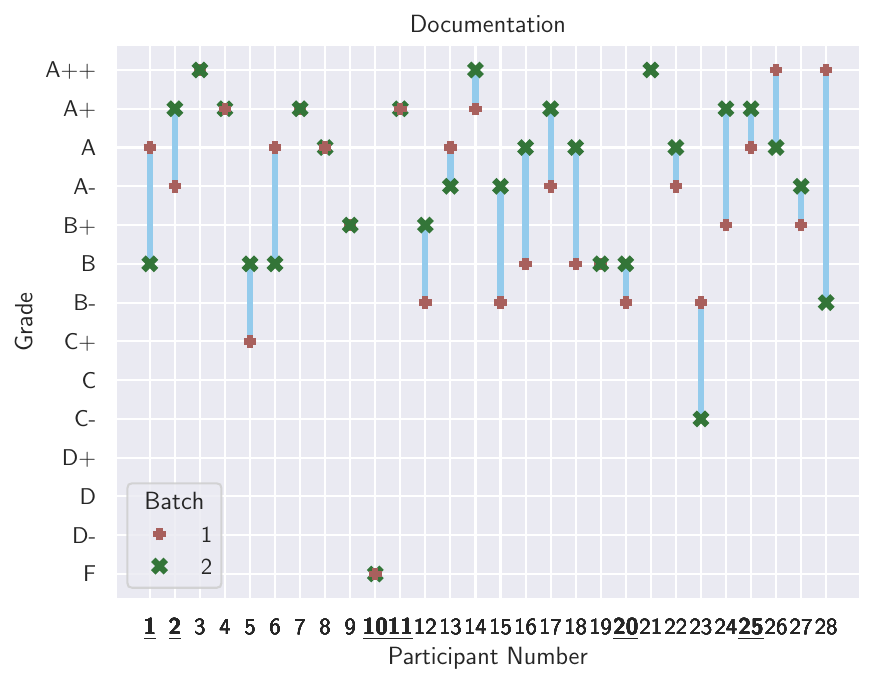}
    \end{minipage}%
     \caption{The distance between the awarded grades for each skill in batch one and batch two for each participant when grading the duplicated assignment. The underlined participants indicated that they noticed the duplicate during the post-study interview. Participant 21 did not supply a grade for documentation in batch two, so it only shows one grade.}
    \label{fig:intra_distance}
\end{figure}

As part of our study, we purposefully duplicated an assignment within the second batch to allow us to evaluate the consistency of the individual graders. 
22 participants did not notice the duplicate in the second batch; six participants (1, 2, 10, 11, 20, 25) did. 
Figure \ref{fig:intra_distance} shows the distance between the 14 awarded grades when regarding the same assignment for a second time, for each participant for each skill.

The average absolute distance between grades when regrading the same assignment a second time was correctness = 1.786, code elegance = 1.250, readability = 1.357 and documentation = 1.571. 
These results show that for the consistency of the individual graders in our study, on average, they were between one and two grades apart between batches one and two, with a higher variability in correctness and documentation, and were more consistent with themselves when grading code elegance or readability.

\subsubsection{Post-Study Interview}
During the post-study interview, 24 out of 28 graders stated that they graded consistently. 
However, our results show that when grading a repeated assignment at a later period, only two participants graded the duplicated assignment the same across all skills. 
Only two participants, three and ten, provided the same grade in both batches for all four skills. 
Participant 10 noticed the duplicate, and during their post-study interview, participant 10 mentioned that they noticed the duplicate and changed the grades for both batches to F as they believed it was plagiarism, while the other five participants followed the instruction that if you notice a duplicate, treat it as you are grading it the first time; as this was a pair programming assignment, duplicates could still exist within the dataset. 
Participant 3 did not notice the duplicate and said during their post-study interview, ``...like some assignments used the same animals... but \textit{[the implementation]} was a little bit different so not like an exact copy of those two assignments...''.

In an effort to improve their consistency, 20 of the participants opted to review their grades. 
However, only 10 participants had a two or less grade difference between grading the first and second batches. 
While many opted to review their grades, many only gave a cursory review or reviewed if they thought they graded an assignment exceptionally high. 
Participant 12 stated, ``... not too much time on all the of the like remarking all of them, just spending a minute on each assignment...'' and Participant 11 said ``...if I graded someone high. I'll go back and see if I graded anyone else at the same level...''.
While the participants' reviewing process may have increased their consistency for the grades they reviewed, it did not improve their overall consistency.

\subsubsection{Grading Experience}
\begin{figure}
     \centering 
    \includegraphics[width=0.65\textwidth]{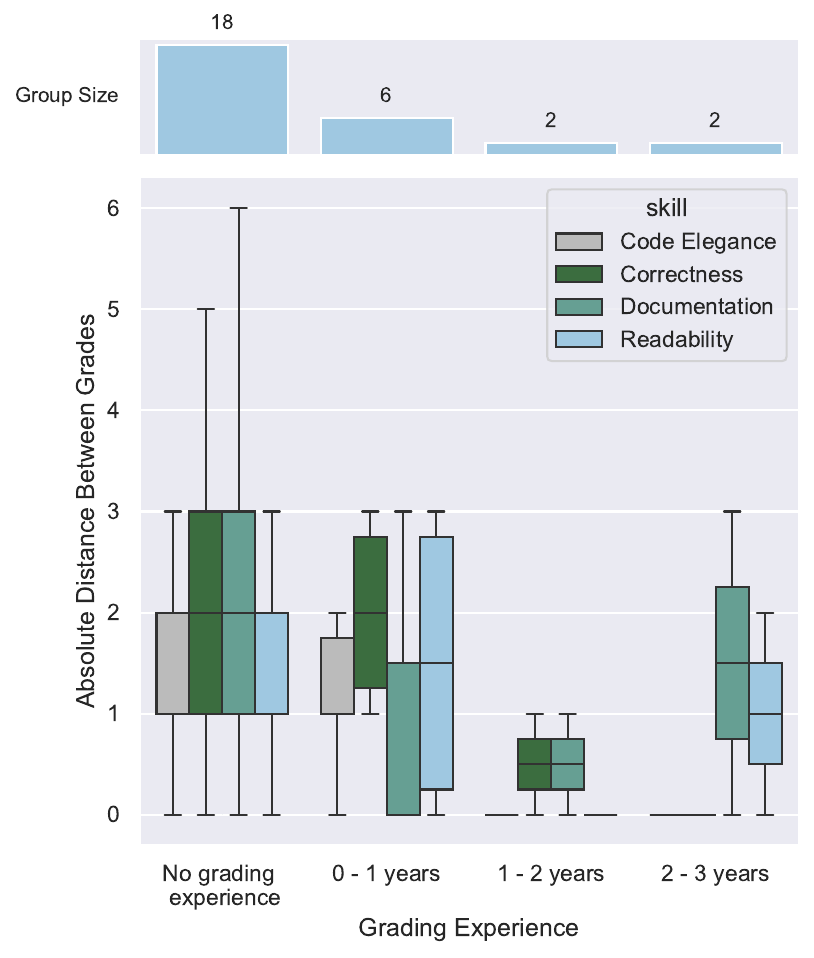}    
    \caption{This figure shows the distribution of the absolute distance between the awarded grades in batch one and batch two as a measure of the participant's self-consistency for the participant's grading experience and all four skills. The box plots' box shows the inter-quartile range, while the whiskers extend to points within 1.5 of the inter-quartile range. The bar plot shows the proportion of participants with the varying levels of grading experience.}
    \label{fig:grading_exp_dist}
\end{figure}

Figure \ref{fig:grading_exp_dist} shows the distribution by grading experience of the absolute distance between grades awarded in batch one and batch two for each of the four skills. 
Those with no prior grading experience have the largest difference between grades across all skills. 
Those with 1-2 years of grading experience have at most a difference of three grades between the first and second time they graded the duplicated assignment, and those with one or more years of grading have a grade difference of less than one for most skills. 
However, readability and documentation for those with two or three years of grading experience had a maximum grade distance of two or three, respectively.

\newpage

\section{Discussion}\label{sect:discussion}
\subsection{How Consistently Does a Group of Graders Apply a Rubric? (RQ1)}
Consistency was very poor overall. 
Even the best values achieved for Krippendorff's $\alpha$ reflect distinctly poor consistency, with none of our seven groups of graders achieving an $\alpha > 0.667$, the lower bound to consider reliable \cite{Krippendorff2013}, with the average $\alpha = 0.105$ across all groups and skills. 

While the correctness of a program is, in most cases, inherently objective, there can be elements of subjectiveness, especially within open-ended coursework assignments. 
The nature of open-ended assignments, where students decide how they implement a set of features using their ideas and creativity, can make it difficult for the groups of assessors to grade consistently. However, open-ended assignments allow instructors to design assignments that allow students to explore their creativity, which is a key educational goal \cite{Chan2014, Richardson2018, Wagner2010}, mimic professional development experiences, and improve students' self-efficacy in programming \cite{Sharmin2019}.

Furthermore, the subjectiveness of open-ended programming assignments extends not just to whether the code is correct and meets the requirements but to the quality of the code itself \cite{Kirk2024a}. 
While some aspects of code quality are shared, such as consistent indentation and meaningful variable names, the nuances of programming style and design are individualised and remain active research topics not just within computer science education \cite{Stegeman2016, Kirk2024}, but within professional practice as well \cite{Liang2023}.

While having experienced graders may increase the consistency of assessment, this is not always possible as the typical demographics of teaching assistants are undergraduate, master's or PhD students \cite{Mirza2019, Riese2022, Riese2021, Wald2020}, and many are new to assessment and have little experience; which is the case in our study.
Having those with many years of grading experience is typically rare, as undergraduate teaching assistants are typically late-stage and are about to graduate, and graduate teaching assistants within Europe typically graduate within three to four years.

However, those with one year of grading experience had similar inter-rater reliability to those with no previous grading experience, indicating that more grading experience alone will not improve the grading consistency to an acceptable level. 
This supports the idea that to conduct a consistent assessment, assessors require adequate training to interpret and apply a rubric \cite{Riese2022, Kristiansen2024}, and that assessment is complex and intuitive \cite{Bloxham2016, Jonsson2021}.

Rubric design plays an important role in the validity and consistency of the rubric \cite{Jonsson2007}. 
The rubric provided to our participants follows the standard practice for rubric design \cite{Andrade2005, Reddy2010} and includes evaluation criteria, the definitions of quality and a scoring strategy. 

However, as with the subjectiveness of programming, especially with open-ended assessment and code quality, as we have previously discussed, aspects of the rubric are subjective. 
Some aspects of the rubric are very clear, especially for the grades at either end of the grade spectrum. 
For example, for correctness, an F is defined as ``Significant details of a task are violated, or the program often exhibits incorrect behaviour.' and an A+ - A++ is defined as ``The application works as described in the assignment; all base tasks are completed; the implementation demonstrates originality, creativity and technical excellence...''.

Meanwhile, for the middle grades, the rubric becomes more subjective. 
For example, for correctness, a D is defined as ``Some tasks are incomplete; the application functions incorrectly on some inputs/actions.'' and a C is defined as ``Minor details of the task(s) are violated; the application functions correctly on the majority inputs/actions''.
Other examples from the rubric include D - code elegance, ``The application is unnecessarily complex...'', D - documentation, ``...the code is overly commented.'', D - readability, ``The application has more than two issues that make the program difficult to understand''.
The lack of specificity in the rubric, especially for the middle grades, could have impacted the consistency of the grading, as different assessors can have differing opinions on what constitutes ``some'' and ``majority'' correct inputs and if the source code is unnecessarily complex, over-documented or difficult to understand.

There is likely to be less disagreement at either end of the grading spectrum, an F or an A+/A++, though rarely do any of these have complete agreement unless the submissions do not compile or do not contain any student code, with most groups being within one or two grades at these grades. Meanwhile, the middle grades are not graded consistently, with most groups having a grade difference of three or more.

While a more explicit rubric may improve the consistency of the grading, it could reduce the validity of the assignment \cite{Jonsson2007}, as a more explicit rubric would require a less open and creative assignment, limiting the students' self-efficacy \cite{Sharmin2019}, and learning opportunities. 
Furthermore, even an explicit and analytic rubric may not infer consistent assessment, as grading consistently can be extremely difficult even with well-designed assessment criteria \cite{Jonsson2021}. 
However, future work should investigate which aspects of code quality can be explicitly defined in the rubric and if defining these objectively improves consistency when grading open-ended assignments. Explicitly defined criteria could include explicitly defining code style conventions students should follow, such as naming conventions, avoiding single-character variable names, indentation, and spaces around operators. 

\subsection{Are Individual Graders Consistent at Applying a Rubric? (RQ2)}\label{sect:rq2_disc}
Compared to the consistency of multiple graders, readability, code elegance, and documentation achieved a higher individual consistency. 
The higher consistency of code elegance, readability, and documentation could be due to our participants being experienced programmers and having a well-defined internal preference for code elegance, readability and documentation \cite{Mohan2004, Li2018}. 

The lower consistency of correctness could be due to many of our participants having not graded before, which is typical when using teaching assistants \cite{Mirza2019, Riese2022, Riese2021, Wald2020}. While some may have completed the assignment as a student, they would have only seen a limited set of other submissions. 
Another aspect could be the participants having to learn the specificity of what makes a submission good for this specific assignment; as the assignment is open-ended many different solutions can meet the requirements to the same standard, but with the specificity of the rubric, that standard is up interpretation by the grader. Similar to multiple grader consistency, a more specific rubric could increase the consistency, but at the cost of the validity \cite{Jonsson2007} and creativity of the assignment.

Having more grading experience can lead to being more internally consistent when grading similar work, and supports the emphasis on providing appropriate training for teaching assistants \cite{Kristiansen2024, Wald2020}.
However, grading experience alone does not completely mitigate errors in consistency with experienced graders \cite{Riese2022}, especially when assessing subjective aspects like documentation and readability. 

\subsection{Implications for Human Grading}
We have found that, in our study,  typical graders are very inconsistent with each other and also are inconsistent with themselves.  This suggests that current grading practices that use multiple assessors are likely to be quite unfair, which can lead to unfair outcomes for the students and demotivation among the students.  Many suggest the importance of training assessors \cite{Kristiansen2024, Wald2020, Borela2023}, and it may be that this is a necessary step to ensure fairness in grading.

We believe that our result warrants further consideration and investigation.  Programming assignments are supposed to be relatively objectively judgeable and yet our research suggests that even with a shared rubric, inconsistency is the default outcome, which supports the idea that grading is `complex, intuitive and tacit' \cite{Bloxham2016}. This also lends support to the idea of investigating alternative grading approaches, such as those suggested by \citet{Decker2024} for making grading more equitable.  They propose changing the grading scales (such as making them coarser) in order to speed up grading and reduce disputes about exact grading scores -- although our study already uses a coarser scale than 0--100 and found inconsistency even on a coarser scale.

\subsection{Implications for Automated Grading}
Part of our motivation for conducting this study was to get a ``gold standard'' baseline human work in order to later investigate the accuracy of automated grading.  
However, our results suggest that the idea of a gold standard of human grading may be flawed, given that human graders cannot agree on the grade to give a piece of student work, even with themselves: inconsistency was high between and ``within'' individual graders.  This has strong implications for the acceptability of automated grading. One could argue that automated grading (including AI-based grading) needs to be within the same range as human grading to be considered acceptable, and since the range for human grading is wide, automated grading is more likely to be considered sufficient. 
However, this is an unconvincing argument. Rather than using the weakness of human grading as a convenient, low bar for assessing auto-graders, we – as a teaching community – should recognise and address the serious problem and try to find ways to improve the fundamental practice of grading student assignments. Automated grading may or may not have a part in this. 
One of the contributions of this paper is a public release of our dataset, including the human grades, which can be used by all researchers for future work on automated grading.

\section{Threats to Validity}\label{sect:threats}
\subsection{Internal}
One limitation of this study is that the participants knew it was a study that was not providing real grades and did not affect student outcomes. 
However, based on the interviews, many participants treated it as if they were marking for real, with many students opting to review their grades. 
The participants were aware that this was a study into grading consistency and, as such, may have altered their behaviour to maximise their internal consistency. 
While the participants may have altered their behaviour, our results still show significant inconsistency within their grades between a group of graders and when they graded an identical assignment later.

The rubric that was used in the original coursework provided generic descriptions for each grade, which may be less clear and well-defined than point-based rubrics that provide a detailed rubric and the points associated with each specific element. 
However, as the rubric was provided to the students when they originally submitted the coursework, we decided to mimic real-world assessment at our institution to provide an accurate insight into the consistency of assessment within this course.

We opted to provide only one identical submission in the second batch, which limits the validity of our evaluation of the internal consistency of human graders. 
We chose to only have one repeat to maximise the number of graded unique submissions, allowing for a more robust dataset when conducting future work. 
While only having one repeated assignment limits the validity of this part of our experiment, the results show that even with our small sample size, with each group of four graders grading a different duplicate, nearly all participants did not provide the same grade when grading the assignment a second time.

As we are using historical assignments from the CS1 programming course at our institution and many of our participants took that course during their undergraduate, there is a remote chance that the participants graded their own work. 
As these assignments were completely anonymised before undertaking this study, we could not tell if we asked the participants to mark their work. 
However, as the dataset of anonymised consists of over 600 submissions, the chances that one of the 272 graded submissions will be graded by its author is very low.

\subsection{External}
Our raw data and participants were all captured from one institution and one course, which limits how these results can be applied to other institutions or courses. Programming courses which use different rubrics, especially those that are more specific than the one provided in Appendix \ref{appx:rubric}, could have increased consistency. Prior work, such as by \citet{Passonneau2023} and \citet{Jonsson2021}, suggests that simply making the rubric more detailed may not necessarily produce more consistent grading. Future work should evaluate how the specificity of the rubric affects consistency and the ability to provide open-ended assignments. 
Furthermore, different groups of graders, whether that be undergraduate, postgraduate teaching assistants, instructors, or a combination of all three, could result in varying grading consistency. Further work could investigate how different combinations of graders, training, and other methods can result in providing grades that are consistently applied between the group members.

As assessment is such a core aspect of education and students' prospects, we believe our results highlight a fundamental issue with how summative assessment is currently conducted, whether that is caused by instructors not providing detailed rubrics or utilising multiple teaching assistants with limited training in assessment.
This fundamental issue is further supported by many institutions implementing rigorous assessment procedures to ensure that the assessment practices within their institutions are fair to all students, and indicates that rubrics should be seen as potentially useful addition to the range of assessment tools available and not a solution for all concerns raised by students in terms for quality, consistency and usefulness \cite{Cockett2018TheReview}.

\section{Conclusion}\label{sect:conclusion}
Inconsistent grading can impact students throughout their education and their post-education opportunities. 
Receiving inconsistent grades can confuse many students about why they received a particular grade, especially when comparing their grades with those of their peers, and affect their progression through their degree, especially if they do not meet course prerequisites, as well as affecting their sense of justice~\cite{Nesbit2006}. 
Post-education, inconsistent grading can affect a student's career prospects, as many companies utilise degree classification or average grade in their hiring process~\cite{Stepanova2021}. 

Our study investigated the consistency of human grading by asking 28 participants to grade 272 authentic student assignments. 
We selected participants from our institution who were PhD or master's students with more than 3 years of programming knowledge; previous grading experience was not a requirement. 
This selection of participants mimics the demographics of the teaching assistants who complete the grading at many institutions, especially those with large cohorts, where the individual module/class leaders can not feasibly grade the cohort in the required timeframe for feedback to be meaningful.

We split our 28 participants into groups of 4; each group graded the same 40 assignments in two batches of 20, with a two-week gap between them. 
To investigate the consistency of using multiple graders, we evaluated the inter-rater reliability using Krippendorff's $\alpha$ \cite{Krippendorff2013}, which factors the ordinal nature of grades by applying weights to the metric calculation. 
We found that our participants provided inconsistent grades across correctness, code elegance, readability and documentation, with the most consistent being correctness with an average $\alpha = 0.2$ and code elegance, readability and documentation all having an average $\alpha < 0.1$ -- where an $\alpha \geq 0.667$ is required to draw tentative conclusions and an $\alpha > 0.8$ suggests acceptable consistency \cite{Krippendorff2013}.  This indicates a very high level of inconsistency of grading between graders.

In addition to investigating the consistency of multiple graders, we analysed the individual graders' self-consistency by duplicating one of the assignments in the first batch to the second batch with a different submission ID and measuring the difference between the grades they awarded in the first batch, and the grade they awarded in the second batch. 
Only one participant did not notice the duplicate assignment and graded consistently for all four skills. 
We found that an individual's self-consistency was, on average, higher for the subjective elements under assessment, code elegance, readability and documentation, compared to correctness, which was more inconsistent. 
The average grade difference was 1.79 for correctness, and the subjective elements all had an average grade difference of less than 1.6. 

While these results are only from one assignment at one institution, which utilised a generic rubric, our study supports the notion that assessment is complex and intuitive \cite{Bloxham2016} and that having a well-designed open-ended assignment may not imply consistent evaluation \cite{Jonsson2007}. Our results suggest that there is a balance to be found between the openness of the assignment and the specificity of the rubric, and future research should investigate how aspects of code quality and correctness can be explicitly defined while allowing for an open-ended assignment.  It also highlights that further research is required to improve the consistency of a group of graders and individual graders when grading open-ended programming assignments, whether it be alternative grading practices or methods of effectively training undergraduate and graduate teaching assistants.

\subsection{Future Work}
As assessment plays a crucial role in a student's education and future career prospects, we recommend that further research be undertaken on how to improve the consistency of grading programming assignments, whether that be investigating rubric design, the importance of assessor training, the use of alternative assessment approaches such as comparative judgement, or automated assessment tools.

In the future, we plan on expanding on this work by evaluating the consistency of the feedback provided as part of this study, including the topics on which the assessors gave feedback and the quality of their feedback. 

\section{Data Availability}\label{sect:data_ava}
All our raw data and data analysis notebooks can be found on GitHub\footnote{Data analysis repository: \url{https://anonymous.4open.science/r/Consistency_In_Grading_Analysis-AC73/README.md}}. 
The \anon{Menagerie} dataset, which includes all the anonymised student submissions and all the human grades gathered during this study, can be found on the Open Science Foundation\footnote{Menagerie: \url{https://osf.io/q8jbt/}}.
\begin{acks}
\anon{We thank the King's College Teaching Fund for providing the funding necessary to undertake this research. We also thank Jeffery Raphael for providing the original data and Zara Lim and Nikolaj Jensen for anonymising the student assignments dataset. We thank Carlos Matos and Nuno Barreiro for their comments and ideas throughout the project and Mark Guzdial and Jarmoir Savelka for proofreading and commenting on the final draft. We thank the reviewers for their detailed and constructive reviews.}
\end{acks}

\bibliographystyle{ACM-Reference-Format}
\bibliography{references}

\appendix
\section{Further Details on the \anon{Menagerie} dataset}\label{appx:Menagerie}
\subsection{Data Processing}
This project was a small-group project, with students in groups of two (or rarely three). 
Each student was required to submit a copy of the assignment, so all submissions were duplicated at least once. 
We first generated hashes to remove these duplicated assignments. 

Additionally, we manually reviewed potential duplicates based on the number of classes and the number of source lines of code; if they were identical, the copy was removed. 
After removing the duplicates, we extracted the students' submissions and removed unnecessary files, such as IDE property files and \texttt{.class} files.

To anonymise the dataset, we automatically removed the JavaDoc \texttt{@author} tag lines in all classes. 
The \texttt{@author} tag is typically used to identify who wrote the class and would be the most likely place for a student's name or identification number to appear in the code. 
To verify that no further personally identifiable information was left in the source code and to minimise duplicated submissions, we generated a list of changes made by the student compared to the template code. 
Multiple research assistants then used the list of changes to review the submitted code and remove any remaining personally identifiable information. 
Our data anonymisation approach limits the potential for students to be identified from the anonymised submission while retaining all other information.

During the anonymisation process, we found that many projects opted to include image files as part of their project. 
As these cannot be removed as they would interfere with how the submission is executed and cannot be anonymised easily, we opted to replace all images with equivalently sized black images.

\subsection{An In-Depth Example Submission}

\begin{figure}
    \centering
    \includegraphics[width=0.85\textwidth]{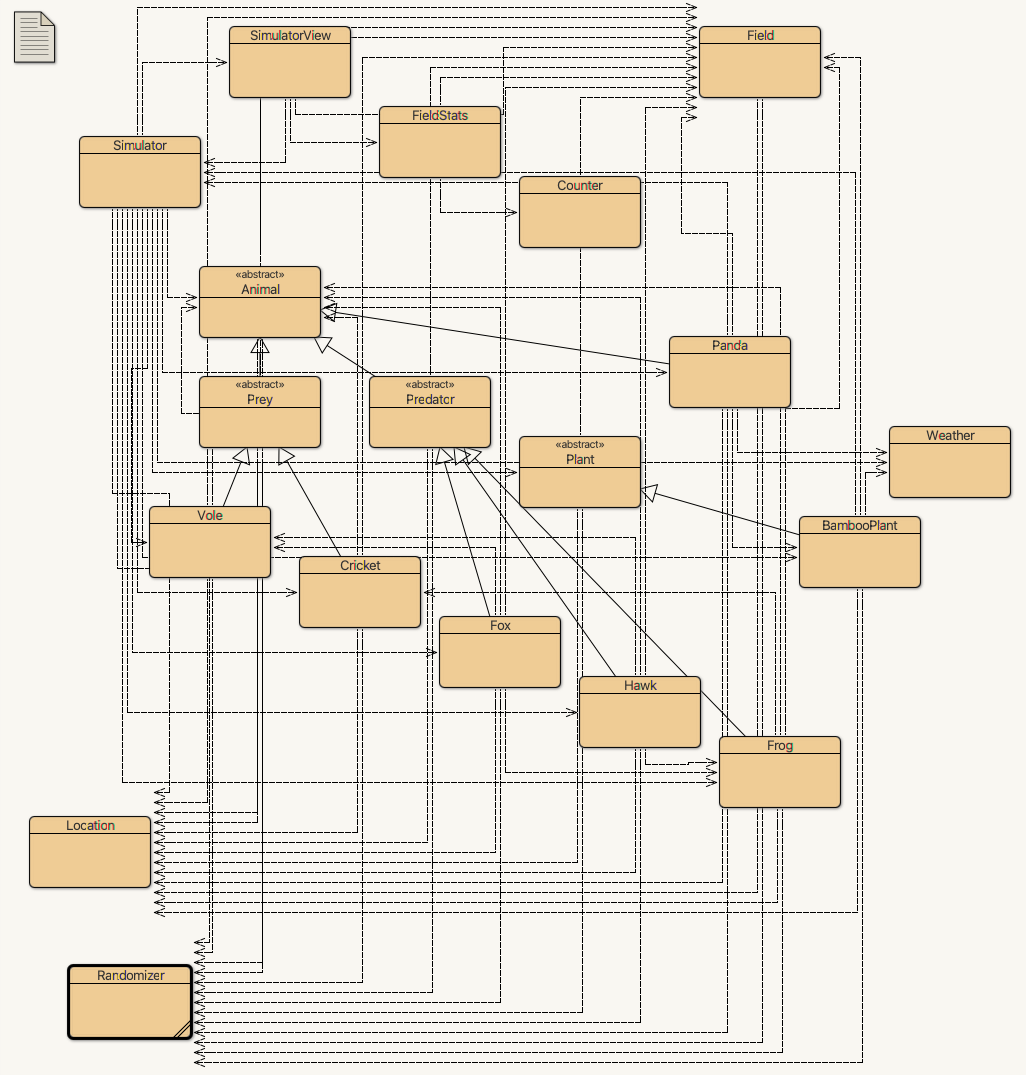}
    \caption{The UML Diagram of the median-sized submission by class size and source lines of code.}
    \label{fig:submission_uml}
\end{figure}

To give a specific example of an assessment submission, we picked submission 189, which is median-sized regarding the number of classes implemented and the number of code source lines, with 19 classes and 1345 source lines of code.
\autoref{fig:submission_uml} is the final UML of the submission and, by comparison to \autoref{fig:template_url}, shows that the students significantly expanded on the template project.

To complete the tasks, as defined in Section \ref{sect:assignment}, the students implemented the five species, with at least two being prey and at least two being predators, using inheritance and abstract classes to implement shared functionality. 
All of the five species classes inherit from the \texttt{Animal} class, with the prey and predators inheriting the \texttt{Animal} class indirectly through their parent classes, \texttt{Prey} and \texttt{Predator} respectively (which were not present in the original code, and have been added as part of the object-oriented design by the students). 
Both the \texttt{Fox} and \texttt{Hawk} species compete for \texttt{Voles} and \texttt{Frogs}, and \texttt{Frogs} hunt for \texttt{Crickets} by implementing the \texttt{abstract} method \texttt{hunt} from the \texttt{Predator} class and using \texttt{instanceof} to validate if the adjacent locations contain one of the prey.

To distinguish between male and female animals, the students added a boolean field and a function to randomly assign male or female in the \texttt{Animal} class. 
In each species class, the students added functions to validate whether the two members of the same species were adjacent and of opposite genders. 
If so, they add a new animal to a space in the two-dimensional array.

To incorporate the time of day into their submission, the students added fields \texttt{hour}, \texttt{day} and \texttt{dayNightStatus} to the \texttt{Simulator} class, and a \texttt{isDay} field to the \texttt{Field} class. 
The \texttt{Cricket} is the only species with different actions during the day or night. 
During the day and night, the \texttt{Cricket}'s age increases, and they handle the disease, as explained later in this section, and at night they also propagate.

For the challenge tasks, the students implemented all three suggestions outlined in Section \ref{sect:assignment}.
To simulate the lifecycle of plants, the students added an abstract \texttt{Plant} class and concrete class \texttt{BambooPlant}. 
The plant propagated based on a probability to produce seeds and was food for the \texttt{Panda} class, which inherited directly from the \texttt{Animal} class.

The students implemented the four seasons to implement weather, which changed every 90 time steps. 
The \texttt{BambooPlant} has different functionality depending on the season. 
In the spring, it gets water, grows, and spreads, whereas in the winter, it loses water and does not grow or spread.

Finally, the students implemented a disease that affects all animals by adding an \texttt{isInfected}, probability of infection and death fields, and functions to catch, spread and check if the animal is infected to the \texttt{Animal} class. 
These functions were called as part of each animal's act methods, which were responsible for simulating each object at each time step.

\newpage

\section{Rubric}\label{appx:rubric}
The rubric below was used in our study. It was provided to the students when they originally wrote their submissions.

\section*{Grade A+ - A++ (80 - 100)}

\subsubsection*{Program Correctness.} The application works as described in the assignment; all base tasks are completed; the implementation demonstrates originality, creativity and technical excellence in the completion of all challenging tasks the student included documentation describing the challenge tasks that were completed and how; all submission instructions were followed correctly.

\subsubsection*{Code Elegance.} Student demonstrates excellent use of classes and functions to produce reusable and maintainable code, where possible, in the base and all of the challenge tasks; the code is efficient without sacrificing readability and understanding

\subsubsection*{Documentation. } The documentation is well written, organised and clearly explains what the code is accomplishing and how; the student documents the contents of each class at the beginning of the file; the student documents the purpose of each function, i.e., the function's parameters and return values at the beginning of each function; the student documents each logical block of code when performing non-obvious operations; Every file includes header information (e.g., student name and id).

\subsubsection*{Readability. } The application is exceptionally well organised and very easy to understand; the student used indentation appropriately and consistently to delineate code blocks; each function performs a single well defined operation; the student used meaningful identifier names, i.e., good function and variables names; the student used white space between logical code blocks; the student uses consistent spacing around operators and variables; classes are self contained with private data hidden and methods are public only when necessary.

\section*{Grade A (70 - 79)}
\subsubsection*{Program Correctness.} The application works as described in the assignment; all base tasks are completed; the student has completed some of the challenge tasks; the student included documentation describing the challenge tasks that were completed and how; all submission instructions were followed correctly.

\subsubsection*{Code Elegance.} All of the base tasks and some of the challenge tasks are implemented in such a way that code can be reused, where possible; the majority of the code is written in such a way that is easy to maintain, i.e., add new or extend features; the code is efficient without sacrificing readability and understanding

\subsubsection*{Documentation. } The documentation consists of comments that are useful in understanding the code and/or structure of the program; each file has header information (e.g., student name and id).

\subsubsection*{Readability. } The code is clean, understandable and organised; the student used meaningful identifier names, i.e., good function and variables names: the student used white space between logical code blocks,

\section*{Grade B (60 - 69)}
\subsubsection*{Program Correctness.} The application works as described in the assignment; all the base tasks are completed, however, no challenge task was at-tempted: all submission instructions were followed correctly.

\subsubsection*{Code Elegance.} The application is implemented in such a way that only a few code segments could be rewritten to increase code re usability; the code is fairly efficient without sacrificing readability and understanding; the student made a poor design choice in a single area.

\subsubsection*{Documentation. } The documentation consists of comments that are somewhat useful in understanding the code; some of the comments state the obvious.

\subsubsection*{Readability. } The application has minor issues such as inconsistent indentation.

\section*{Grade C (50 - 59)}
\subsubsection*{Program Correctness.} Minor details of a task(s) are violated; the application functions correctly on the majority inputs/actions.

\subsubsection*{Code Elegance.} The application is implemented in such a way that many code segments (or functions) could be rewritten to increase code re-usability: some of the code is unnecessarily complex or poorly designed.

\subsubsection*{Documentation. } The documentation is simply comments embedded in the code but does not help the reader understand the code.

\subsubsection*{Readability. } The application has one or two issues that makes the program difficult to understand such as poorly named identifiers and disorganised code.

\section*{Grade D (40 - 49)}
\subsubsection*{Program Correctness.} Some tasks are incomplete; the application functions incorrectly on some inputs/ actions.

\subsubsection*{Code Elegance.} The application is unnecessarily complex and/or uses brute force; the application contains many instances where the code could have been written in an easier, faster or better fashion.

\subsubsection*{Documentation. } One or more code segments could benefit from comments or the code is overly commented.

\subsubsection*{Readability. } The application has more than two issues that makes the program difficult to understand.

\section*{Grade F (0 - 39)}
\subsubsection*{Program Correctness.} Significant details of a task are violated, or the program often exhibits incorrect behaviour; the applications does not open or run in RueJ; Java files are missing

\subsubsection*{Code Elegance.} The application is not organised for re-usability, e.g., all the code is in a single class (or function); no effort is made to create reusable code; the application is excessively long and poorly organised.

\subsubsection*{Documentation. } The application has no comments, not even the header with the student's name or id.

\subsubsection*{Readability. } The code is readable only by someone who knows what it is supposed do; the application has several issues that makes the program difficult to understand.

\section{Grades Per Assignment}\label{appx:grades_per_assignment}

This section shows all our results for the minimum, maximum and mean grade per group and per skill:
\begin{itemize}
    \item Figure \ref{fig:all_groups_correctness} shows the minimum, maximum and mean grades for each of the seven groups when grading correctness.
    \item The mean, maximum, and minimum grades for each of the seven groups when grading code elegance are displayed in Figure \ref{fig:all_groups_code_elegance}.
    \item Figure \ref{fig:all_groups_readability} shows the minimum, maximum and mean grades for each of the seven groups when grading readability.
    \item The mean, maximum, and minimum grades for each of the seven groups when grading documentation are displayed in Figure \ref{fig:all_groups_documentation}.
\end{itemize}
\def\figsize{0.35}
\begin{figure}[!htb]
    \begin{minipage}[t]{\figsize\textwidth}
         \centering
        \includegraphics[width=\linewidth]{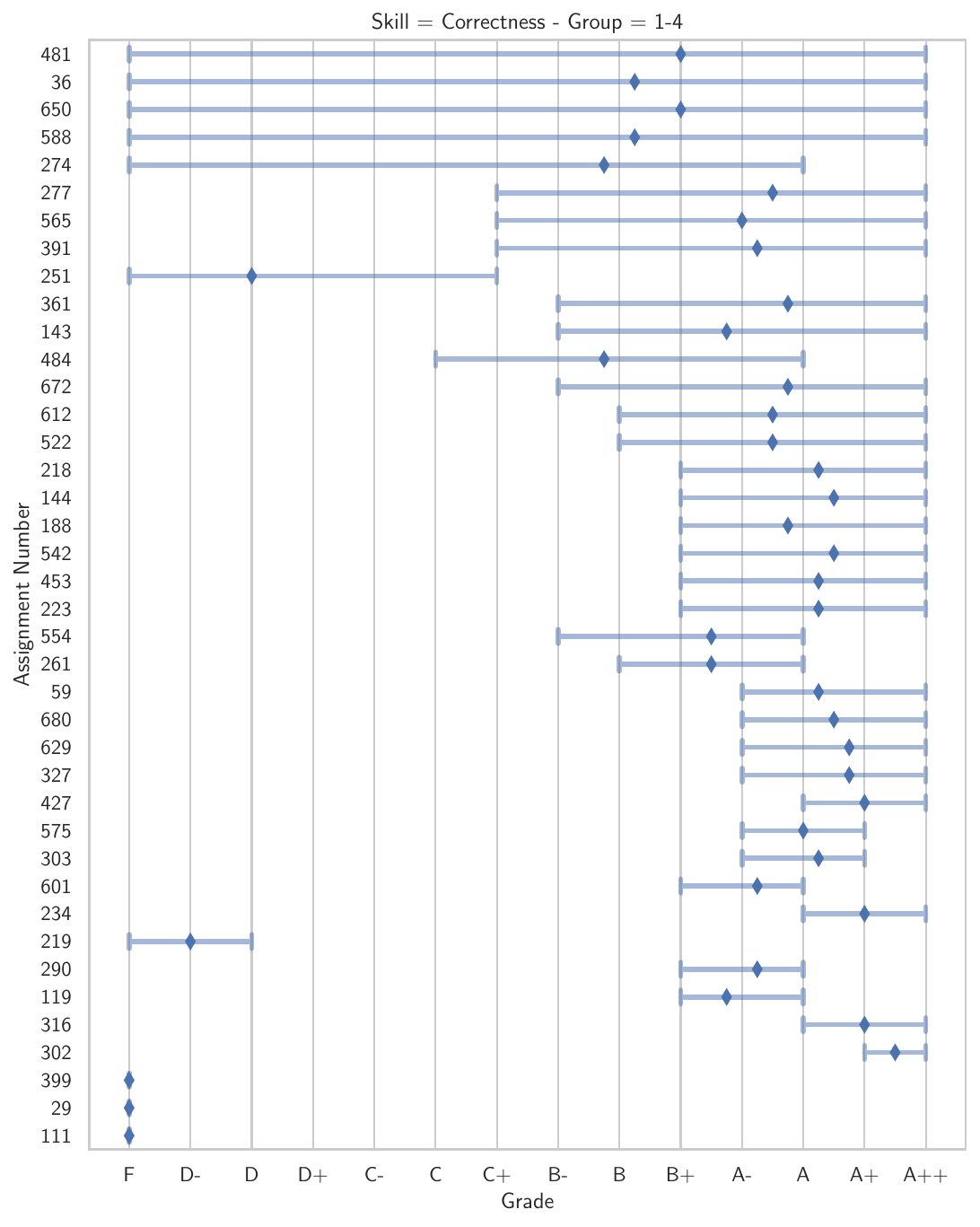}
    \end{minipage}%
    \begin{minipage}[t]{\figsize\textwidth}
       \centering
        \includegraphics[width=\linewidth]{figures/grade_per_assignment/grade_per_assignment_Correctness_5-8.pdf}
    \end{minipage}%
    \begin{minipage}[t]{\figsize\textwidth}
       \centering
        \includegraphics[width=\linewidth]{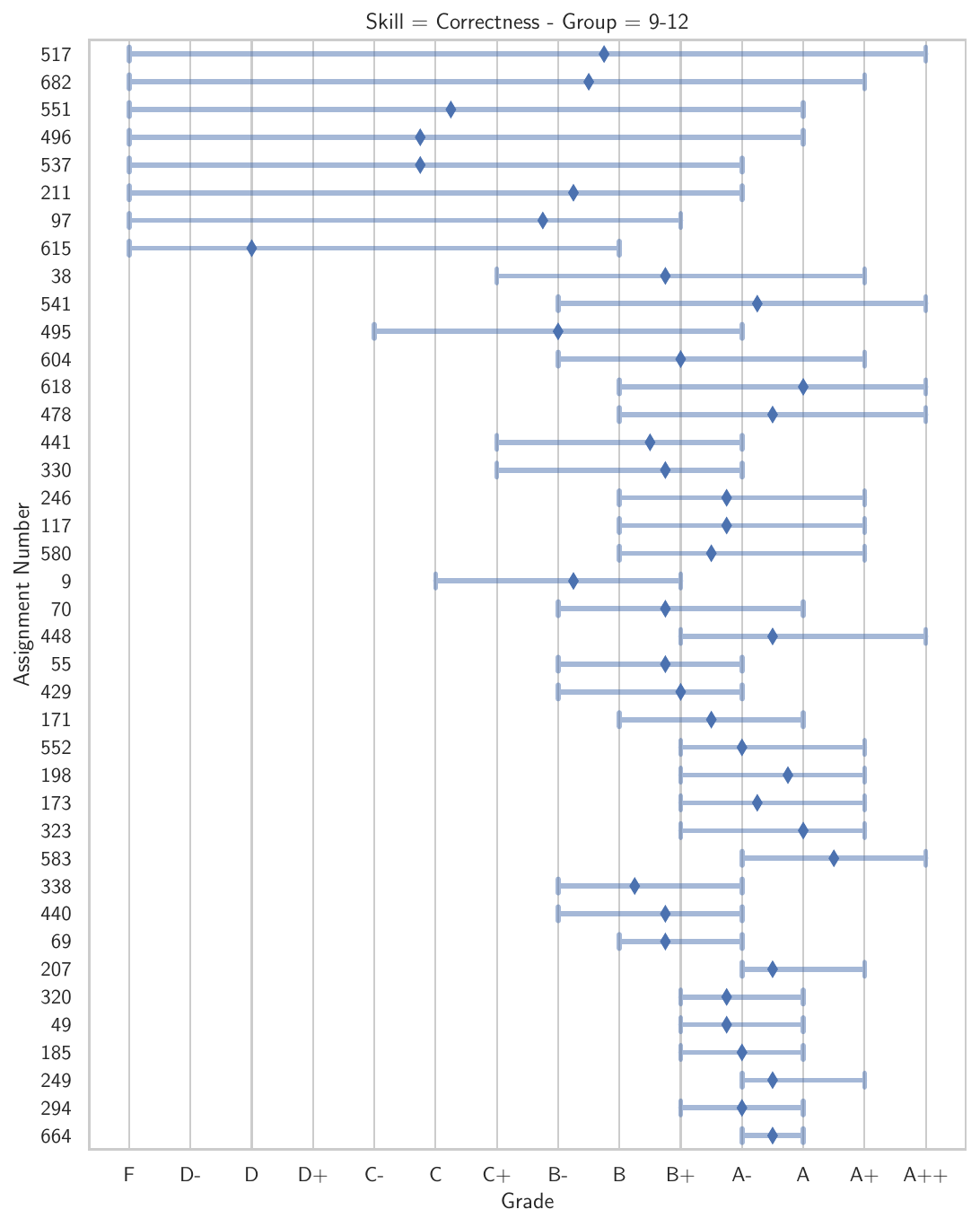}
    \end{minipage}
    \begin{minipage}[t]{\figsize\textwidth}
       \centering
        \includegraphics[width=\linewidth]{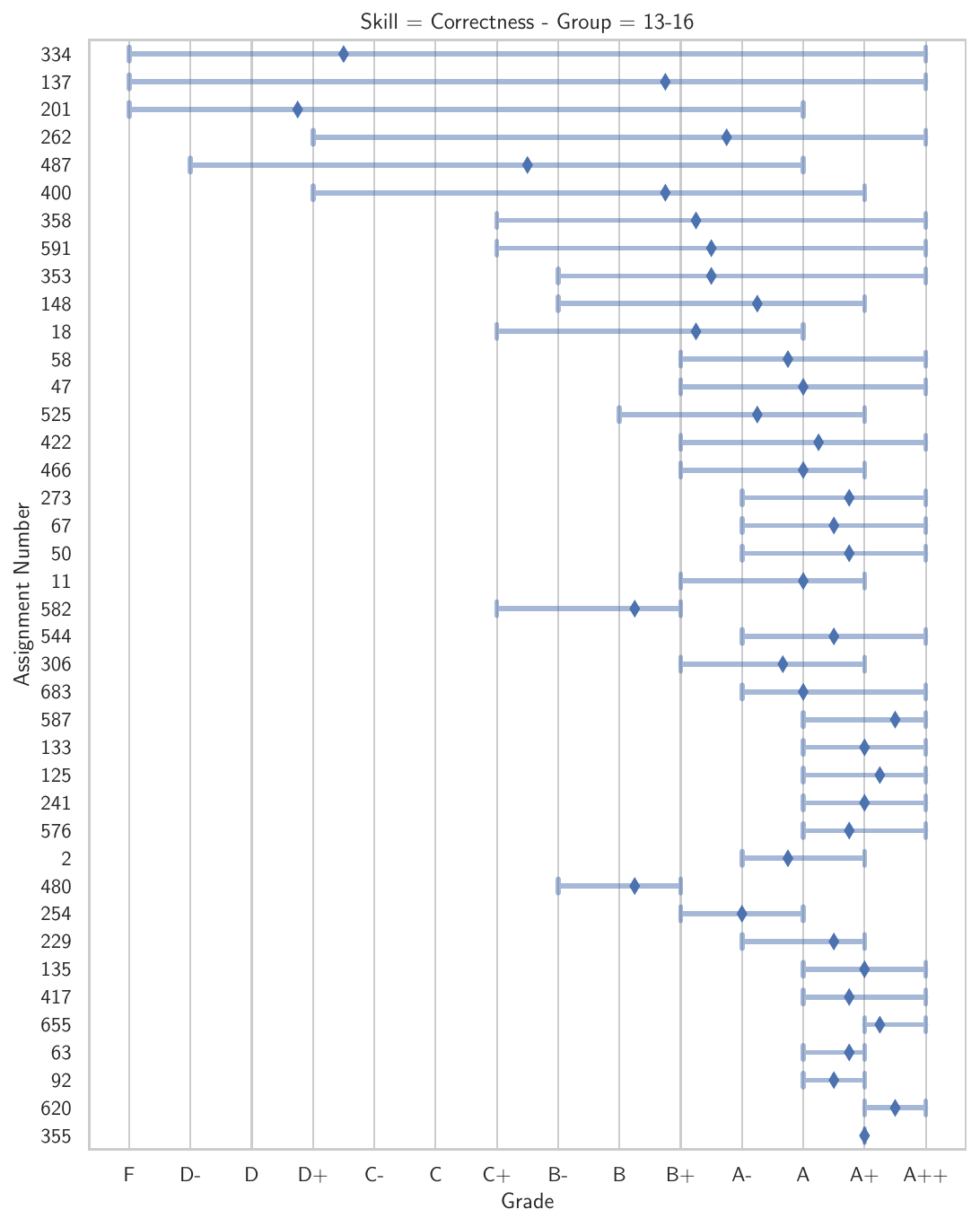}
    \end{minipage}%
    \begin{minipage}[t]{\figsize\textwidth}
       \centering
        \includegraphics[width=\linewidth]{figures/grade_per_assignment/grade_per_assignment_Correctness_17-20.pdf}
    \end{minipage}%
    \begin{minipage}[t]{\figsize\textwidth}
       \centering
        \includegraphics[width=\linewidth]{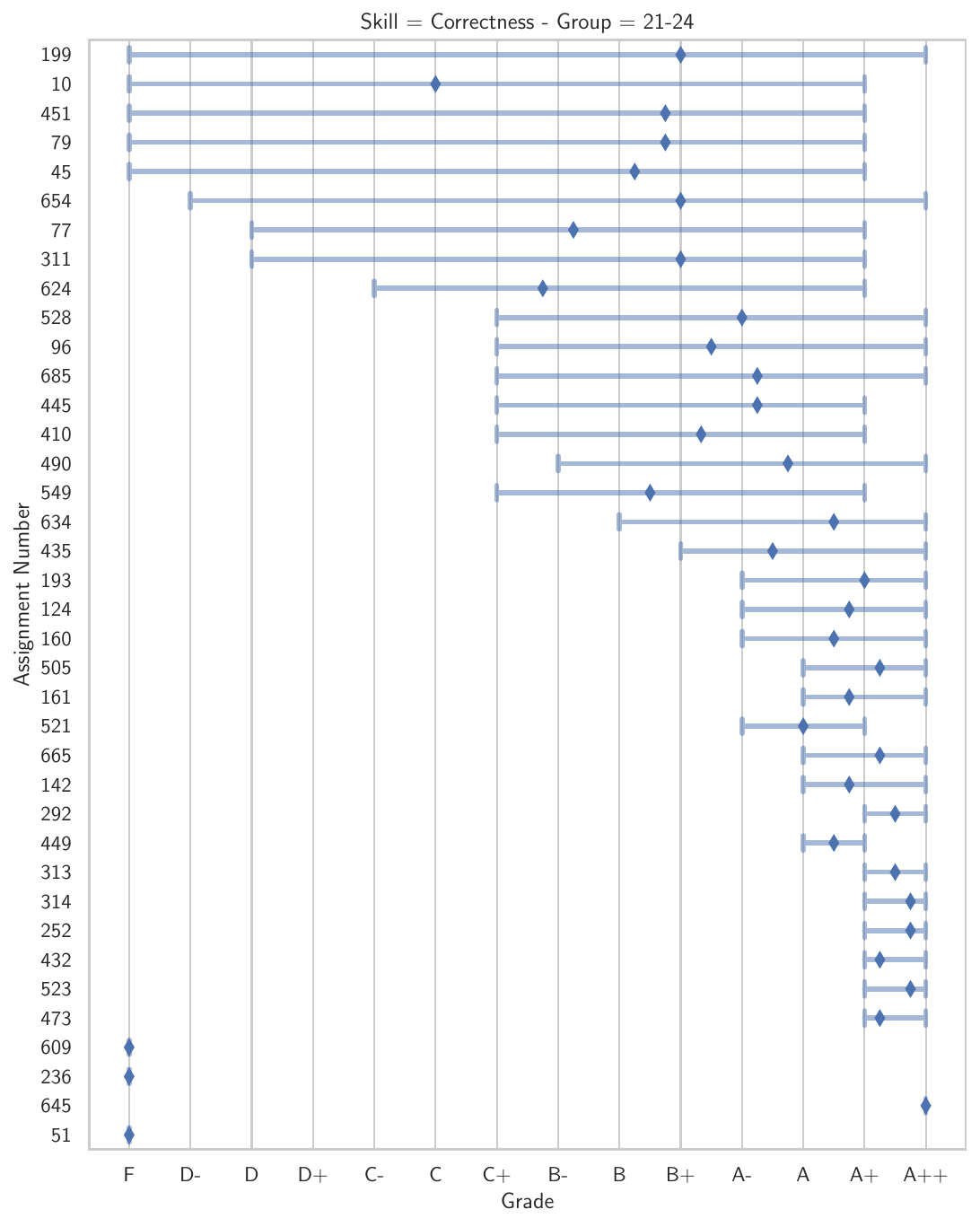}
    \end{minipage}
    \begin{minipage}[t]{\figsize\textwidth}
       \centering
        \includegraphics[width=\linewidth]{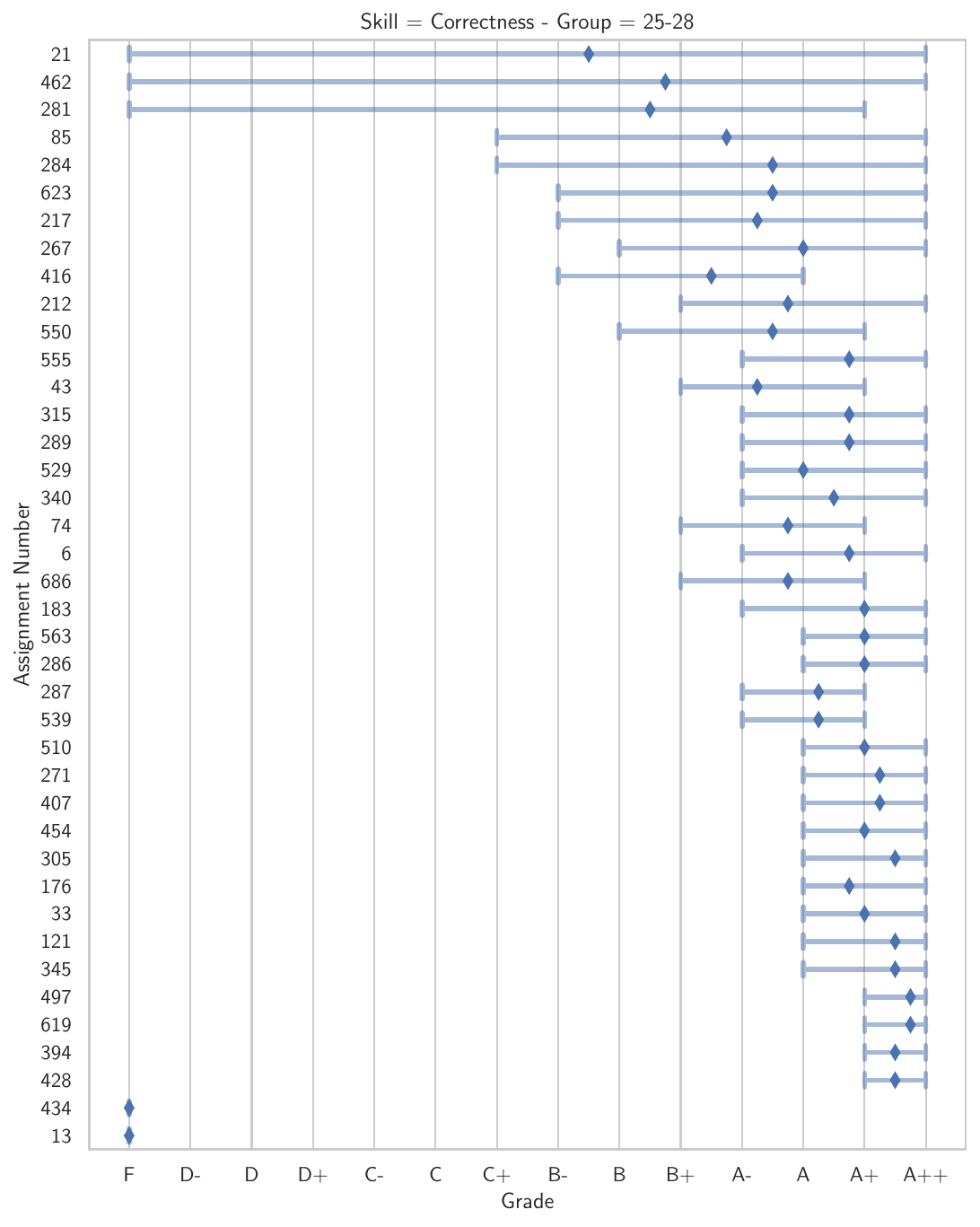}
    \end{minipage}
     \caption{These figures show the minimum, maximum and mean correctness grade awarded by the participants for each assignment.}
     \label{fig:all_groups_correctness}
\end{figure}

\begin{figure}[!htb]
    \begin{minipage}[t]{\figsize\textwidth}
         \centering
        \includegraphics[width=\linewidth]{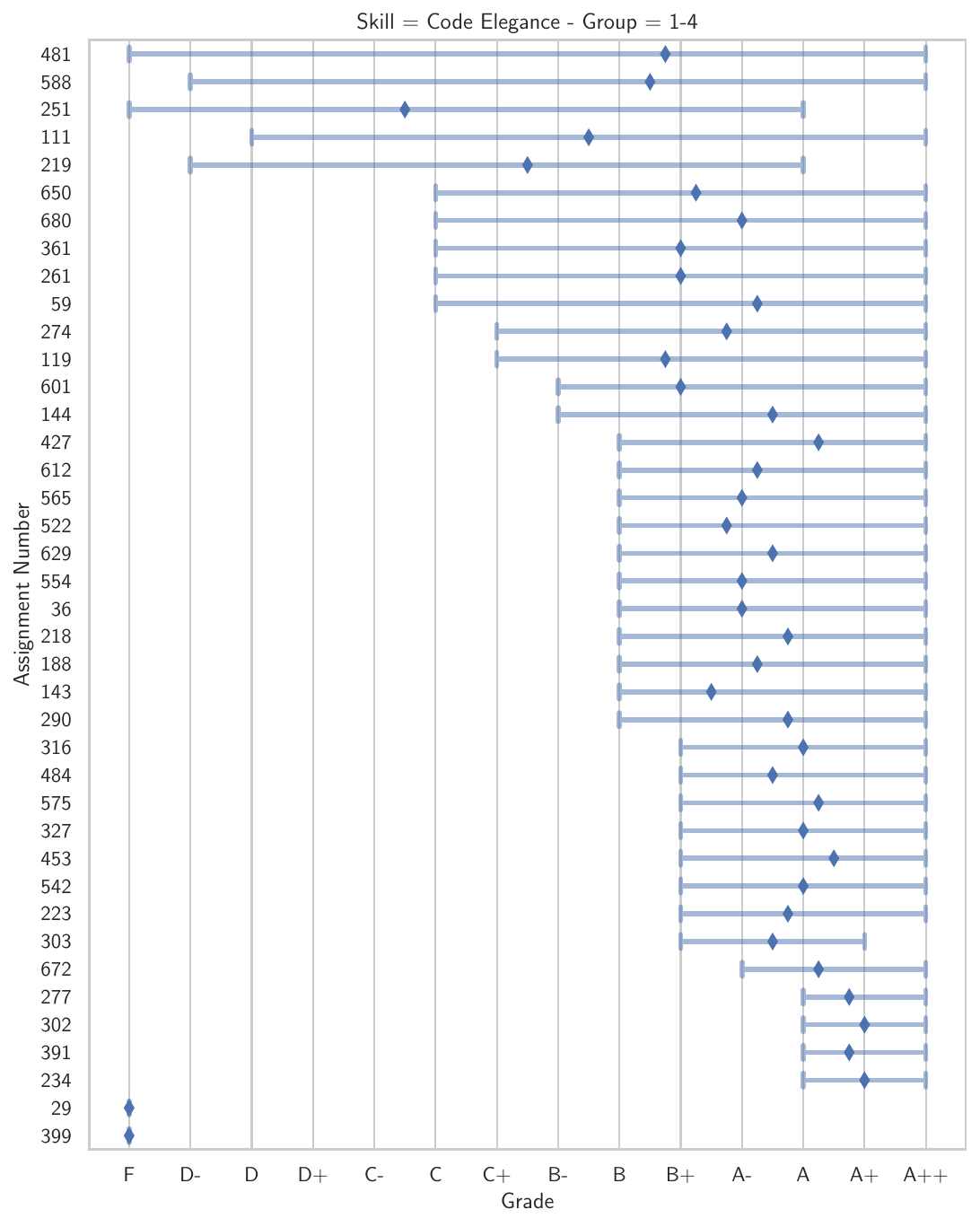}
    \end{minipage}%
    \begin{minipage}[t]{\figsize\textwidth}
       \centering
        \includegraphics[width=\linewidth]{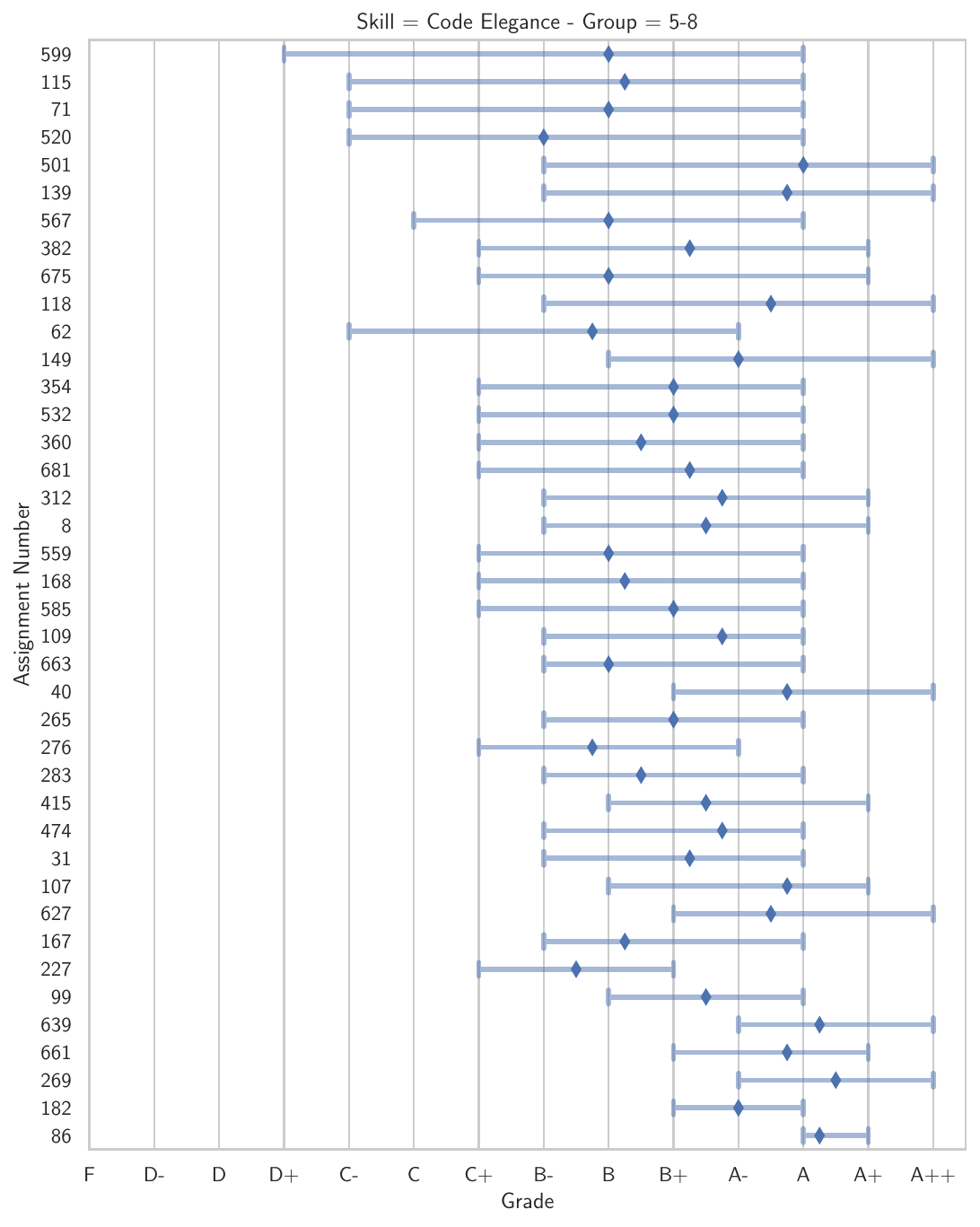}
    \end{minipage}%
    \begin{minipage}[t]{\figsize\textwidth}
       \centering
        \includegraphics[width=\linewidth]{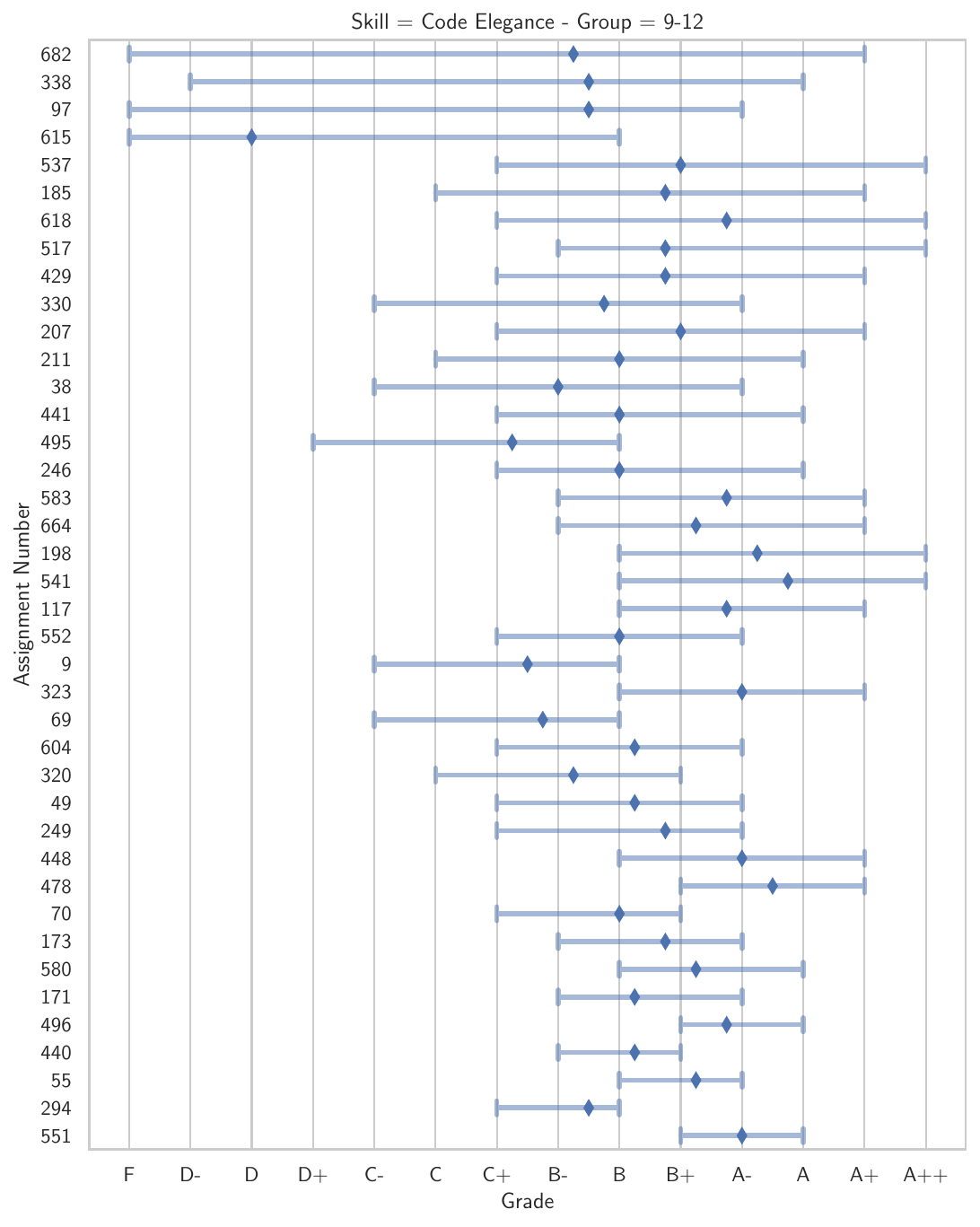}
    \end{minipage}
    \begin{minipage}[t]{\figsize\textwidth}
       \centering
        \includegraphics[width=\linewidth]{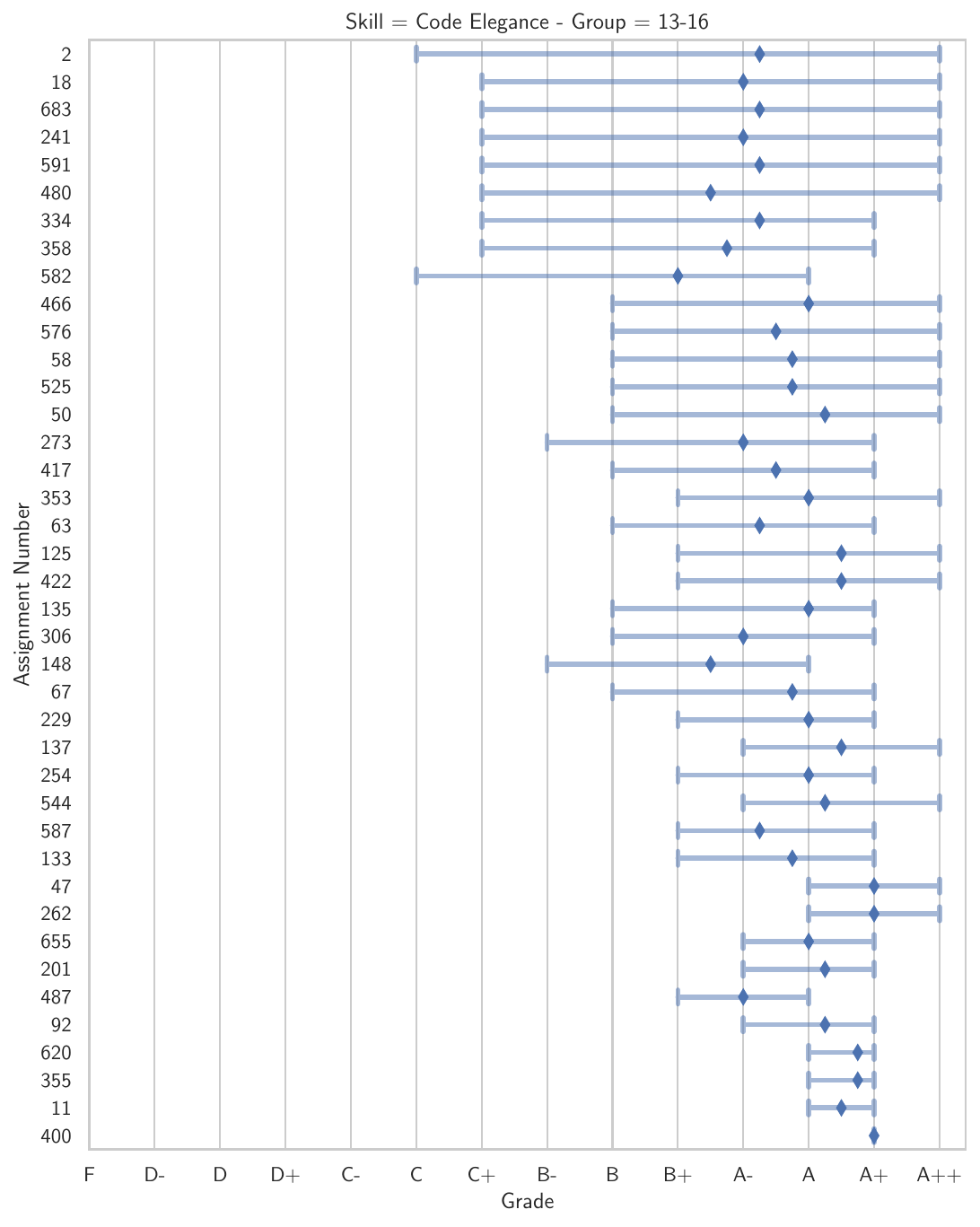}
    \end{minipage}%
    \begin{minipage}[t]{\figsize\textwidth}
       \centering
        \includegraphics[width=\linewidth]{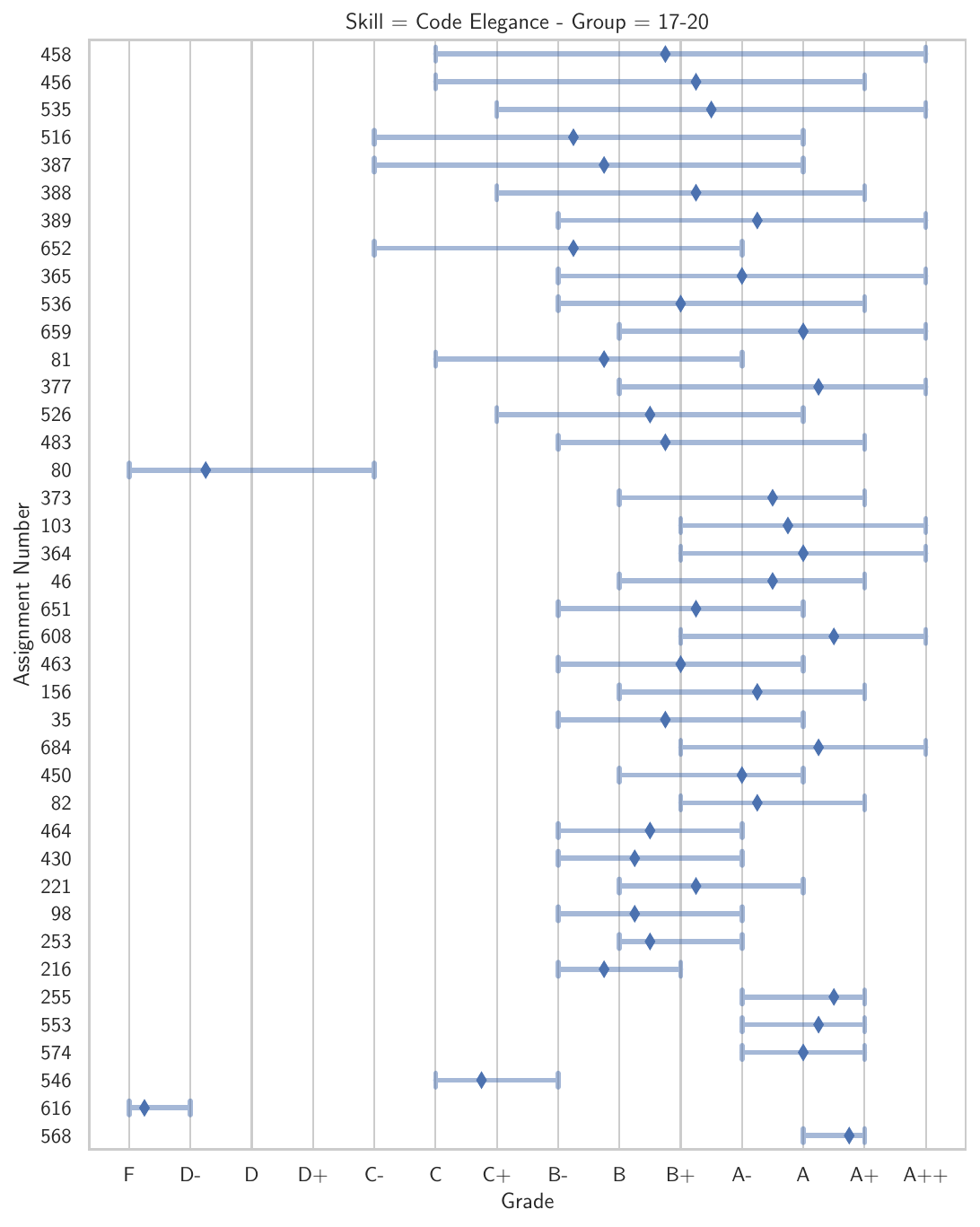}
    \end{minipage}%
    \begin{minipage}[t]{\figsize\textwidth}
       \centering
        \includegraphics[width=\linewidth]{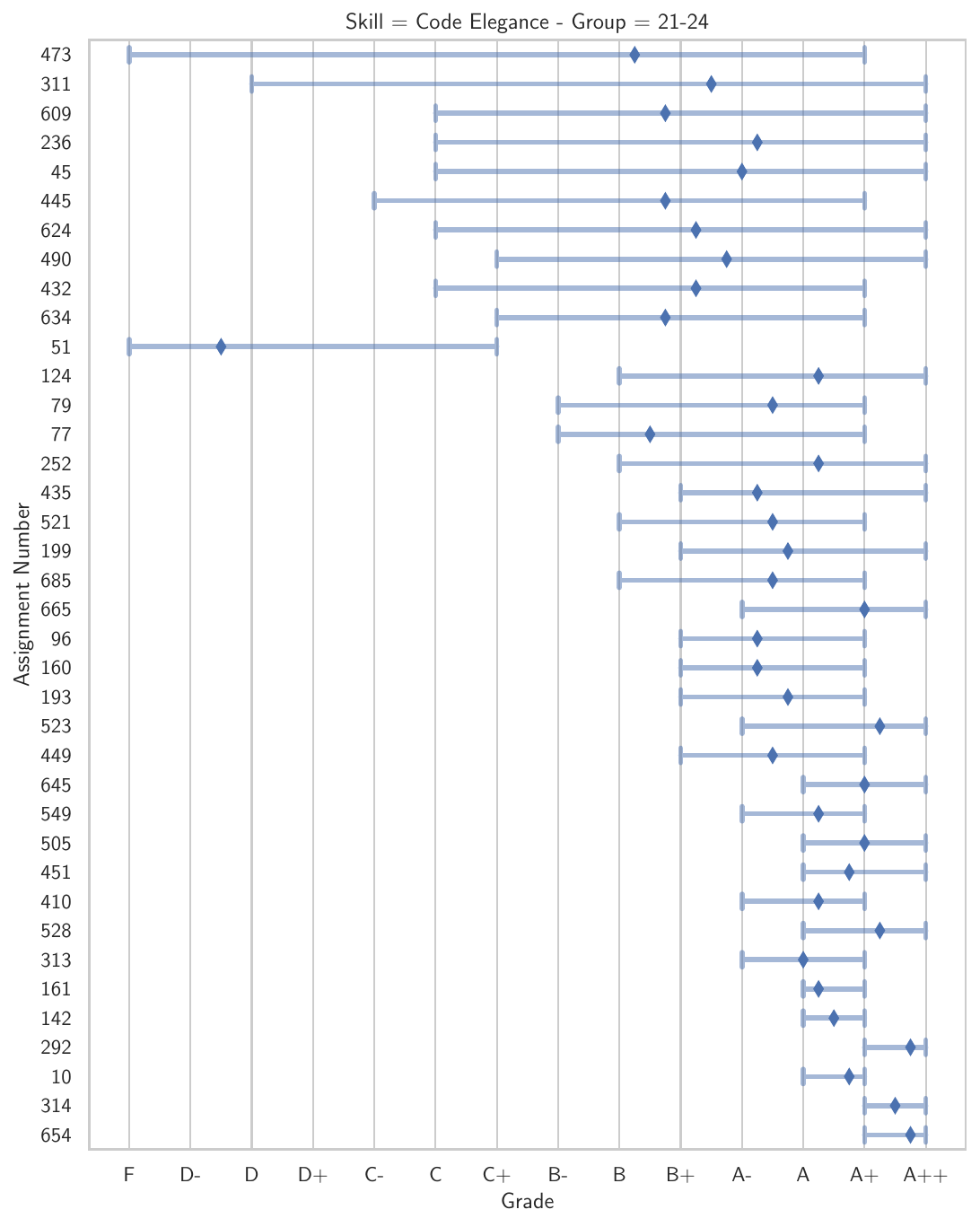}
    \end{minipage}
    \begin{minipage}[t]{\figsize\textwidth}
       \centering
        \includegraphics[width=\linewidth]{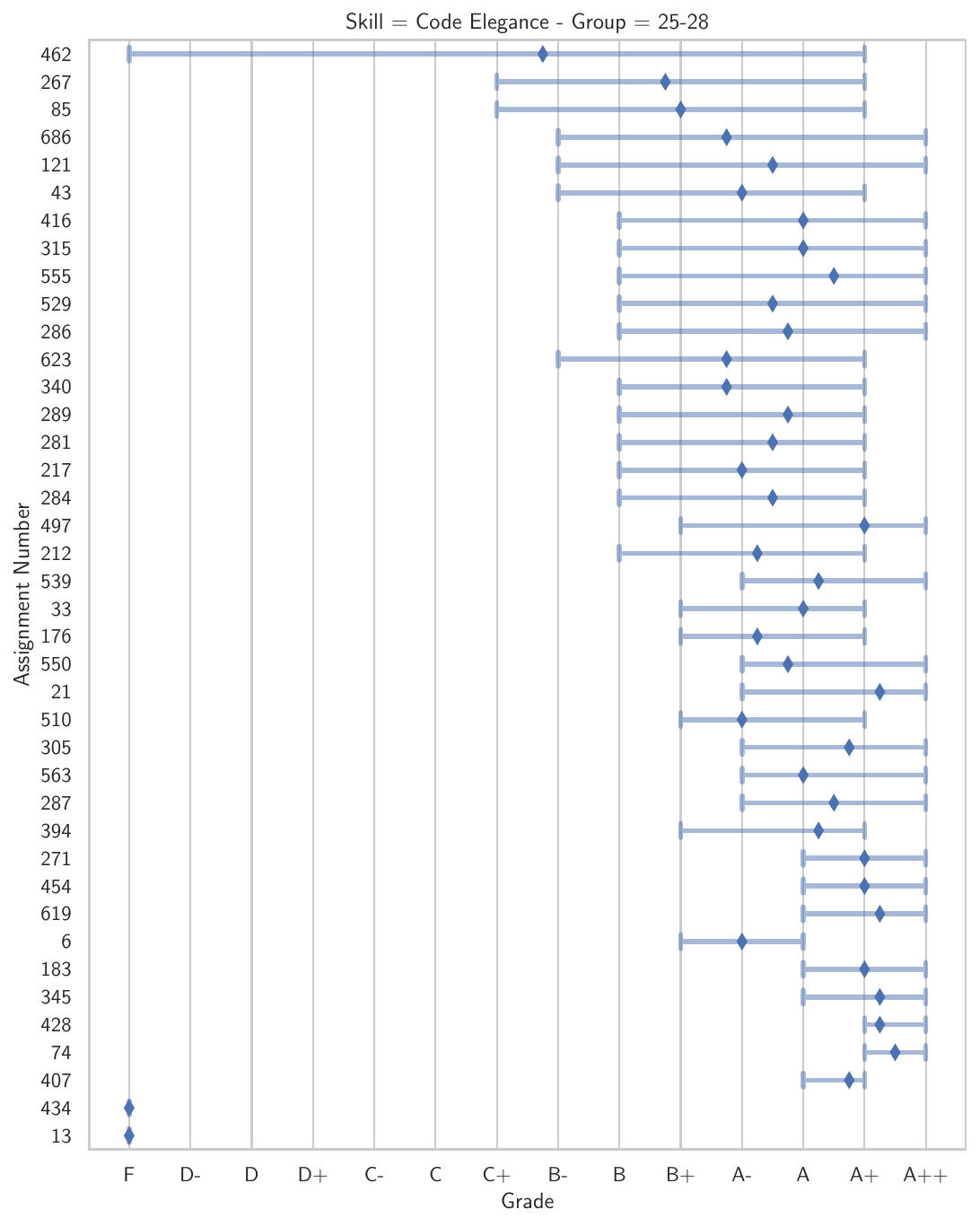}
    \end{minipage}
     \caption{These figures show the minimum, maximum and mean code elegance grade awarded by the participants for each assignment.}
     \label{fig:all_groups_code_elegance}
\end{figure}

\begin{figure}[!htb]
    \begin{minipage}[t]{\figsize\textwidth}
         \centering
        \includegraphics[width=\linewidth]{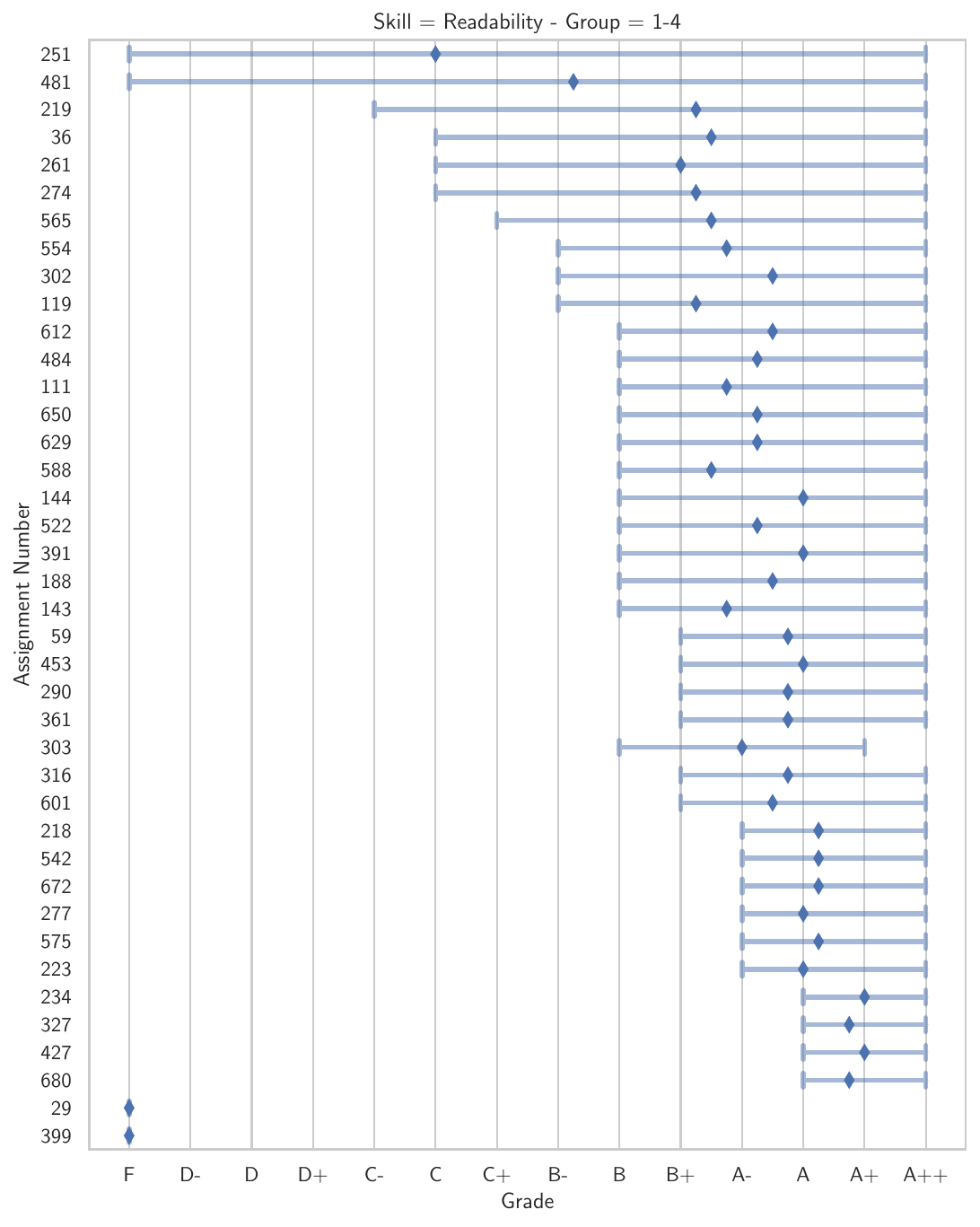}
    \end{minipage}%
    \begin{minipage}[t]{\figsize\textwidth}
       \centering
        \includegraphics[width=\linewidth]{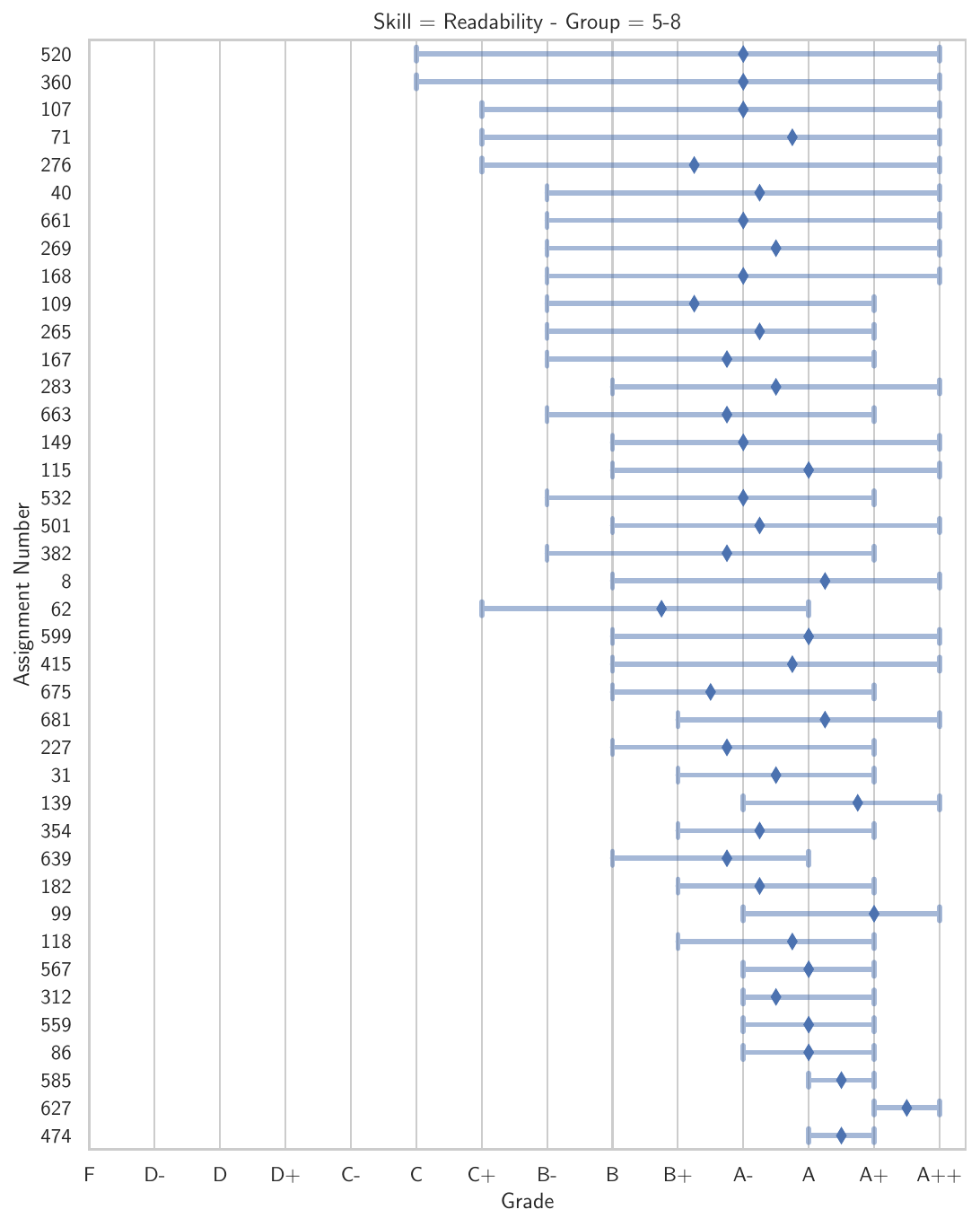}
    \end{minipage}%
    \begin{minipage}[t]{\figsize\textwidth}
       \centering
        \includegraphics[width=\linewidth]{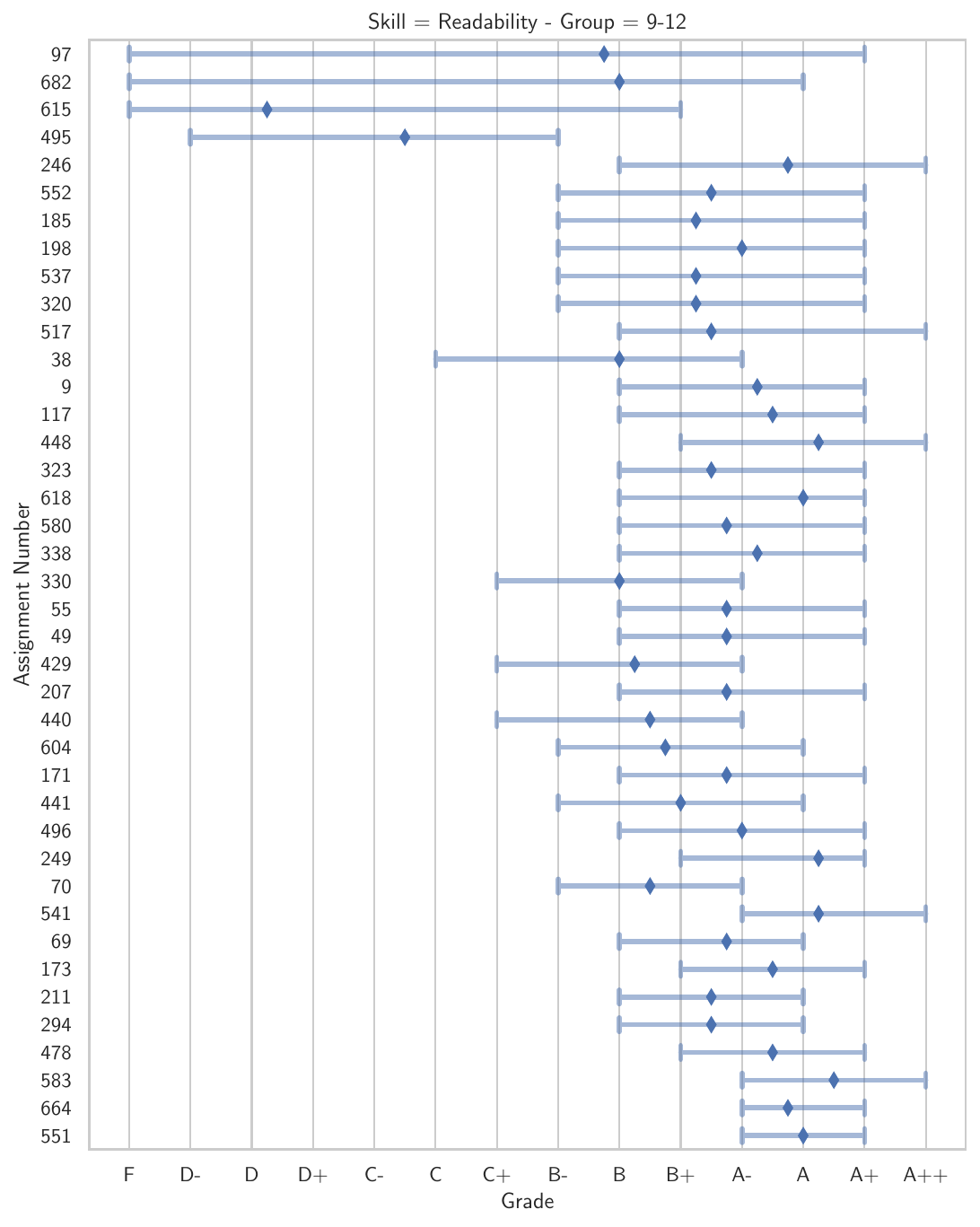}
    \end{minipage}
    \begin{minipage}[t]{\figsize\textwidth}
       \centering
        \includegraphics[width=\linewidth]{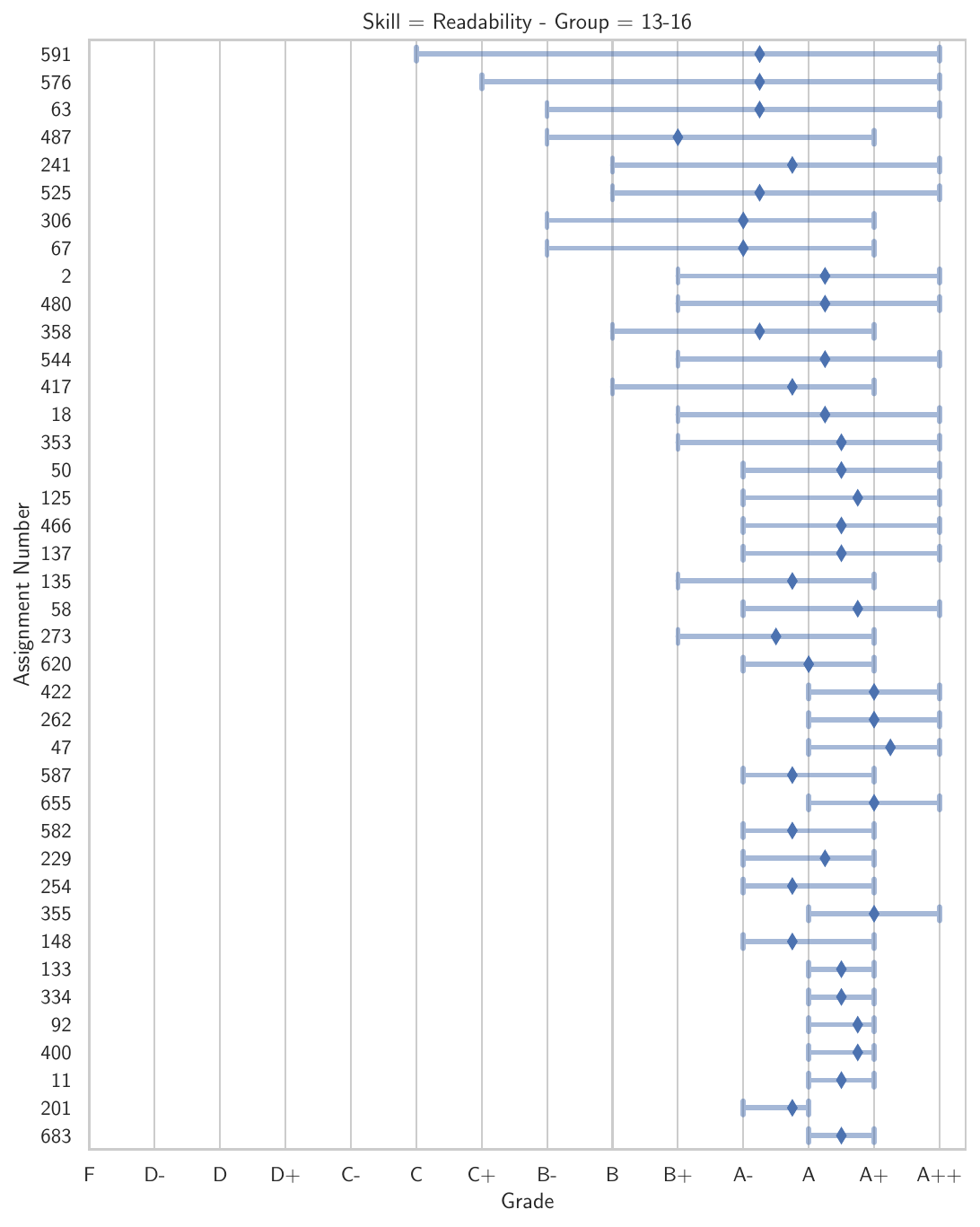}
    \end{minipage}%
    \begin{minipage}[t]{\figsize\textwidth}
       \centering
        \includegraphics[width=\linewidth]{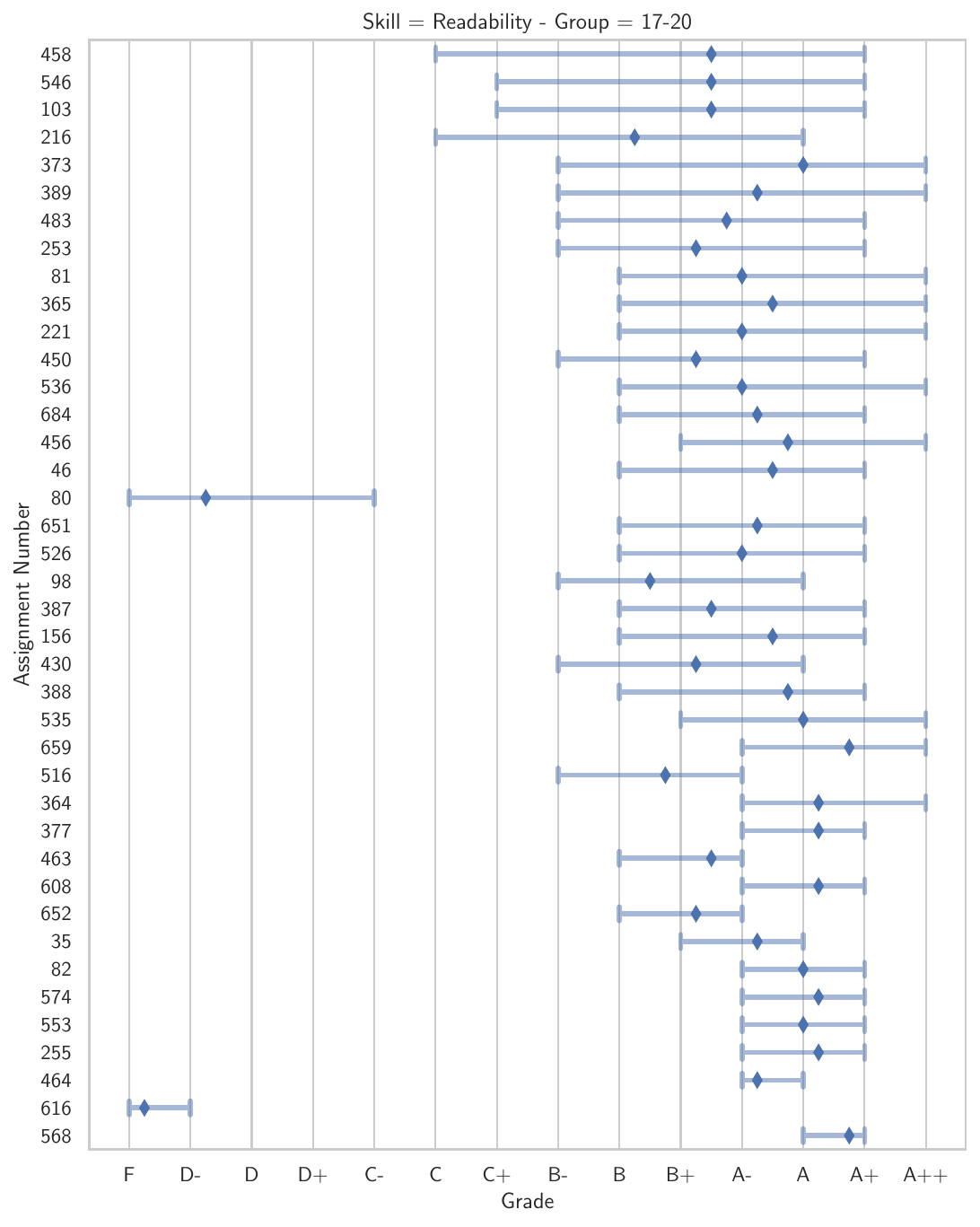}
    \end{minipage}%
    \begin{minipage}[t]{\figsize\textwidth}
       \centering
        \includegraphics[width=\linewidth]{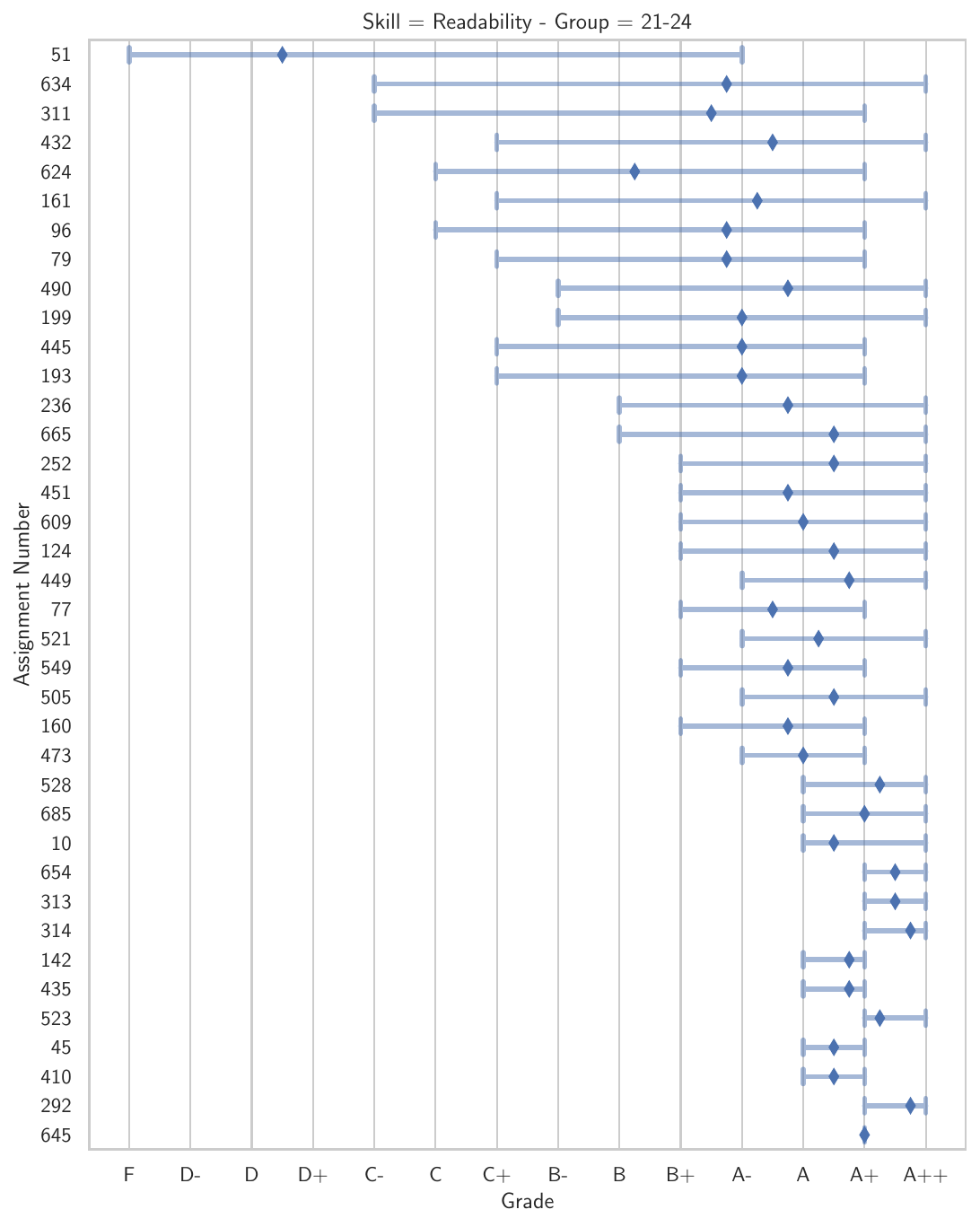}
    \end{minipage}
    \begin{minipage}[t]{\figsize\textwidth}
       \centering
        \includegraphics[width=\linewidth]{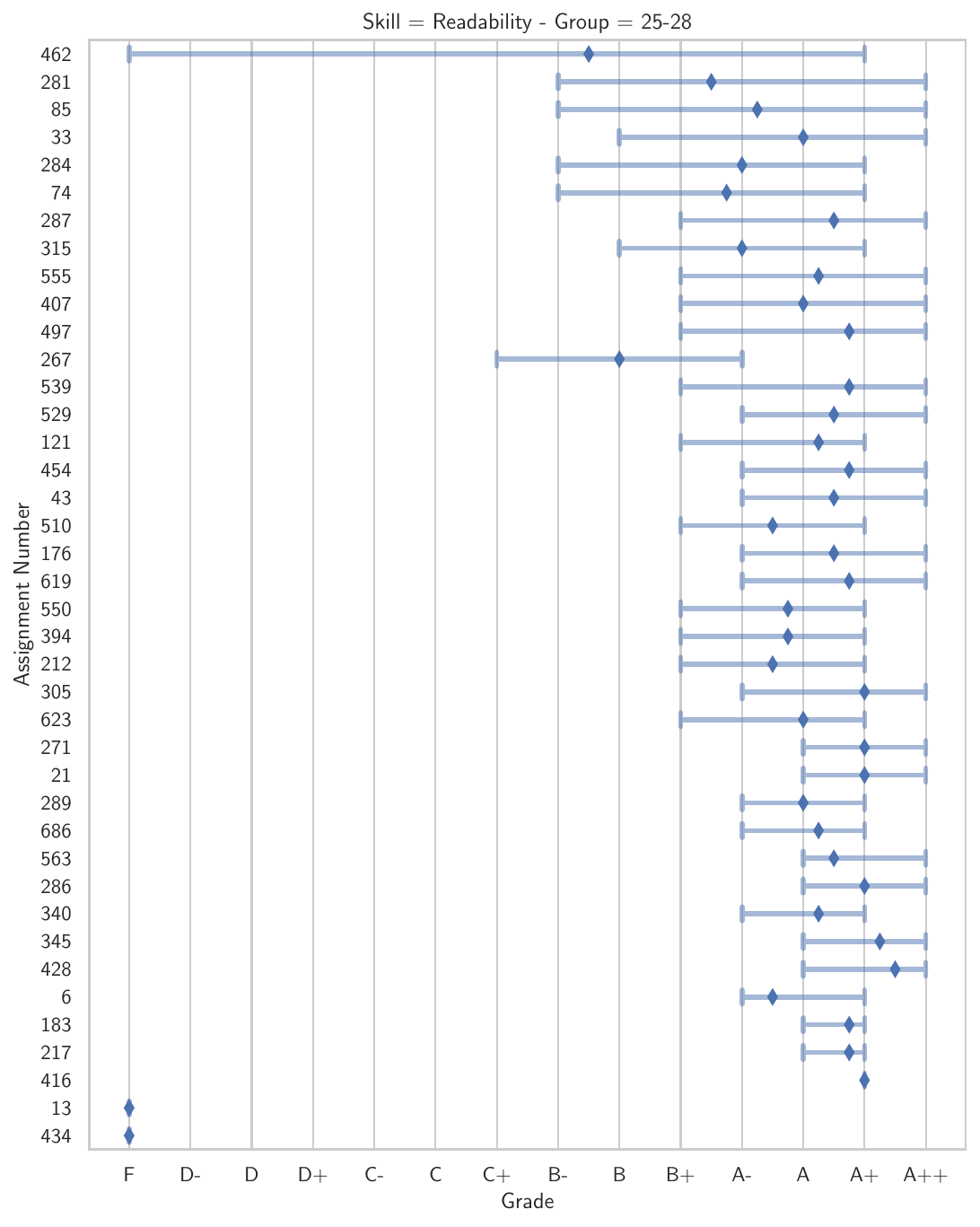}
    \end{minipage}
     \caption{These figures show the minimum, maximum and mean readability grade awarded by the participants for each assignment.}
    \label{fig:all_groups_readability}
\end{figure}

\begin{figure}[!htb]
    \begin{minipage}[t]{\figsize\textwidth}
         \centering
        \includegraphics[width=\linewidth]{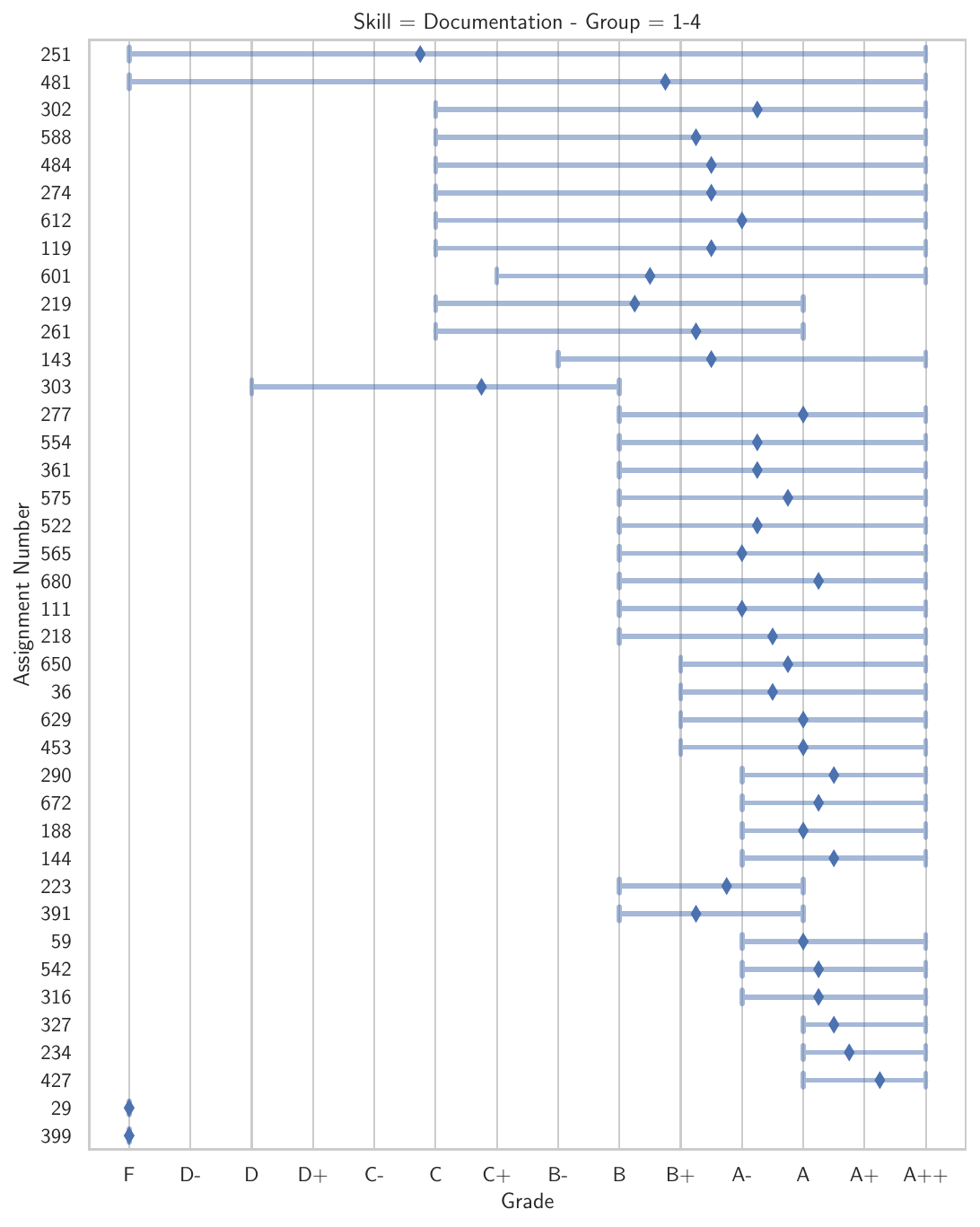}
    \end{minipage}%
    \begin{minipage}[t]{\figsize\textwidth}
       \centering
        \includegraphics[width=\linewidth]{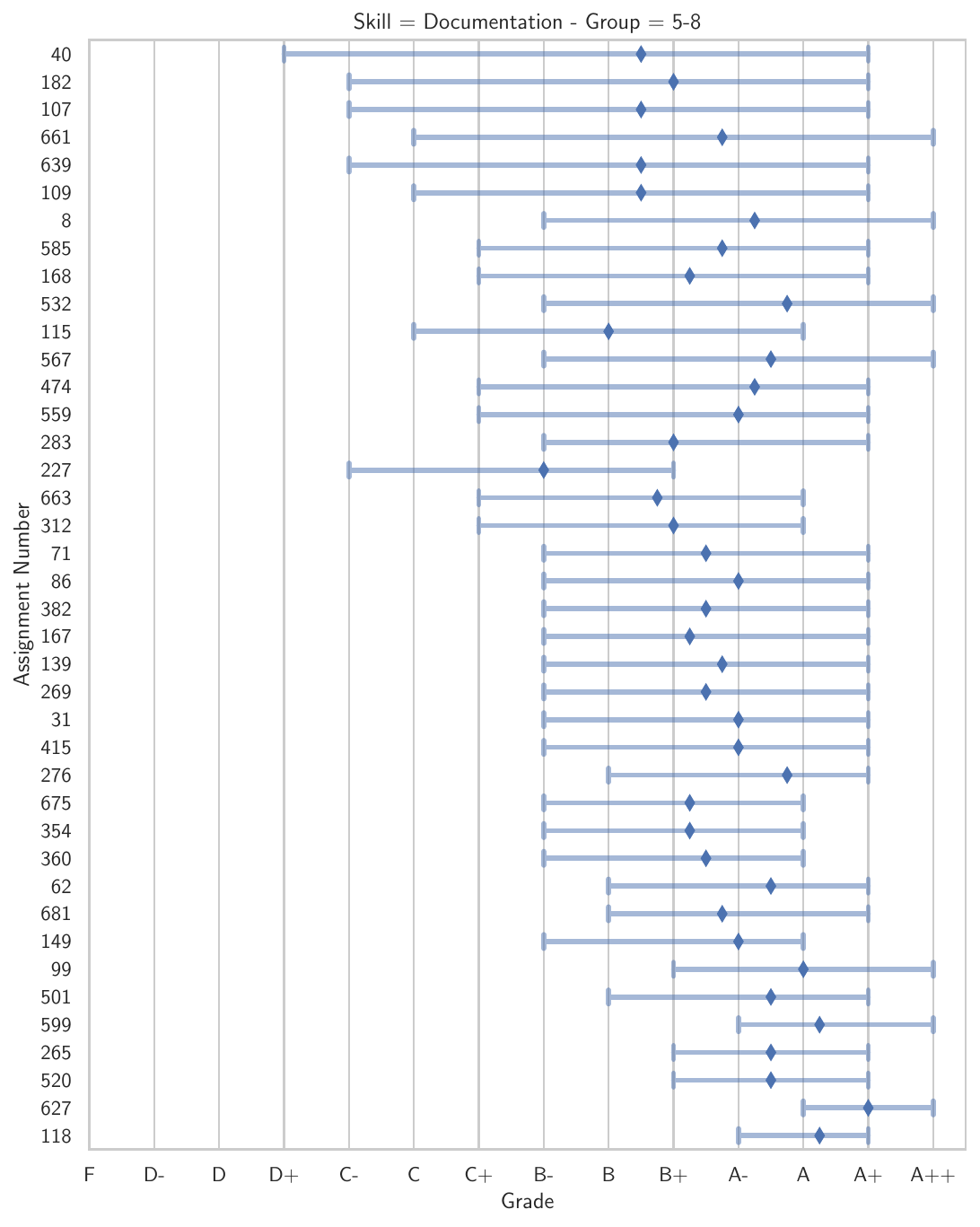}
    \end{minipage}%
    \begin{minipage}[t]{\figsize\textwidth}
       \centering
        \includegraphics[width=\linewidth]{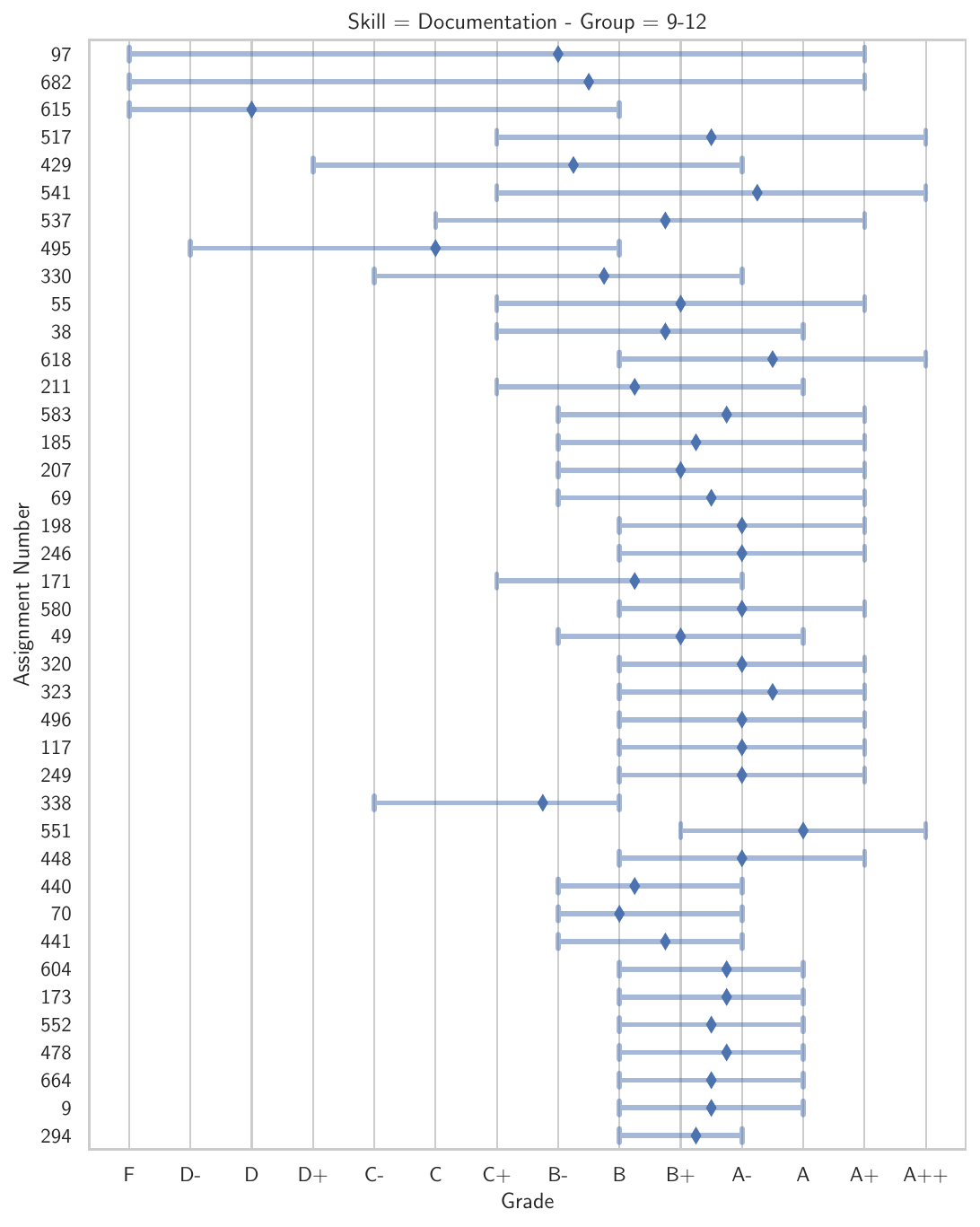}
    \end{minipage}
    \begin{minipage}[t]{\figsize\textwidth}
       \centering
        \includegraphics[width=\linewidth]{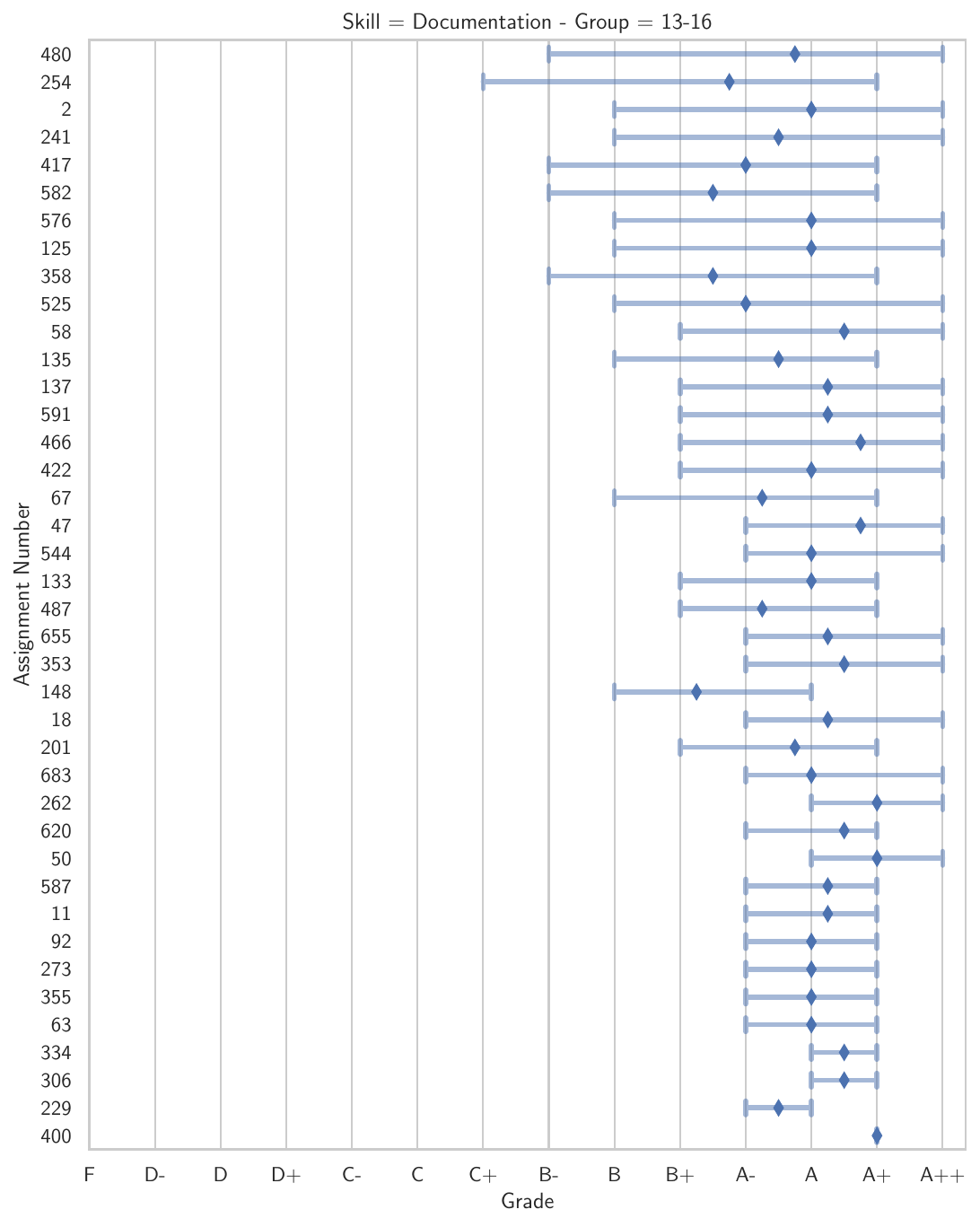}
    \end{minipage}%
    \begin{minipage}[t]{\figsize\textwidth}
       \centering
        \includegraphics[width=\linewidth]{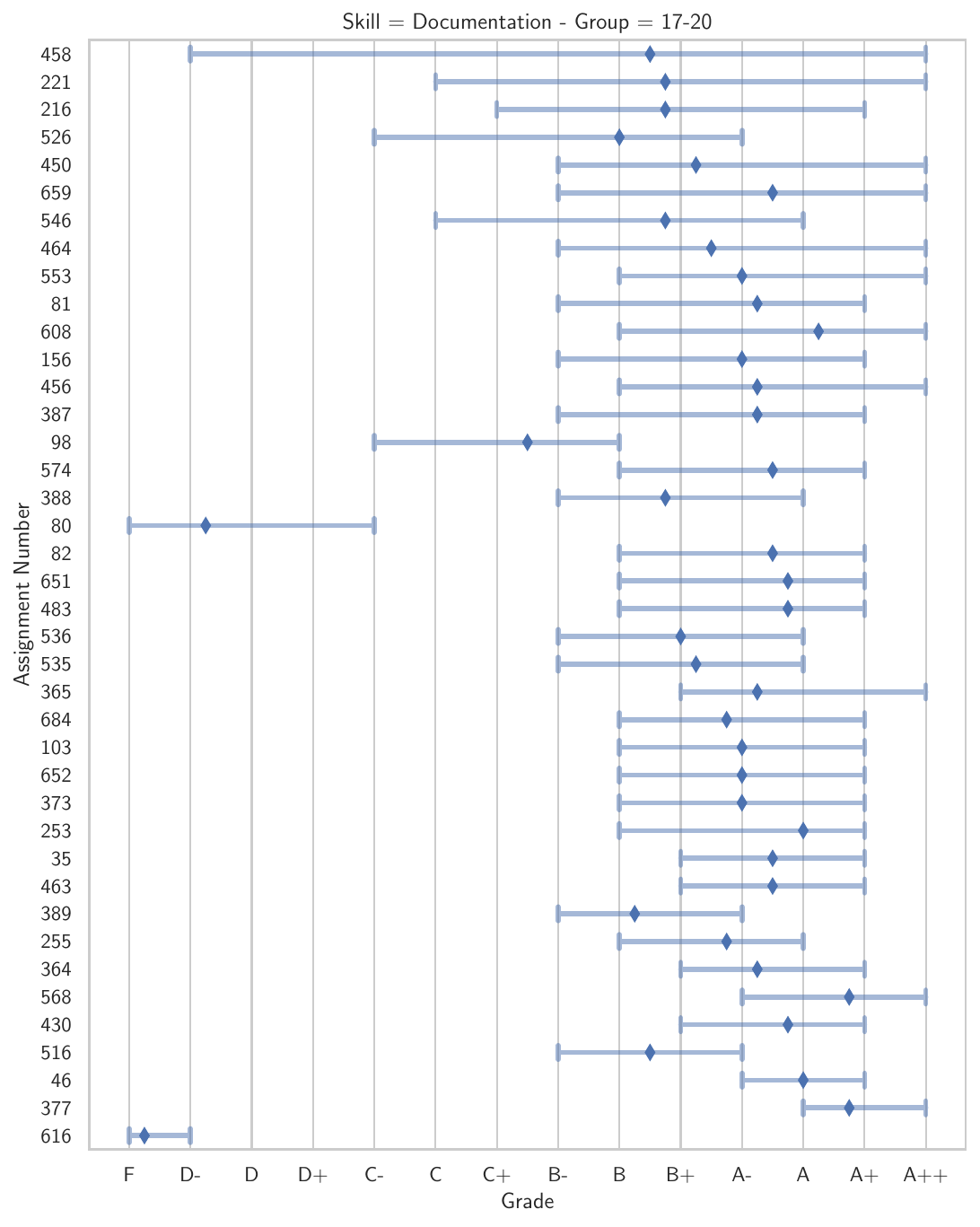}
    \end{minipage}%
    \begin{minipage}[t]{\figsize\textwidth}
       \centering
        \includegraphics[width=\linewidth]{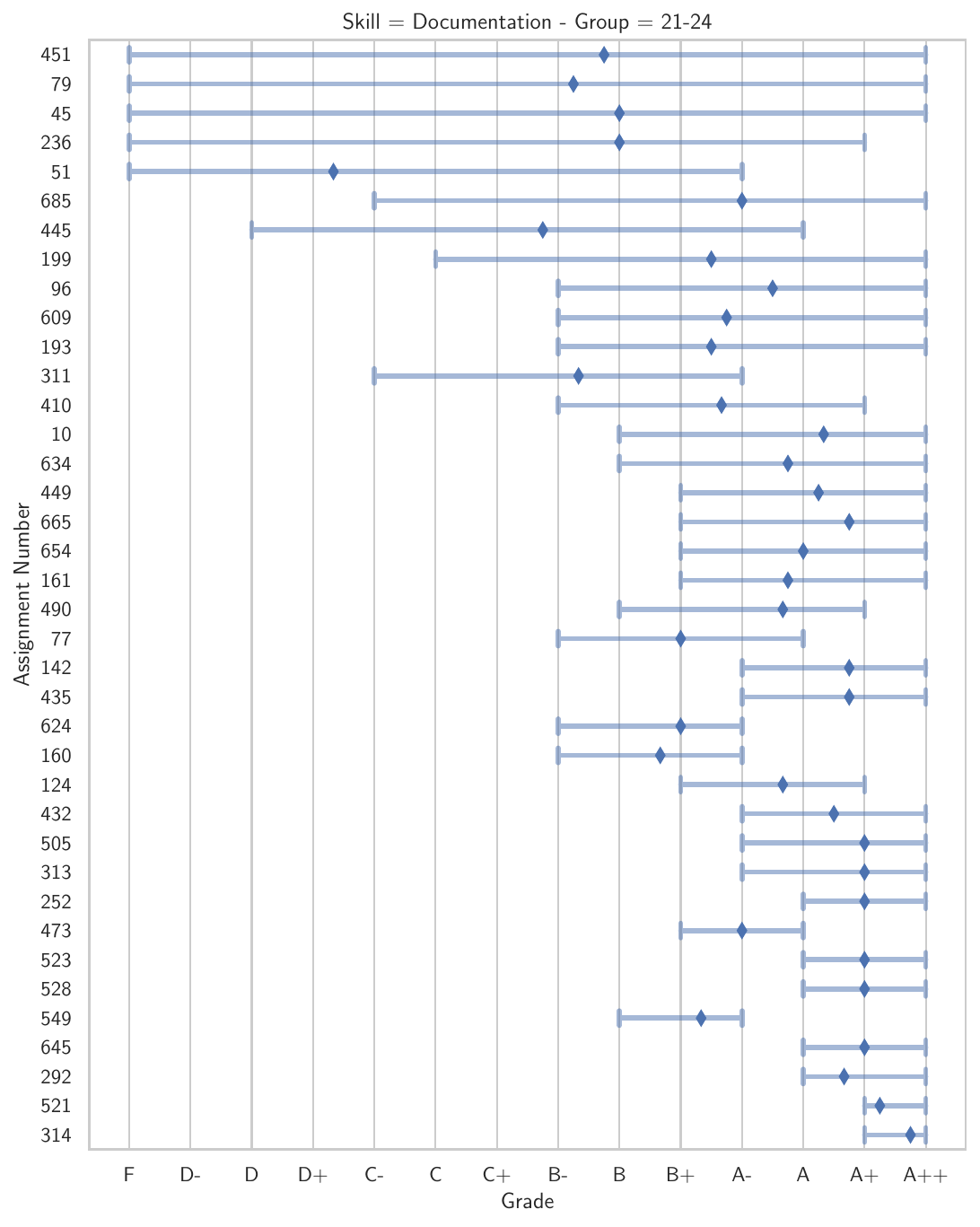}
    \end{minipage}
    \begin{minipage}[t]{\figsize\textwidth}
       \centering
        \includegraphics[width=\linewidth]{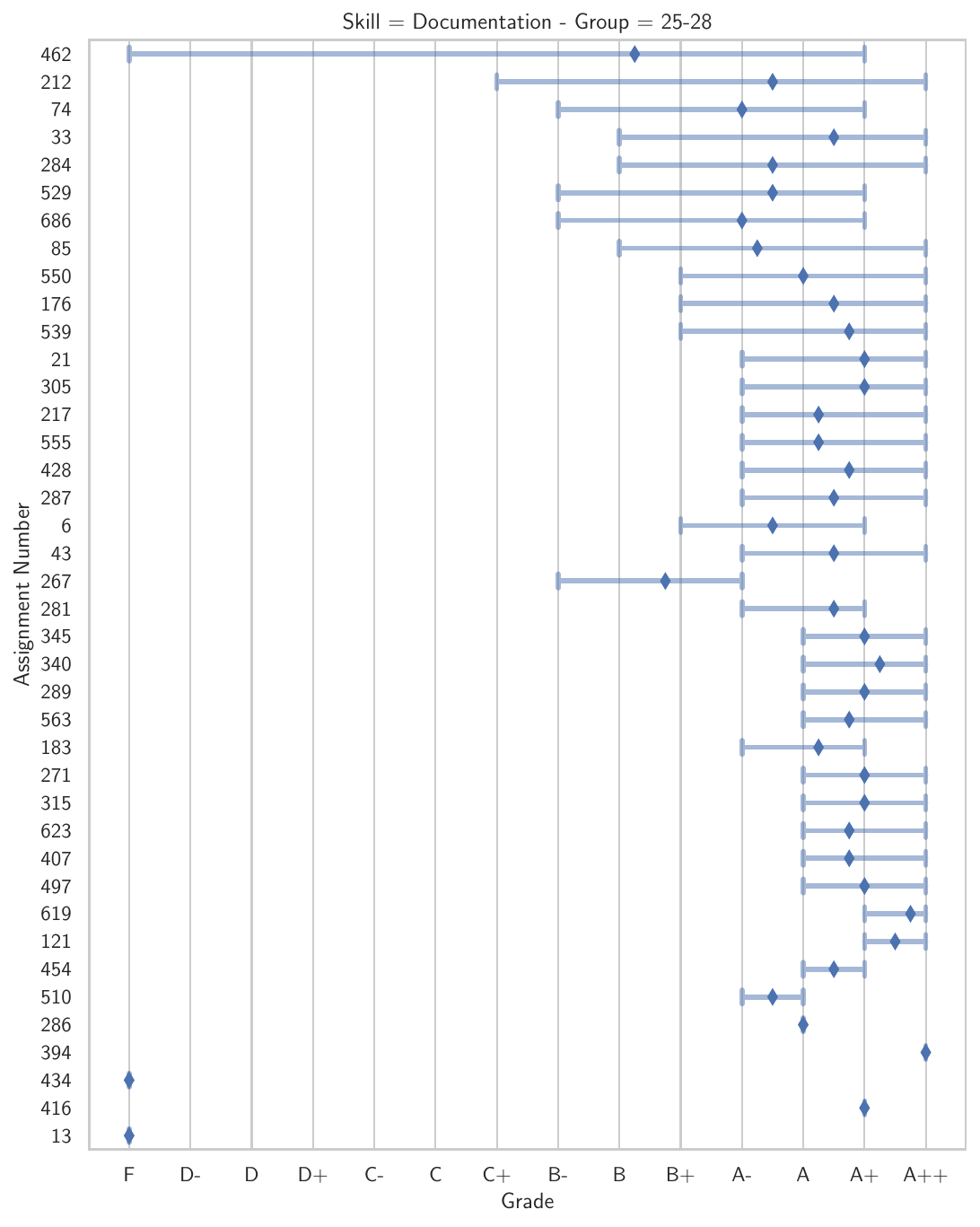}
    \end{minipage}
     \caption{These figures show the minimum, maximum and mean documentation grade awarded by the participants for each assignment.}
    \label{fig:all_groups_documentation}
\end{figure}

\end{document}